\documentclass[10pt]{article}
\input amssym.def
\input amssym.tex
\usepackage{mathrsfs}
\usepackage{enumerate}

\begin{document}

\hfuzz=100pt
\title{{\bf Deformation of Codimension-2 Surface and Horizon Thermodynamics}}
\author{Li-Ming Cao\footnote{e-mail address:
caolm@phys.kindai.ac.jp, caolm@itp.ac.cn}
\\
Department of Physics, Kinki University,
\\ Higashi-Osaka, Osaka
577-8502, Japan,} \maketitle

\begin{abstract}

The deformation equation of a spacelike submanifold with an
arbitrary codimension is given by a general construction without
using local frames. In the case of codimension-1, this equation
reduces to the evolution equation of the extrinsic curvature of a
spacelike hypersurface. In the more interesting case of
codimension-2, after selecting a local null frame, this deformation
equation reduces to the well known (cross) focusing equations. We
show how the thermodynamics of trapping horizons is related to these
deformation equations in two different formalisms: with and without
introducing quasilocal energy. In the formalism with the quasilocal
energy, the Hawking mass in four dimension is generalized to higher
dimension, and it is found that the deformation of this energy
inside a marginal surface can be also decomposed into the
contributions from matter fields and gravitational radiation as in
the four dimension. In the formalism without the quasilocal energy,
we generalize the definition of slowly evolving future outer
trapping horizons proposed by Booth to past trapping horizons. The
dynamics of the trapping horizons in FLRW universe is given as an
example. Especially, the slowly evolving past trapping horizon in
the FLRW universe has close relation to the scenario of slow-roll
inflation. Up to the second order of the slowly evolving parameter
in this generalization, the temperature (surface gravity) associated
with the slowly evolving trapping horizon in the FLRW universe is
essentially the same as the one defined by using the quasilocal
energy.

\end{abstract}

\vspace{.3cm}
\newpage

\section{Introduction}
Quantum mechanics together with general relativity (GR) predicts
that black hole behaves like a black body, emitting thermal
radiations, with a temperature proportional to its surface gravity
at the black hole horizon and with an entropy proportional to its
horizon area~\cite{Bek,Haw}. The Hawking temperature and the horizon
entropy together with the black hole mass obey the first law of
black hole thermodynamics~\cite{BCH}. Since these seminal works in
the 1970s, the relation among thermodynamics, quantum theory and
spacetime geometry has been widely discussed, and recent status  can
be found in a nice review~\cite{Wald:1999vt}.

Most of the studies of the black hole thermodynamics have been
focused on the event horizons of stationary black holes. For
example, Kerr-Newmann solution family in GR. However, this kind of
horizon strongly depends on the global structure of the spacetime.
For example, to define an event horizon, we have to know the future
null infinity of the spacetime. The so called apparent horizon
defined by Hawking~\cite{HawkingEllis} also depends on the slicing
(or $3+1$ decomposition) of the spacetime. To classify the two
surfaces embedded in some slice (for example, trapped, untrapped and
marginal trapped) of the spacetime, of course, one has to study the
extrinsic properties of the two surfaces embedded in this slice. To
define the apparent horizon (a hypersurface), one has to repeat this
classification in each slice of the spacetimes. Recent years, based
on the Hawking's definition of the apparent horizon, people have
given some definitions of the so called quasilocal horizons, see a
review~\cite{Ashtekar:2004cn}. Further, from a more broad view, some
local defined horizon has also been proposed by Jacobson {\it et
al}, and profound connection between gravitation and thermodynamics
has been revealed~\cite{Jac,Jac1}.

The pioneer work on the quasilocal horizon is the {\it trapping
horizon} defined by Hayward more than ten years ago~\cite{Hayward,
Hayward1, Hayward2}. Roughly speaking, this kind of horizon is a
hypersurface foliated by marginal surfaces of the spacetime.
However, here, the so called marginal is different from the one
given by Hawking: The former is a two dimension surface embedded in
the spacetime ~\cite{Penrose:1964wq}, while the later is a kind of
two dimension surface embedded in some slice of the spacetime. The
trapping horizon can be null, spacelike or timelike according to
different spacetime structures. At the end of 1998, Ashtekar {\it et
al} proposed a new horizon which is called {\it isolated
horizon}~\cite{Ashtekar:1998sp,Ashtekar:1999wa,Ashtekar:1999yj,Ashtekar:2000hw,Ashtekar:2000sz}.
This kind of horizon is null, and does not exchange matter and
energy with it's environment. About two years later, {\it dynamical
horizon} is also proposed by Ashtekar and
Krishnan~\cite{Ashtekar:2002ag,Ashtekar:2003hk}.  It's a spacelike
hypersurface and dynamically evolves. Another interest object is the
so called {\it slowly evolving horizon} developed by Booth {\it et
al}~\cite{Booth:2003ji,Kavanagh:2006qe,Booth:2006bn}. In this
proposal, the marginal surface (and associated physical quantities)
slowly evolves on the future trapping horizon (In fact, in this
theory, one can also consider the event
horizon~\cite{Booth:2006bn}). The slowly evolving horizon describes
some near equilibrium state of the thermodynamics of the spacetime.

To study the dynamics of the quasilocal horizon, usually, there are
two formalisms: One of them heavily depend on some quasilocal energy
inside a given two dimension surface. By using Einstein equation and
the quasilocal energy, one may directly gets some first law like
equation. Actually, to study the dynamics of the trapping horizon in
general spherically symmetric spacetime, Hayward has translated the
Einstein equation into a simple first law like equation by selecting
the quasilocal energy to be Misner-Sharp
energy~\cite{Misner:1964je}. Without the spherical symmetry, the
problem becomes complicated. Based on the Hawking mass
(energy)~\cite{Hawking:1968} and  focusing and cross focusing
equations, the dynamics of the trapping horizon is also studied by
Hayward {\it et
al}~\cite{Hayward3,Hayward:2004dv,Hayward:2004fz,Bray:2006pz}. In
this formalism, the quasilocal energy plays a key role, while the
(cross) focusing equations link the variation of the quasilocal
energy, matter fields and some possible gravitational radiation
together.  Another formalism is independent of the quasilocal
energy. In some sense, the most important implement in this method
is the (cross) focusing  equations (or generalized
version)\footnote{The Einstein equation is always used in any
formalism.}. From these focusing (cross focusing) equations,
firstly, one can defines some energy flux which is naturally related
to the variation of the area of the cross section of the quasilocal
horizon. This is a Clausius like equation if we simply regard that
the area of the cross section corresponds to the entropy associated
with the horizon. Secondly, one can also study the variation of the
angular momentum of the horizon. Further, by assuming the first law
of thermodynamics is still valid on the quasilocal horizon, from the
Clausiu like equation and the variation of the angular momentum, one
finally gets some energy of the horizon. In this formalism, we need
not introduce some quasilocal energy in advance. Contrarily, the
energy of the horizon can be regarded as a byproduct of the
theory~\cite{Ashtekar:2004cn,Ashtekar:2002ag,Ashtekar:2003hk,
Booth:2003ji,Kavanagh:2006qe,Booth:2006bn}. However, these two
formalism are both useless for the isolated horizon. In fact, there
is no dynamical version first law of thermodynamics (or dynamical
version Clausius relation) associated with the isolated horizon. To
study this object, one has to consider phase space
method~\cite{Ashtekar:2004cn,Ashtekar:1998sp,Ashtekar:1999wa,Ashtekar:1999yj,Ashtekar:2000hw,Ashtekar:2000sz}.

So the (cross) focusing  equations are very important to study the
dynamics of the quasilocal horizons. Actually, to study the dynamics
of local Rindler horizons, Jacobson {\it et al} also apply the so
called Raychaudhuri equation (corresponding to one of the focusing
equations) to relate the variation of the area of the horizon and
the matter flux~\cite{Jac,Jac1}. Therefore, these equations
inevitably appear whenever we hope to study the dynamics of the
horizon quasilocally or locally. On the other hand, the (cross)
focusing equations can be get from the study of the deformation of
the codimension-2 spacelike submanifold. In the light of the
importance of these equations, in this paper, we study the
deformation of a spacelike submanifold with an arbitrary
codimension, and a local frame independent equation will be given.
In the case of codimension-1, this equation reduces to the evolution
equation of the extrinsic curvature of a spacelike hypersurface in
an $n$-dimensional Einstein theory. In the more interesting case of
codimension-2, after selecting a local null frame, this equation
naturally reduces to the well known (cross) focusing equations.

According to the two formalisms mentioned in previous paragraph, in
this paper, we show how the thermodynamics of the trapping horizons
are related to these deformation equations in two different
approaches: with and without introducing the quasilocal energy. In
the first formalism, we generalize the Hawking mass (energy) in four
dimension to higher dimension and study the deformation of this
energy. We find the deformation of this energy inside a marginal
surface (More precisely, the evolution of the energy inside the
marginal surface on the horizon) can be also decomposed into the
contributions from  the matter fields and the gravitational
radiation as in the case of the four dimension. When the marginal
surfaces are closed Einstein manifolds, we also study the first law
like equation of the trapping horizon. Roughly speaking, it has a
similar form as the one with the spherical symmetry. However,
generally, it's impossible to define a surface gravity which is a
constant on the marginal surface. Further, the surface gravity also
evolves on the trapping horizon even in the spherically symmetric
case. This means the system is generally nonequilibrium (even far
from some equilibrium point) if we regard the temperature is
proportional to the surface gravity. To make the problem easy to
understand, it's necessary to study some near equilibrium state at
first. The slowly evolving horizon is just the object which
describes the near equilibrium state of the horizon thermodynamics.
However, most of the studies of the slowly evolving horizon have
been focused on the future outer trapping
horizons~\cite{Booth:2003ji,Kavanagh:2006qe,Booth:2006bn}, and it's
quite necessary to study a slowly evolving past trapping horizon if
we hope to study the near equilibrium state of some spacetime which
only contains the past trapping horizon (for example, our
cosmology). Therefore, in the formalism without introducing the
quasilocal energy, we generalize the definition of the slowly
evolving future outer trapping horizon to the past trapping horizon.

The dynamics of the trapping horizon in FLRW universe is studied as
an example. We find the slowly evolving past trapping horizon in the
FLRW universe has close relation to the scenario of the slow-roll
inflation. With the slowly evolving conditions, we also find: up to
the second order of the corresponding slowly evolving parameter, the
temperature (surface gravity) associated with the slowly evolving
past trapping horizon in the FLRW universe is essentially the same
of the temperature (surface gravity) defined by using the quasilocal
energy.

This paper is organized as follows: In Sec.2, a preliminary of
submanifold theory is given. In Sec.3, we deduce the deformation
equation of a spacelike submanifold with an arbitrary codimension,
and the cases of codimension-1 and codimension-2 are given as
examples. In Sec.4, By selecting some local frame, we show our
equation reduces to the (cross) focusing equations (in the case of
codimension-2). We also give the deformation equation of the
$SO(1,1)$ connection associated with the local frame.  The general
definition of the trapping horizon is provided in Sec.5. In Sec.6,
we generalize the Hawking mass of the four dimension to the higher
dimension, and study the deformation of this quasilocal energy by
using the general deformation equation we have obtained. The
dynamics of the trapping horizon is discussed based on the
deformation of this generalized Hawking energy. In Sec.7, Firstly,
we study the null trapping horizon which corresponds to some
equilibrium state, and then we generalize the slowly evolving future
outer trapping horizon to the past cases. In Sec.8, the tapping
horizon in the FLRW universe is investigated. We give the slowly
evolving conditions of the trapping horizon in the FLRW universe.
The surface gravity associated with this kind of trapping horizon is
also studied. The Sec.9 is conclusion and discussion.

\section{Theory of Submanifold and Geometry of Codimension -2 Surface}
To define the trapping horizon, one has to study the intrinsic and
the extrinsic geometries of the codimension-2 spacelike surface,
$S$, embedded in the spacetime $(\mathcal{M},g)$, and then define
the union of all the codimension-2 surfaces which satisfy some
conditions (be marginal) to be the horizon  of the spacetime. So
it's important to study the geometry of the codimension-2 surface.

To describe the geometry of the codimension-2 surface, in this
section, we list some important formula in the submanifold theory.
We use abstract index notation to make the formula have similar
style as the theory of the hypersurface in GR~\cite{Wald:1984}. More
details can be found in the papers of Carter~\cite{Carter:1992vb,
Carter:1997pb, Carter:2000wv}. The readers who are familiar with the
submanifold theory can skip this subsection.

For a spacelike codimension-2 surface, from the submanifold theory,
one can always decompose the metric of the spacetime into
\begin{equation}
\label{decompositionofmetric} g_{ab}=h_{ab}+q_{ab}\, ,
\end{equation}
where $q_{ab}$ is the induced metric of the surface $S$. The induced
metric $q_{ab}$ is Riemannian because that $S$ is spacelike, while
the transverse part, i.e., $h_{ab}$, is Lorentzian. By this
decomposition, the corresponding projection operators are given by
$q_{a}^{~b}$ and $h_{a}^{~b}$, and they satisfy
\begin{equation}
q_a^{~c}q_{c}^{~b}=q_a^{~b}\, ,\qquad h_a^{~c}h_{c}^{~b}=h_a^{~b}\,
,\qquad h_a^{~c}q_{c}^{~b}=0\, .
\end{equation}
Since $h_{ab}$ is Lorentzian, it's natural to introduce two future
directed null vector fields $\ell$ and $n$, and express $h_{ab}$ as
\begin{equation}
\label{hab} h_{ab}=-\ell_an_b-n_a\ell_b=\varepsilon_{IJ}e^I_ae^J_b\,
.
\end{equation}
where $I$ and $J$ take values $\{1,2\}$, and $e^1=\ell$, $e^2=n$.
The symbol $\varepsilon_{IJ}$ represents a constant matrix given by
$\varepsilon_{11}=\varepsilon_{22}=0,
~\varepsilon_{12}=\varepsilon_{21}=-1$. According to pointing to
singularity or not, the vectors $n$ and $\ell$ are called  inward
(ingoing) and outward (outgoing) respectively. Obviously, there are
some freedoms to choose  $\ell$ and $n$. However, if we require
$\ell_an^a=-1$, the only remainder freedom is just the rescaling of
the null vectors, i.e., $\ell\rightarrow \lambda \ell$,
$n\rightarrow n/\lambda$ with some positive regular function
$\lambda$.

Certainly, one can also introduce an orthogonal frame such that
$h_{ab}$ can be expressed as
\begin{equation}\label{haborthogonal}h_{ab}=-u_au_b+v_av_b\, ,
\end{equation}
where $u_a$ and $v_a$ satisfy: $u^au_a=-1\, ,$ $v^av_a=1$ and
$u^av_a=0$. Now, in eq.(\ref{hab}), we can take $e^1=u$, $e^2=v$,
$\varepsilon_{11}=-\varepsilon_{22}=-1$ and
$\varepsilon_{12}=\varepsilon_{21}=0$. Similar to the null frame,
there are also some freedoms (for example, the $SO(1,1)$ rotation of
the frame) to take different $u_a$ and $v_a$.

Assuming the covariant derivative of the spacetime $(\mathcal{M},g)$
is given by $\nabla$, then, {\it second fundamental tensor}
$K_{ab}^{~~c}$ is defined as~\cite{Carter:1992vb}
\begin{equation}
K_{ab}{}{}^c=q_a^{~d}q_b^{~e}\nabla_{d}q_e^{~c}\, .
\end{equation}
This is an important extrinsic quantity of the surface $S$,  and it
can be defined without introducing any local frame of the spacetime.
It is easy to find that the second fundamental tensor satisfies
\begin{equation}
K_{ab}{}{}^c=K_{ba}{}{}^c\, ,\quad
h_a^{~d}K_{db}{}{}^c=K_{ab}{}{}^dq_{d}{}^c=0\, .
\end{equation}
This tensor can be decomposed into a traceless part ($C_{ab}^{~~c}$)
and a trace part ($K^c$), i.e.,
\begin{equation}
\label{generalshear} K_{ab}{}{}^c=\frac{1}{n-2}q_{ab}
K^c+C_{ab}{}{}^c\, ,
\end{equation}
where $K^c=g^{ab}K_{ab}{}{}^c$ is called {\it extrinsic curvature
vector} or {\it mean curvature vector}, which is an important tensor
in the submanifold theory. By using the null frame, one gets these
extrinsic quantities along the directions of $\ell$ and $n$:
\begin{equation}
\label{extrinsicnullframe}
K_{ab}^{(\ell)}=-K_{ab}{}{}^c\ell_c=q_a{}^cq_b{}^d\nabla_c\ell_d\,
,\qquad K_{ab}^{(n)}=-K_{ab}{}{}^cn_c=q_a{}^cq_b{}^d\nabla_cn_d\, .
\end{equation}
Similarly, the extrinsic vector is also decomposed as
\begin{equation}
\label{extrinsic-expansion}
\theta^{(\ell)}=-K^c\ell_c=q^{ab}\nabla_a\ell_b\,
,\quad\theta^{(n)}=-K^cn_c=q^{ab}\nabla_an_b\, .
\end{equation}
These two quantities are called the expansions along $\ell$ and $n$
respectively. The traceless part is decomposed as
\begin{eqnarray}
\label{extrinsic-shear}
&&\sigma_{ab}^{(\ell)}=-C_{ab}{}{}^c\ell_c=\left(q_a{}^cq_b{}^d-\frac{1}{n-2}q_{ab}q^{cd}\right)\nabla_{c}\ell_d\,
,\nonumber\\
&&\sigma_{ab}^{(n)}=-C_{ab}{}{}^cn_c=\left(q_a{}^cq_b{}^d-\frac{1}{n-2}q_{ab}q^{cd}\right)\nabla_{c}n_d\,
.
\end{eqnarray}
These are just the usual shear tensors along the directions of
$\ell$ and $n$. For the orthogonal frame $\{u,v\}$, we can also get
$K^{(u)}_{ab}=-K_{ab}^{~~c}u_c$, $K^{(v)}_{ab}=-K_{ab}^{~~c}v_c$ and
the corresponding expansions and shear tensors. Actually, for an
arbitrary normal vector $X$, we can define
$$K^{(X)}_{ab}=-K_{ab}^{~~c}X_c=q_a^{~c}q_{b}^{~d}\nabla_cX_d\, ,$$ and
the expansion and the shear tensor are respectively given by
$$\theta^{(X)}=-K^cX_c\, ,\qquad \sigma^{(X)}_{ab}=-C_{ab}^{~~c}X_c\, .$$

To study the intrinsic geometry of  $S$, it's necessary to introduce
the corresponding connection or covariant derivative $D_a$ on $S$.
For the tensor field which is invariant under projection operator
$q_{a}^{~b}$, for example, $T_{a}^{~b}$, the corresponding covariant
derivative is defined by
\begin{equation}
D_cT_a^{~b}=q_{a}^{~d}q_e^{~b}q_c^{~f} \nabla_fT_d^{~e}\, ,
\end{equation}
and more general cases are similar. With this definition of the
covariant derivative, for vectors $\xi$ and $\eta$ which are tangent
to $S$ (invariant under the projection operator $q_{a}^{~b}$), it's
easy to find
\begin{equation}
\eta^c\nabla_c\xi^b=\eta^cD_c\xi^b+K_{a}{}_c{}{}^b\eta^a\xi^c\, .
\end{equation}
This is just {\it Gauss's formula}. By using this covariant
derivative $D_a$, from the usual definition
\begin{equation}
R_{abcd}\xi^d=(D_{a}D_b-D_bD_a)\xi_c\, ,
\end{equation}
one gets the intrinsic Riemann curvature tensor $R_{abcd}$. The
relation between this intrinsic curvature of $S$ and the curvature
of the spacetime is encoded in {\it Gauss equation}:
\begin{equation}
\label{Gauss} R_{abcd}=K_{ca}{}{}^eK_{bde}-K_{cb}{}{}^eK_{ade}
+q_a{}^eq_{b}{}^fq_{c}{}^gq_d{}^h~\mathscr{R}_{efgh}\, ,
\end{equation}
where $\mathscr{R}_{abcd}$ is the Riemann curvature of the
spacetime. This equation can be easily found from the definitions of
$D_a$ and $R_{abcd}$.

Similar to the covariant $D_a$ for the intrinsic geometry of $S$,
for a mixed tensor $T_{abc\cdots}$ with tangent indices $a,b$ and a
normal index $c$, it's convenient to define a covariant derivative
$\tilde{D}_a$ as follows:
\begin{equation}
\label{normalcovariant} \tilde{D}_eT_{abc\cdots}=h_{c}^{~d}\cdots
q_{a}^{~g}q_{b}^{~h}\cdots q_{e}^{~f}\nabla_f T_{ghd}\, .
\end{equation}
Obviously, for tangent tensor (which is invariant under $q_a^{~b}$),
this covariant derivative reduces to the derivative $D_a$. For an
arbitrary normal vector $X$ and a tangent vector $\xi$, by using
above definition, it's easy to find
\begin{equation}
\xi^c\nabla_cX^b=-K_{a}^{~b}{}{}_{c}\xi^aX^c+\xi^c\tilde{D}_cX^b\, .
\end{equation}
This is just {\it Weingarten's formula}. So, for normal vectors,
$\tilde{D}_a$ is just the usual normal covariant derivative. Based
on this covariant derivative, by calculating
\begin{equation}
\label{normalcurvature}
\Omega_{abcd}X^d=(\tilde{D}_{a}\tilde{D}_{b}-\tilde{D}_{b}\tilde{D}_{a})X_c
\end{equation}
for an arbitrary normal vector $X$, we get the corresponding
curvature tensor $\Omega_{abcd}$, which has form
\begin{equation}
\label{Ricci}
\Omega_{abcd}=q_{a}^{~e}q_{b}^{~f}h_{c}^{~g}h_{d}^{~h}\mathscr{R}_{efgh}+K_{aed}K_{b}^{~e}{}_{c}-K_{bed}K_{a}^{~e}{}_{c}\,
.
\end{equation}
This is {\it Ricci equation}. Obviously, this curvature has property
of Weyl tensor, in fact, after some rearrangement, it can be
expressed as
\begin{equation}
\Omega_{abcd}=q_{a}^{~e}q_{b}^{~f}h_{c}^{~g}h_{d}^{~h}\mathscr{C}_{efgh}+C_{aed}C_{b}^{~e}{}_{c}-C_{bed}C_{a}^{~e}{}_{c}\,
,
\end{equation}
where $\mathscr{C}_{efgh}$ is the Weyl tensor of the spacetime,
while $C_{abc}$ is the traceless part of the second fundamental
tensor.

Further, in our codimension-2 cases, from
eq.(\ref{normalcovariant}), for an arbitrary normal vector
$X_a=\alpha \ell_a+\beta n_a$,  it's easy to find
\begin{equation}
\label{DtildeX}
\tilde{D}_{a}X_b=(D_a\alpha+\omega_a\alpha)\ell_b+(D_a\beta-\omega_a\beta)n_b\,
,
\end{equation}
where $\omega_a$ is defined as
\begin{equation}
\label{omegaa} \omega_a=-q_a^{~e}n_d\nabla_e\ell^d\, .
\end{equation}
This is just the normal covariant derivative given in some
references (for example~\cite{Booth:2006bn}).  Sometime, the
$\omega_a$ is called the $SO(1,1)$ connection of the $SO(1,1)$
normal bundle (see, for example, ~\cite{Szabados:1994fe}). Of
course, this definition of the connection on the normal bundle
depends on the null frame (so it's gauge dependant). Similarly, if
we consider the orthogonal frame $\{u,v\}$, for $X_a=\alpha
u_a+\beta v_a$, it's easy to find
\begin{equation}
\label{DtildeXuv}
\tilde{D}_{a}X_b=(D_a\alpha+\omega_a\beta)u_b+(D_a\beta+\omega_a\alpha)v_b\,
,
\end{equation}
and now  $\omega_a$ is defined as
\begin{equation}
\label{omegaaorthogonal} \omega_a=-q_a^{~c}u^b\nabla_cv_b\, .
\end{equation}
It should be noted here: we have used the same notation $\omega_a$
as in the case of the null frame, but their values are usually
different from each other.

Generally, the connection is defined to be
\begin{equation}
\label{Omegaabc}
\omega_{abc}=\varepsilon_{IJ}e^{I}_c\tilde{D}_{a}e^J_b=\omega_a\epsilon_{bc}\,
,
\end{equation}
where $ \epsilon_{ab}=n_a\ell_b-\ell_an_b$ for the null frame
$\{\ell,n\}$, and $\omega_a$ is given in eq.(\ref{omegaa}). While
for the orthogonal frame $\{u,v\}$,  $ \epsilon_{ab}=u_av_b-v_au_b$
and $\omega_a$ can be found in eq. (\ref{omegaaorthogonal}) (The
properties of the tensor $\epsilon_{ab}$ is given in Appendix B.).
It's easy find this $\omega_{abc}$ satisfies standard relations
\begin{equation}
\label{omegaln} \omega_{abc}\ell^c=\tilde{D}_a\ell_b\, ,\qquad
\omega_{abc}n^c=\tilde{D}_{a}n_b\, .
\end{equation}
for the null frame, and
\begin{equation}
\label{omegauv}\omega_{abc}u^c=\tilde{D}_au_b\, ,\qquad
\omega_{abc}v^c=\tilde{D}_{a}v_b\, .
\end{equation}
for the orthogonal frame. For the null frame $\{\ell,n\}$ or the
orthogonal frame $\{u,v\}$, from eq.(\ref{normalcurvature}) and
above two relations, it's easy to find that the curvature tensor
$\Omega_{abcd}$ now can be put into
\begin{equation}
\label{normalcurvature1}
\Omega_{abcd}=\tilde{D}_a\omega_{bcd}-\tilde{D}_b\omega_{acd}+\omega_{ac}^{~~e}\omega_{bed}-\omega_{bc}^{~~e}\omega_{aed}\,
.
\end{equation}
Considering the definition of $\tilde{D}_a$, this curvature is just
the one proposed by Carter~\cite{Carter:1992vb, Carter:1997pb,
Carter:2000wv}:
\begin{equation}
\label{normalcurvature2}
\Omega_{abcd}=\left(h_{c}^{~e}h_{d}^{~f}q_{a}^{~g}q_{b}^{~h}\nabla_g
\omega_{hef} +
\omega_{ac}^{~~e}\omega_{bed}\right)-\left(a\leftrightarrow
b\right)\, .
\end{equation}
Actually, the expressions (\ref{normalcurvature1}) and
(\ref{normalcurvature2}) are valid in the case with an arbitrary
codimension. In the special case of codimension-2 in this paper,
after a short calculation, we find
\begin{equation}
\Omega_{abcd}=(D_a\omega_b-D_b\omega_a)\epsilon_{cd}\, .
\end{equation}
Here, $\Omega_{ab}=D_a\omega_b-D_b\omega_a$ is the curvature
associated with the $SO(1,1)$ connection (\ref{omegaa}).

 Another important formula is {\it Codazzi
equation}. This equation can be obtained from applying the covariant
derivative $\tilde{D}_a$ on the second fundamental tensor. From
equation
\begin{equation}
\tilde{D}_d(K_{abc}X^c)=\tilde{D}_d(K_{abc})X^c+K_{abc}\tilde{D}_dX^c\,
,
\end{equation}
and the relation $K_{ab}^{~~c}X_c=-q_a^{~e}q_b^{~f}\nabla_eX_f$,
it's not hard to find:
\begin{equation}
\label{codazzi}
\tilde{D}_aK_{bcd}-\tilde{D}_{b}K_{acd}=-q_a^{~e}q_{b}^{~f}q_{c}^{~h}h_{d}^{~g}\mathscr{R}_{efhg}\,
.
\end{equation}
This is just the Codazzi equation. For an arbitrary normal vector
$Y$, it gives
\begin{equation}
\label{modifiedcodazi}
\left(\frac{n-3}{n-2}\right)D_a\theta^{(Y)}-D_{b}\sigma^{(Y)b}_{~a}
+K_{d}\tilde{D}_aY^d-K_{a}^{~b}{}{}_{d}^{}\tilde{D}_{b}Y^d=q_a^{~e}q^{bc}Y^d\mathscr{R}_{ebcd}\,
.
\end{equation}
In the case of codimension-1, this equation is just the so called
momentum constraint equation in Hamiltonian formalism in GR if we
select $Y^a=u^a$, where $u^a$ is the unit normal vector of some
spacelike hypersurface. In the case of codimension-2, by using
eq.(\ref{omegaln}), immediately, we get
\begin{equation}
\label{codazziL}
\left(\frac{n-3}{n-2}\right)\left(D_a-\omega_a\right)\theta^{(\ell)}-\left(D_{b}-\omega_b\right)\sigma^{(\ell)b}_{~a}
=q_a^{~e}q^{bc}\ell^d\mathscr{R}_{ebcd}
\end{equation}
and
\begin{equation}
\label{codazziN}
\left(\frac{n-3}{n-2}\right)\left(D_a+\omega_a\right)\theta^{(n)}-\left(
D_{b}+\omega_b\right)\sigma^{(n)b}_{~a}
=q_a^{~e}q^{bc}n^d\mathscr{R}_{ebcd}\, .
\end{equation}
The right hands of above equations can also be transformed into the
form composed by the Weyl tensor $\mathscr{C}_{abcd}$ and Einstein
tensor $\mathscr{G}_{ab}$:
\begin{equation}
\label{RiemanntoWeylEinstein}
q_a^{~e}q^{bc}Y^d\mathscr{R}_{ebcd}=q_a^{~e}q^{bc}Y^d\mathscr{C}_{ebcd}-\left(\frac{n-3}{n-2}\right)q_a^{~e}Y^b\mathscr{G}_{eb}\,
.
\end{equation}
We can get similar equations
\begin{equation}
\left(\frac{n-3}{n-2}\right)\left(D_a\theta^{(u)}-\omega_a\theta^{(v)}\right)
-\left(D_{b}\sigma^{(u)b}_{~a}-\omega_b\sigma^{(v)b}_{~a}\right)
=q_a^{~e}q^{bc}u^d\mathscr{R}_{ebcd}\, ,
\end{equation}
\begin{equation}
\left(\frac{n-3}{n-2}\right)\left(D_a\theta^{(v)}-\omega_a\theta^{(u)}\right)
-\left(D_{b}\sigma^{(v)b}_{~a}-\omega_b\sigma^{(u)b}_{~a}\right)
=q_a^{~e}q^{bc}v^d\mathscr{R}_{ebcd}
\end{equation}
by using eq.(\ref{omegauv}) if the orthogonal frame is considered.
Eqs.(\ref{Gauss}), (\ref{Ricci}) and (\ref{codazzi}) are important
relations in the submanifold theory. They are valid in the case with
an arbitrary codimension. In the four dimension, some of the formula
for the codimension-2 surfaces we have listed here can also be found
in~\cite{Booth:2006bn}. It should be noted here: Some definition
might be not well defined if we only consider a single surface. For
example, $\nabla_a$ is not well defined if the support of the tensor
field is confined to the single surface. Actually, in this case,
only $q_a^{~b}\nabla_b$ is well defined on the
surface~\cite{Carter:1997pb}. However, in this paper, we do not
consider this possibility. In our setting, we can always imagine
that $S$ is a leaf of the foliation of some neighborhood of $S$. By
this consideration, $\nabla_a$ is always well defined.


\section{Deformation of Submanifold}
\subsection{Cases with Arbitrary Codimensions}

Now, let's consider the deformation of some spacelike surface $S$
embedded in the spacetime with an arbitrary codimension. Assume
$X^a$ is a normal vector, then, generally speaking, the Lie
derivative of the projection operator $q_a^{~b}$ along  $X^a$ is not
vanished. However, we can consider a foliation of some neighborhood
of $S$ (Certainly, $S$ is  a leaf of this foliation), and require
that $X$ has some relation to the structure of this foliation such
that the projection operator is Lie dragging along $X$, i.e.,
\begin{equation}
\label{lieprojection} \mathcal{L}_{X}\left(q_{a}^{~b}\right)=0\, ,
\end{equation}
where $\mathcal{L}_X$ is the Lie derivative along $X$. This relation
just means: any tangent tensor (invariant under $q_a^{~b}$)
preserves to be a tangent tensor under the Lie derivative along $X$.
For example, for any tangent vector $\xi^a$, we have $\mathcal{L}_X
\xi^a=\mathcal{L}_X( \xi^bq_{b}^{~a})=q_{b}^{~a}\mathcal{L}_X
\xi^b$. So the resulting vector is still a tangent vector. The
relation (\ref{lieprojection}) also means that
\begin{equation}
\mathcal{L}_X\left(h_{a}^{~b}\right)=0\, .
\end{equation}
This equation can be easily found from the relation
$q_{a}^{~b}+h_a^{~b}=\delta_a^{~b}$.

Of course, now $X$ is constrained by above two equations (only one
of them is independent). The situation is very similar to the $3+1$
decomposition in GR. In that case, for some foliation parameter
$\tau$, the evolution vector $X$ is required to satisfy
$\mathcal{L}_X \tau=1$, and the three dimension projection operator
is also Lie dragging along the evolution vector $X$. Here, we have
generalized this relation to the case with an arbitrary codimension.
For the case of codimension-1,  more details can be found in
reference~\cite{Gourgoulhon:2007ue}.

 Contrarily, for a given normal
vector $X$, we can find an appropriate foliation of some
neighborhood of the given codimension-2 surface such that the
corresponding projection operator is Lie dragging along $X$.

The relation (\ref{lieprojection}) automatically implies following
two equations:
\begin{equation}
\label{constraint} q_a^{~c}\mathcal{L}_Xq_c^{~b}=0\, ,\qquad
h_a^{~c}\mathcal{L}_X q_c^{~b}=0\, .
\end{equation}
After straightforward calculation, the first equation implies
\begin{equation}
\label{XnablaY1} X^dq_{a}^{~c}\nabla_dq_{cb}= -X^dh_{bc}\nabla_d
h_{a}^{~c} =h_{bd}q_{a}^{~c}\nabla_cX^d=\tilde{D}_{a}X_b\, .
\end{equation}
So, for another normal vector $Y$ (which is arbitrary), we arrive at
\begin{equation}
\label{XnablaY2} q_{a}^{~c}X^d\nabla_d Y_c=-Y_c \tilde{D}_a X^c\, .
\end{equation}
This equation just provides the expression of the tangent part of
the vector $X^d\nabla_d Y_c$. Generally, it may has a normal part.
The second equation in eq.(\ref{constraint}) gives result
\begin{equation}
\label{condition2}
q_c^{~b}X^d\nabla_dq_a^{~c}=\tilde{D}^{b}X_a=-h_a^{~c}q_d^{~b}\nabla_cX^d
\end{equation}
where we have used eq.(\ref{XnablaY1}). So, by considering the
definition of $\tilde{D}_a$ in eq.(\ref{normalcovariant}), we have
\begin{equation}
h_b^{~d}q_{a}^{~c}(\nabla_cX_d+\nabla_dX_c)=h_b^{~d}q_{a}^{~c}\mathcal{L}_Xg_{cd}=0\,
,
\end{equation}
This relation just means: after the deformation, the spacetime
metric still preserves the orthogonal property of the projection
operators. Furthermore, for the (arbitrary) normal vector $Y$,
eq.(\ref{condition2}) also implies relation
\begin{equation}
[X,Y]^eq_{eb}=0\, ,
\end{equation}
This just tells us that the Lie bracket of $X$ and $Y$ is also a
normal vector. Actually, if this condition is regarded as a primary
assumption, then, one can easily get the equation (\ref{XnablaY2}).
This  logic has been used in the paper by Booth~\cite{Booth:2006bn}.
Now, it's easy to find that equations
\begin{equation}
q_b^{~d}q_a^{~c}\mathcal{L}_Xq_c^{~b}=h_b^{~d}q_a^{~c}\mathcal{L}_Xq_c^{~b}=q_b^{~d}h_a^{~c}\mathcal{L}_X
q_c^{~b}= h_b^{~d}h_a^{~c}\mathcal{L}_X q_c^{~b}=0
\end{equation}
are trivially satisfied. So there are no further useful constraint
conditions provided by eq.(\ref{lieprojection}).

\vspace{.3cm}

By using the property that projection operator is Lie dragging along
$X$, we can get a simple expression for the second fundamental
tensor:
\begin{equation}
\mathcal{L} _{X}q_{ab}=q_a^{~c}q_b^{~d}\mathcal{L}
_{X}q_{cd}=-2K_{ab}^{~~c}X_c=2K^{(X)}_{ab}\, .
\end{equation}
Similarly, one finds the expansion along $X$ can be expressed as
\begin{equation}
\label{expansionlieD}
\mathcal{L}_{X}\epsilon_q=\theta^{(X)}\epsilon_q\, ,
\end{equation}
where $\epsilon_q$ is the area element of the $(n-2)$-dimension
submanifold $S$. Actually, once some quantity  is invariant under
the projection operator, we can use (\ref{lieprojection}) to
simplify the Lie derivative of the quantity as above two examples.

Let's consider the deformation of the second fundamental tensor
$K_{ab}^{~~c}$ (projecting along the direction of $Y$). From above
discussion, we have
\begin{equation}
\label{LXKYAB}
\mathcal{L}_XK^{(Y)}_{ab}=q_{a}^{~c}q_{b}^{~d}\mathcal{L}_{X}K^{(Y)}_{cd}=q_{a}^{~c}q_{b}^{~d}X^e\nabla_eK^{(Y)}_{cd}
+K^{(X)}_{~~a}{}^{c}K^{(Y)}_{cb}+K^{(X)}_{~~b}{}^{c}K^{(Y)}_{ca}\, .
\end{equation}
After a little bit complicated calculation (especially for the first
term in right hand of above equation), we gets
\begin{eqnarray}
\label{DeltaXKY} \mathcal{L}_XK^{(Y)}_{ab}&=&
q_{a}^{~c}q_{b}^{~d}X^eY^f\mathscr{R}_{ecdf}+K^{(Y)}_{a}{}^cK^{(X)}_{bc}-Y^c\tilde{D}_a\tilde{D}_bX_c
\nonumber\\
&&+K_{acb}\left(Y_d\tilde{D}^cX^d\right)-K_{abc}\left(X^d\nabla_{d}Y^c\right)\,
.
\end{eqnarray}
The details to get above equation are given in Appendix A.
Considering $$(\mathcal{L}_X q^{ac})q_{cb}=-2q^{ac}K^{(X)}_{cb}\quad
\mathrm{and} \quad q_{a}^{~c}\mathcal{L}_X h_{cb}=0\, ,$$ we have
\begin{eqnarray}
\mathcal{L}_X\theta^{(Y)}&=&\mathcal{L}_X(K^{(Y)}_{ab}g^{ab})=\left(\mathcal{L}_X K^{(Y)}_{ab}\right)g^{ab}+K^{(Y)}_{ab}\mathcal{L}_X g^{ab}\nonumber\\
&&=\left(\mathcal{L}_XK^{(Y)}_{ab}\right)g^{ab}+K^{(Y)}_{ab}\mathcal{L}_X\left(q^{ab}+h^{ab}\right)\nonumber\\
&&=\left(\mathcal{L}_XK^{(Y)}_{ab}\right)g^{ab}-2K^{(Y)}_{ab}K^{(X)ab}\,
.
\end{eqnarray}
After substituting this result into eq.(\ref{DeltaXKY}), we get the
deformation equation of the expansion $\theta^{(Y)}$ (The normal
vector $Y$ is arbitrary.),
\begin{equation}
\label{FocusXY} \mathcal{L}_X\theta^{(Y)}=
q^{cd}X^eY^f\mathscr{R}_{ecdf}-K^{(Y)}{}^{ab}K^{(X)}_{ab}-Y^c\tilde{D}_a\tilde{D}^aX_c
-K_{c}\left(X^d\nabla_{d}Y^c\right)\, .
\end{equation}

The results (\ref{DeltaXKY}) and (\ref{FocusXY}) are valid in the
case with an arbitrary codimension. In fact, in above discussions,
we have not impose any requirement on the dimension of the part
associated with $h_{ab}$.

In above discussions, we only consider the deformation of the
submanifold along a normal vector $X^a$. For a tangent vector, for
example, $\phi^a$, the Lie derivative of $\theta^{(Y)}$ along
$\phi^a$ is constrained by the Codazzi equations (\ref{codazzi}) and
(\ref{modifiedcodazi}):
\begin{eqnarray}
\label{lieDphi}
&&\left(\frac{n-3}{n-2}\right)\mathcal{L}_{\phi}\theta^{(Y)}=\phi^aD_{b}\sigma^{(Y)b}_{~a}-\left(\frac{n-3}{n-2}\right)\phi^aK_{d}\tilde{D}_aY^d\nonumber\\
&&+~\phi^aC_{a}^{~b}{}{}_{d}^{}\tilde{D}_{b}Y^d+q^{fg}\phi^eY^h\mathscr{R}_{efgh}\,
.
\end{eqnarray}
Eqs.(\ref{DeltaXKY}) and (\ref{FocusXY}) are main results of this
subsection. To understand them, in following two subsections, we
will apply them to the cases of codimension-1 and codimension-2
respectively.

\subsection{Codimension-1 Cases}

In the case of codimension-1, we can set $h_{ab}=-u_au_b$, where
$u^a$ is an unit timelike normal vector of the hypersurface (the
observer associated with the vector $u^a$ is just the so called
Euler observer). So the extrinsic curvature is simply given by
$K_{abc}=K_{ab}u_c$. In this case, $X$ is just the evolution vector
$X_a=Nu_a$ with lapse function $N$. We can select $Y_a=u_a$ such
that $\theta^{(Y)}$ is given by $\theta^{(Y)}=K=-K^au_a$.
Remembering the definition of $\tilde{D}_a$ in
eq.(\ref{normalcovariant}), we have
\begin{eqnarray}
&&\quad Y^c\tilde{D}_a\tilde{D}^aX_c=-D_aD^aN\, ,\qquad
K^{(Y)ab}K^{(X)}_{ab}=NK^{ab}K_{ab}
\, ,\nonumber\\
&&q^{fg}X^eY^h\mathscr{R}_{efgh}=-N\mathscr{R}_{ab}u^au^b\, ,\qquad
K_{c}\left(X^d\nabla_{d}Y^c\right)=0\, .
\end{eqnarray}
So eq.(\ref{FocusXY}) is transformed into a very familiar form
\begin{equation}
\label{codimension11}
-\frac{1}{N}\mathcal{L}_{X}K=\mathscr{R}_{ab}u^au^b+K^{ab}K_{ab}-\frac{1}{N}D^aD_aN\,
.
\end{equation}
This is nothing but the evolution equation of $K$ in the $3+1$
formulism of GR (Actually, now, it's $(n-1)+1$ decomposition of the
$n$-dimensional Einstein theory). It should be noted here: Usually,
in GR, people use the definition of the extrinsic curvature as
$K_{ab}=-q_a^{~c}q_{b}^{~d}\nabla_cu_d$. So it has a sign difference
from our definition. If we use this definition, the minus in front
of $\mathcal{L}_{X}K/N$ will disappear. Similarly, we can get the
evolution equation of $K_{ab}$ from eq.(\ref{DeltaXKY}) and the
Gauss equation (\ref{Gauss}):
\begin{equation}
\label{codimension12}
-\frac{1}{N}\mathcal{L}_{X}K_{ab}=-q_a^{~c}q_b^{~d}\mathscr{R}_{cd}+R_{ab}+KK_{ab}-2K_{ac}K_{b}^{~c}-\frac{1}{N}D_aD_bN\,
.
\end{equation}
From this simple example, it's very clear how that the vector $X$ is
adapted with the foliation of the neighborhood of the hypersurface
such that $q_{a}^{~b}$ is Lie dragging along it. Of course, for a
different lapse function $N$ (or a different evolution vector $X$),
one has to consider a different foliation structure of the
neighborhood of the hypersurface~\cite{Gourgoulhon:2007ue}.

\subsection{Codimension-2 Cases}

Let's consider the more interesting case of codimension-2. Noted
that in this case we have
\begin{eqnarray}
\label{Riemannpart} &&q^{fg}X^eY^h\mathscr{R}_{efgh}=
-X^eY^h\mathscr{R}_{eh}-h^{fg}X^eY^h\mathscr{R}_{efgh}\nonumber\\
&&=-X^eY^h\mathscr{R}_{eh}-\frac{1}{2}h^{fg}h^{eh}\mathscr{R}_{efgh}\left(X_bY^b\right)\,
.
\end{eqnarray}
However, from the Gauss equation (\ref{Gauss}), we get
\begin{equation}
\label{contractGauss}
\mathscr{R}-2\mathscr{R}_{ab}h^{ab}+h^{ac}h^{bd}\mathscr{R}_{abcd}=R+K_{abc}K^{abc}-K^eK_e\,
.
\end{equation}
After substituting eqs.(\ref{Riemannpart}) and (\ref{contractGauss})
into eq.(\ref{FocusXY}), we find
\begin{eqnarray}
\label{FocusXY2}
\mathcal{L}_X\theta^{(Y)}&=&-\mathscr{G}_{ab}\left[X^aY^b-h^{ab}\left(X_eY^e\right)\right]-K^{(Y)}{}^{ab}K^{(X)}_{ab}\nonumber\\
&&+\frac{1}{2}\left(R+K_{abc}K^{abc}-K^aK_a\right)\cdot\left(X_eY^e\right)
\nonumber\\
&&-Y^c\tilde{D}_a\tilde{D}^aX_c
-K_{c}\left(X^d\nabla_{d}Y^c\right)\, ,
\end{eqnarray}
where $\mathscr{G}_{ab}$ is the Einstein tensor of the spacetime. We
can rearrange the terms in above equation, and transform it into
another more covariant form
\begin{eqnarray}
\label{FocusXY1}
\mathcal{L}_X\theta^{(Y)}&=&-\left(\mathscr{G}_{ab}+K_{cda}K^{cd}{}_{b}\right)\left[X^aY^b-h^{ab}\left(X_eY^e\right)\right]\nonumber\\
&&+\frac{1}{2}\left(R-K_{abc}K^{abc}-K_cK^c\right)\cdot\left(X_eY^e\right)
\nonumber\\
&&-Y^e\tilde{D}_c\tilde{D}^cX_e-K_{c}\left(X^e\nabla_{e}Y^c\right)\,
.
\end{eqnarray}
This is the main result of this subsection. Eq.(\ref{FocusXY1}) is
very important to study the thermodynamics of the horizons. Of
course, here, this deformation equation is valid for an arbitrary
codimension-2 surface (not only for the case which has closed
relation to the horizons).

\vspace{.3cm}

During above reductions, the local frames $e^I=\{\ell, n\}$ or
$e^I=\{u, v\}$ are not necessary. However, to find the explicit
expression for the term $Y^e\tilde{D}_c\tilde{D}^cX_e$, it's better
to introduce some local frame. After some calculation (see Appendix
C), we find
\begin{eqnarray}
\label{YtDtDX}
Y^e\tilde{D}_c\tilde{D}^cX_e&=&\varepsilon_{IJ}\left(Y^ID^cD_cX^J\right)+2\varepsilon_{IJ}\left(\omega^{c}
\epsilon_{bd}Y^be^{Id}D_cX^J\right)\nonumber\\
&+&D_c\omega^{c}\left(\epsilon_{bd}Y^bX^d\right)+\omega_c\omega^c(X_eY^e)\,
,
\end{eqnarray}
where $Y^I$ is the component of $Y^a$ along the vector $e^I_a$,
i.e., $Y^I=Y^ae^I_a$. $D_a$ is the covariant derivative on the
codimension-2 surface. For the null frame $\{\ell, n\}$, the matrix
$\varepsilon_{IJ}$ and tensor $\epsilon_{ab}$ are given in
eqs.(\ref{hab}) and (\ref{Omegaabc}) respectively, and the $SO(1,1)$
connection $\omega_a$ can be found in eq.(\ref{omegaa}). While for
the orthogonal frame $\{u, v\}$, $\varepsilon_{IJ}$ and
$\epsilon_{ab}$ are given just bellow eqs.(\ref{haborthogonal}) and
(\ref{Omegaabc}) respectively, and the connection $\omega_{a}$ can
be found in eq.(\ref{omegaaorthogonal}).

\vspace{.3cm} From above results, it is easy to find that the
operator $\mathcal{L}_{X}$ satisfies
$\mathcal{L}_{X}+\mathcal{L}_{Z}=\mathcal{L}_{X+Z}$. However, due to
the existence of the term  $Y^e\tilde{D}_c\tilde{D}^cX_e$, it has no
property as the usual Lie derivative. For example, the action of the
operator on the scalar $\theta^{(Y)}$ generally has property
\begin{equation}
\label{nonlinearbehavior} \mathcal{L}_{(fX)}\theta^{(Y)}\ne
f\mathcal{L}_{X}\theta^{(Y)}\, ,
\end{equation}
where $f$ is a function on the spacetime. This behavior comes from
the fact that we require that the projection operator is Lie
dragging along $X$. Of course, in the case that  $f$ is a constant
on the codimension-2 surface, from eq.(\ref{YtDtDX}), one finds the
operator $\mathcal{L}_X$ reduces to the usual Lie derivative. We
will find this point in the spherically symmetric cases, for
example, the FLRW universe. For $Y$ vector, we have
\begin{equation}
\label{linearforY}\mathcal{L}_{X}\theta^{(fY)}=f\mathcal{L}_{X}\theta^{(Y)}+\theta^{(Y)}\mathcal{L}_{X}f\,
,
\end{equation}
This is because that the derivative of the $Y$ variable is first
order in eq.(\ref{FocusXY1}). To make the problem easy to
understand, in next section, we give the explicit form of the
deformation equation in some local frames.

The property  (\ref{nonlinearbehavior}) of the operator
$\mathcal{L}_X$ is very similar to the operator $``\delta"$ defined
by Andersson, Mars and Simon~\cite{Andersson:2005gq,
Andersson:2007fh}. Here,  to compare this operator $\delta$  with
our approach, we give (the notations are a little bit different) the
original definition of $\delta$~\cite{Andersson:2007fh}: Assume
$0\in I\subset R$ and $X$ be an arbitray normal vector at $S$. Let
$\Phi_X: S\times I \rightarrow \mathcal{M}$ be a differential map
such that for each fixed $\tau\in I$, $\Phi_X(\cdot,\tau)$ is an
immersion and for fixed $p\in S$, $\Phi_X(p,\tau)$ is a curve
starting at $p$ with tangent vector $X(p)$. This way, a family of
codimension-2 surfaces $S_{\tau}=\Phi_X(S,\tau)$ is defined. Assume
$Y_{\tau}$ be a nowhere zero normal vector of $S_{\tau}$ which is
differentiable with respect to $\tau$, let $\theta^{(Y_{\tau})}$ be
the expansion of $S_{\tau}$ along $Y_{\tau}$, then, the variation of
$\theta^{(Y)}$ is defined as $\delta_{X}\theta^{(Y)}\equiv
\partial_{\tau}\theta^{(Y_{\tau})}|_{\tau=0}$. This is a kind of geometric
variation and we can also assume that $X$ has a  part
$X^{\parallel}$ which is tangent to $S$. By using this definition,
and assuming $X=X^{\parallel}+A\ell-B n$, one
gets~\cite{Andersson:2007fh}:
\begin{eqnarray}
\label{deltaAndersson} \delta_{X}\theta^{(\ell)}&=&
X^{\parallel}\big(\theta^{(\ell)}\big)+ \mathrm{a} \theta^{(\ell)} -
\Delta_{S}B + 2 \omega^a
D_aB-A\Big(K^{(\ell)}_{ab}K^{(\ell)ab}+\mathscr{G}_{ab}\ell^a\ell^b\Big)\nonumber\\
&&+ \frac{B}{2}\Big(-2\omega_a\omega^a+ 2D_a\omega^a +R + K_aK^a
-2\mathscr{G}_{ab}\ell^a n^b \Big)\, ,
\end{eqnarray}
where we have chosen $Y=\ell$. $\omega_a$ is defined in
eq.(\ref{omegaa}).  $\Delta_{S}=D^aD_a$ is the Laplacian on
$(S,q_{ab})$, and
$\mathrm{a}=-n_a\partial_{\tau}\ell^a_{\tau}|_{\tau=0}$.  If we
change notations as: $X\rightarrow q$, $\ell\rightarrow l$,
$n\rightarrow k/2 $,  $\omega\rightarrow s$,
$\theta^{(\ell)}\rightarrow \theta$, $K_aK^a\rightarrow -H^2$,
$A\rightarrow b$ and $B\rightarrow u$, then, (\ref{deltaAndersson})
is just the original result in the Lemma 3.1
in~\cite{Andersson:2007fh}.

Our approach is different from the definition of $\delta$: For an
arbitrary normal vector $X$ of $S$, we assume that there is a
foliation of some neighborhood of $S$ such that $q_{a}^{~b}$ is Lie
dragging along $X$. So our operator $\mathcal{L}_{X}$ is just the
usual Lie derivative  constrained by (\ref{lieprojection}). By this
definition, we can get the value of $\mathcal{L}_{X}\theta^{(Y)}$ on
$S$ for each $X$. However, it will be found soon in the next section
that the result given by $\delta_X$ is very similar to our result by
$\mathcal{L}_X$ (see the discussion  bellow eq.(\ref{KXNC})). In our
approach, the action of $X^{\|}$ is encoded in eq.(\ref{lieDphi}).
To get the result of $\delta_X\theta^{(Y)}$, we need not know the
details of the map $\Phi_X$. A different $X$ may correspond a
different map $\Phi$. Similarly, to get
$\mathcal{L}_{X}\theta^{(Y)}$, we also need not know the details of
the foliation structure of some neighborhood of $S$, and a different
$X$ corresponds to a different foliation of a (maybe different)
neighborhood of $S$.

\vspace{.3cm}

\section{Deformation Equations with Local Frames}
\subsection{Expressions in Null Frame}

\subsubsection{Focusing and Cross Focusing Equations}
Eqs.(\ref{FocusXY2}) and (\ref{FocusXY1}) in last section are
independent of any local frame. In this subsection, by choosing the
null frame $\{\ell, n\}$, we give two important examples. The
components  of the extrinsic curvature along these two null vectors
can be found in eq.(\ref{extrinsicnullframe}), and the corresponding
expansions and shear tensors are defined in
eqs.(\ref{extrinsic-expansion}) and (\ref{extrinsic-shear})
respectively.

Firstly, by setting $Y_a=\ell_a$ and $X_a=A\ell_a-Bn_a$, we have
$X_eY^e=B$. So eqs.(\ref{FocusXY1}) and (\ref{YtDtDX}) give result
\begin{eqnarray}
\label{deltaXthetaL}
&&\mathcal{L}_X\theta^{(\ell)}=\kappa_X\theta^{(\ell)}-D_cD^cB+2\omega^cD_cB
-B\Big{[}\omega_c\omega^c-D_c\omega^c+\mathscr{G}_{ab}\ell^an^b\nonumber\\
&&-\frac{1}{2}R-\theta^{(\ell)}\theta^{(n)}\Big{]}-A\Big{[}\mathscr{G}_{ab}\ell^a\ell^b+\sigma^{(\ell)}_{ab}\sigma^{(\ell)ab}
+\frac{1}{n-2}\theta^{(\ell)}\theta^{(\ell)}\Big{]}\, .
\end{eqnarray}
Here, we have introduced an important quantity---``surface gravity"
\begin{equation}
\label{surfacegravityX} \kappa_X=-n^cX^e\nabla_e\ell_c\, ,
\end{equation}
and then the last term in eq.(\ref{FocusXY1}) becomes
\begin{eqnarray}
\label{KXNC}
&&-K^{c}\left(X^e\nabla_{e}\ell_c\right)=-K^dh_d^{~c}\left(X^e\nabla_{e}\ell_c\right)
\nonumber\\
&&=K^d(\ell^cn_d+n^c\ell_d)\left(X^e\nabla_{e}\ell_c\right)=\kappa_X\theta^{(\ell)}\,
.
\end{eqnarray}
Since we have not discuss any horizon until now, so, in
eq.(\ref{surfacegravityX}), $\kappa_X$ associated with the vector
$X$ is not the real surface gravity of the black hole horizon. It
should be noted here: if we identify $``\kappa_X"$ with
$``\mathrm{a}"$ in (\ref{deltaAndersson}), then
$\delta_X\theta^{(\ell)}$ in (\ref{deltaAndersson}) is the same of
our result (\ref{deltaXthetaL}).

Similarly, by setting $Y_a=n_a$ and $X_a=A\ell_a-Bn_a$, we get
\begin{eqnarray}
\label{deltaXthetaN}
&&\mathcal{L}_X\theta^{(n)}=-\kappa_X\theta^{(n)}+D_cD^cA+2\omega^cD_cA
+A\Big{[}\omega_c\omega^c+D_c\omega^c+\mathscr{G}_{ab}n^a\ell^b\nonumber\\
&&-\frac{1}{2}R-\theta^{(\ell)}\theta^{(n)}\Big{]}+B\Big{[}\mathscr{G}_{ab}n^an^b+\sigma^{(n)}_{ab}\sigma^{(n)ab}
+\frac{1}{n-2}\theta^{(n)}\theta^{(n)}\Big{]}\, .
\end{eqnarray}
In the four dimension, eqs.(\ref{deltaXthetaL}) and
(\ref{deltaXthetaN}) are just the equations given by
Booth~\cite{Booth:2006bn}. In the case where $A=1, B=0$ (or $A=0,
B=-1$), eq.(\ref{deltaXthetaL}) ( or eq.(\ref{deltaXthetaN})) is
just the so called focusing equation. In the case where $A=0, B=-1$
(or $A=1, B=0$), eq.(\ref{deltaXthetaL}) (or
eq.(\ref{deltaXthetaN})) gives the cross focusing equation (see
eqs.(\ref{Focusln}) in next
section)~\cite{Hayward3,Hayward:2004dv,Hayward:2004fz}.

\subsubsection{$Y$ is dual to $X$}

The second  interest example is the case where $X^eY_e=0$. Under
this requirement, eq.(\ref{FocusXY1}) becomes very simple:
\begin{equation}
\mathcal{L}_X\theta^{(Y)}=-\left(\mathscr{G}_{ab}+K_{cda}K^{cd}{}_{b}\right)X^aY^b-Y^e\tilde{D}_c\tilde{D}^cX_e-K_{c}\left(X^e\nabla_{e}Y^c\right)\,
.
\end{equation}
For simplicity, we select $X$ and $Y$ as
\begin{equation}
\label{XYDual} X_a=A\ell_a-Bn_a\, ,\qquad Y_a=A\ell_a+Bn_a\, ,
\end{equation}
where $A$ and $B$ are  functions on the spacetime. It should be
noted here: above expressions are not the most general forms of $X$
and $Y$ (which satisfy the relation $X_eY^e=0$). Actually, here, $X$
and $Y$ are dual to each other, i.e., $Y_a=\epsilon_{ab}X^b$. The
tensor $\epsilon_{ab}$ has been defined in eq.(\ref{Omegaabc}).
After a short calculation, eq.(\ref{YtDtDX}) reduces to
\begin{equation}
\label{totalderivative}
Y^e\tilde{D}_c\tilde{D}^cX_e=D_e(AD^eB-BD^cA-2AB\omega^e )\, .
\end{equation}
It's also easy to find
\begin{equation}
K_{c}\left(X^e\nabla_{e}Y^c\right)=-\kappa_X\theta^{(X)}-\theta^{(\ell)}\mathcal{L}_{X}A-\theta^{(n)}\mathcal{L}_{X}B\,
.
\end{equation}
So the deformation equation of the expansion $\theta^{(Y)}$ becomes
\begin{eqnarray}
\label{LXexpansionXY}
&&\mathcal{L}_X\theta^{(Y)}=\kappa_X\theta^{(X)}-\mathscr{G}_{ab}X^aY^b-\sigma^{(X)}_{ab}\sigma^{(Y)ab}-\frac{1}{n-2}\theta^{(X)}\theta^{(Y)}\nonumber\\
&&-D_e(AD^eB-BD^cA-2AB\omega^e
)+\theta^{(\ell)}\mathcal{L}_{X}A+\theta^{(n)}\mathcal{L}_{X}B\, ,
\end{eqnarray}
where $\kappa_X$ is defined in eq.(\ref{surfacegravityX}). In four
dimension, by setting $A=1$, this equation is just the one given
in~\cite{Booth:2006bn,Gourgoulhon:2005ch,Gourgoulhon:2006uc}.
Consider eq.(\ref{linearforY}), we can translate above equation into
\begin{eqnarray}
&&\kappa_X\theta^{(X)}=\mathscr{G}_{ab}X^aY^b+\sigma^{(X)}_{ab}\sigma^{(Y)ab}+\frac{1}{n-2}\theta^{(X)}\theta^{(Y)}\nonumber\\
&&+D_e(AD^eB-BD^cA-2AB\omega^e
)+A\mathcal{L}_{X}\theta^{(\ell)}+B\mathcal{L}_{X}\theta^{(n)}\, ,
\end{eqnarray}
For $n=4$, this result can also be found
in~\cite{Booth:2006bn,Gourgoulhon:2005ch,Gourgoulhon:2006uc}. This
is a very important equation to study the thermodynamics of the
horizon. If the codimension-2 surface is compact without boundary,
then, it's easy to find
\begin{eqnarray}
\label{XYdualDeformation}
\int\kappa_X\mathcal{L}_X\epsilon_{q}&=&\int\epsilon_q\left[\mathscr{G}_{ab}X^aY^b+\sigma^{(X)}_{ab}\sigma^{(Y)ab}+\frac{1}{n-2}\theta^{(X)}\theta^{(Y)}\right]\nonumber\\
&&+\int\epsilon_q\left[A\mathcal{L}_{X}\theta^{(\ell)}+B\mathcal{L}_{X}\theta^{(n)}\right]\,
,
\end{eqnarray}
where $\epsilon_q$ is the area element of the codimension-2 surface.
Here we have used the relation
$\theta^{(X)}\epsilon_q=\mathcal{L}_X\epsilon_q$ in
eq.(\ref{expansionlieD}).

\subsubsection{Damour-Navier-Stokes like Equation}
The reason to select the null frame $\{\ell,n\}$ is not simply to
make the formula easy to understand. In fact, without the local null
frame, some important physical quantities can not be defined. For
example, $\omega_a$ and $\kappa_X$, which have close relation to the
angular momentum and the surface gravity of the horizon. So another
important equation is the deformation equation of $\omega_a$. From
the definition of $\omega_a$ in eq.(\ref{omegaa}), by using the
relations (\ref{omegaln}) and (\ref{XnablaY2}), it's easy to find
\begin{equation}
\label{deformofomega}
\mathcal{L}_X\omega_a=K_{a}^{~b}{}^{}_c\tilde{D}_b(\epsilon^{cd}X_d)+D_a\kappa_X-\frac{1}{2}q_a{}^bX^d\epsilon^{ce}\mathscr{R}_{dbce}\,
.
\end{equation}
We can also express the right hand of above equation by the Weyl
tensor $\mathscr{C}_{abcd}$ and the Einstein tensor
$\mathscr{G}_{ab}$:
\begin{equation}
\label{RiemanntoWeylEinstein1}
q_a{}^bX^d\epsilon^{ce}\mathscr{R}_{dbce}=q_a{}^bX^d\epsilon^{ce}\mathscr{C}_{dbce}-\frac{2}{n-2}q_a^{~b}\epsilon^{cd}X_d\mathscr{G}_{bc}\,
.
\end{equation}
From the generalized Codazzi equation (\ref{modifiedcodazi}) and
eq.(\ref{RiemanntoWeylEinstein}), we have
\begin{eqnarray}
\label{modifiedcodazzixbar}
&&\left(\frac{n-3}{n-2}\right)D_a\theta^{(Y)}-D_{c}\sigma^{(Y)c}_{~a}
+K_{c}\tilde{D}_aY^c-K_{a}^{~b}{}{}_{c}^{}\tilde{D}_{b}Y^c\nonumber\\
&&=-q_a^{~b}h^{ce}Y^d\mathscr{C}_{bced}-\left(\frac{n-3}{n-2}\right)q_a^{~b}Y\mathscr{G}_{bc}\,
,
\end{eqnarray}
where $Y_a$ is  the dual vector of $X_a$, i.e.,
$Y_a=\epsilon_{ab}X^b$. It's easy to find that
$$q_a^{~b}h^{ce}\epsilon^{df}X_f\mathscr{C}_{bced}=\frac{1}{2}q_a{}^bX^d\epsilon^{ce}\mathscr{C}_{dbce}$$
always holds. Thus, by combining eqs.(\ref{deformofomega}),
(\ref{RiemanntoWeylEinstein1}) and (\ref{modifiedcodazzixbar}), we
have
\begin{equation}
\label{deformationomegaln}
\mathcal{L}_X\omega_a=D_a\kappa_X+\left(\frac{n-3}{n-2}\right)D_a\theta^{(Y)}-D_{c}\sigma^{(Y)c}_{~a}
+K_{c}\tilde{D}_aY^c+q_a^{~b}Y^c\mathscr{G}_{bc}\, .
\end{equation}
The term $K_{c}\tilde{D}_aY^c$ can be calculated from
eqs.(\ref{XYDual}) and (\ref{DtildeX}), and above equation becomes
\begin{eqnarray}
\label{LXOmegaa}
&&\mathcal{L}_X\omega_a+\theta^{(X)}\omega_a=D_a\kappa_X+\left(\frac{n-3}{n-2}\right)D_a\theta^{(Y)}-D_{c}\sigma^{(Y)c}_{~a}\nonumber\\
&&~+q_a^{~b}Y^c\mathscr{G}_{bc}-\theta^{(\ell)}D_aA-\theta^{(n)}D_aB\,
.
\end{eqnarray}
In the case where $X$ is self-dual or anti-self-dual, i.e., $X=\pm
Y$,  by considering the Einstein equation, this equation is a kind
of {\it Damour-Navier-Stokes equation}~\cite{Damour:1978cg}.  In
this case, $X$ or $Y$ can be identified to be the evolution vector
of the event horizon of the spacetime (Actually, when $X=\pm Y$,
eq.(\ref{LXOmegaa}) can be explained to be the Damour-Navier-Stokes
equation on an arbitrary null hypersurface of the spacetime). If
eq.(\ref{LXOmegaa}) is applied to the trapping horizons instead of
the event horizon, one gets {\it generalized Damour-Navier-Stokes}
equations~\cite{Gourgoulhon:2005ch,Gourgoulhon:2008pu}. In this
case, $X$ is identified to be the evolution vector of the trapping
horizon, and $Y$ is just the normal vector of the trapping horizon.
Certainly, now, $X$ need not to be self-dual or anti-self-dual,
i.e., $X$ need not to be null. More details on the physical meaning
of the terms in eq.(\ref{LXOmegaa}) can be found in the papers by
Gourgoulhon {\it
et.al.}~\cite{Gourgoulhon:2005ch,Gourgoulhon:2008pu}.

Let $\phi^a$ be a tangent vector, then, we have
\begin{eqnarray}
&&\mathcal{L}_X\int\epsilon_q\left(\phi^a\omega_a\right)=
\int\epsilon_q\left[\theta^{(X)}\phi^a\omega_a+\mathcal{L}_{X}\phi^a\omega_a+\phi^a\mathcal{L}_X\omega_a\right]\nonumber\\
&&=\int\epsilon_q\Bigg{\{}\mathcal{L}_{X}\phi^a\omega_a+\phi^a\Bigg{[}D_a\kappa_X+\left(\frac{n-3}{n-2}\right)D_a\theta^{(Y)}-D_{c}\sigma^{(Y)c}_{~a}\nonumber\\
&&~+q_a^{~b}Y^c\mathscr{G}_{bc}-\theta^{(\ell)}D_aA-\theta^{(n)}D_aB\Bigg{]}\Bigg{\}}\,
.
\end{eqnarray}
Assume that  $\phi^a$ also satisfies $\mathcal{L}_X\phi^a=0$ and
$D_a\phi^a=0$,  then we get
\begin{eqnarray}
\label{deformangularmomentum}
\mathcal{L}_X\int\epsilon_q\left(\phi^a\omega_a\right)&=&\int\epsilon_q\Bigg{\{}
\frac{1}{2}\left(D^a\phi^b+D^{b}\phi^a\right)\sigma^{(Y)}_{ab}+\phi^aY^b\mathscr{G}_{ab}\nonumber\\
&&+A{\phi}^aD_a\theta^{(\ell)}+B{\phi}^aD_a\theta^{(n)}\Bigg{\}}\, .
\end{eqnarray}
Here, we have assumed that the codimension-2 surface is compact
without boundary. It's obvious this equation does not depend on the
dimension of the spacetime. The left hand of above equation has
closed relation to the angular momentum of the
spacetime~\cite{Brown:1992br}. Actually, the angular momentum can be
defined as $J_{\phi}=\int\epsilon_q(\phi^a\omega_a)$. From above
equation, it's not hard to find the balance equation of the angular
momentum~\cite{Gourgoulhon:2005ch}.

Eqs.(\ref{LXexpansionXY}) and (\ref{LXOmegaa}) have important
applications in the theory of membrane
paradigm~\cite{MembraneParadigm,Parikh:1997ma}. Recent progress on
this topic can be found
in~\cite{Gourgoulhon:2005ch,Eling:2009pb,Eling:2009sj,Booth:2009ct}
and references therein.

\subsection{Expressions in Orthogonal Frame}
Besides the null frame $\{\ell, n\}$, sometime, it's useful to
express the deformation equations in the orthogonal frame $\{u, v\}$
. For example, in the studies of the dynamical horizon, it's
convenient to choose some orthogonal frame. Assume $X_a=Au_a+Bv_a$,
and $Y_a=u_a$, then we have
\begin{equation}
K_c\left(X^e\nabla_eY^c\right)=-\theta^{(v)}\kappa_X\, .
\end{equation}
Similar to the ``surface gravity" defined in
eq.(\ref{surfacegravityX}), here, we have defined
\begin{equation}
\label{surfacegravityXorth} \kappa_{X}=v_cX^e\nabla_eu^c\,
.\end{equation}
From eq.(\ref{YtDtDX}), it's also easy to find
\begin{equation}
Y^e\tilde{D}_c\tilde{D}^cX_e=-D^cD_cA-2\omega^cD_cB-BD_c\omega^c-A\omega_c\omega^c\,
.
\end{equation}
Substituting above results into eq.(\ref{FocusXY1}), we get
\begin{eqnarray}
&&\mathcal{L}_{X}\theta^{(u)}=\theta^{(v)}\kappa_X+D^cD_cA+2\omega^cD_cB+BD_c\omega^c\nonumber\\
&&-B\left(\mathscr{G}_{ab}v^au^b+K^{(u)}_{ab}K^{(v)ab}\right)+A\Big{[}\omega_c\omega^c-\mathscr{G}_{ab}v^av^b\nonumber\\
&&-\frac{1}{2}\Big{(}R+
K^{(u)}_{ab}K^{(u)ab}+K^{(v)}_{ab}K^{(v)ab}+\theta^{(u)}\theta^{(u)}-\theta^{(v)}\theta^{(v)}\Big{)}\Big{]}\,
.
\end{eqnarray}
By similar calculation, for $Y_a=v_a$, we have
\begin{eqnarray}
&&\mathcal{L}_{X}\theta^{(v)}=\theta^{(u)}\kappa_X-D^cD_cB-2\omega^cD_cA-AD_c\omega^c\nonumber\\
&&-A\left(\mathscr{G}_{ab}u^av^b+K^{(u)}_{ab}K^{(v)ab}\right)-B\Big{[}\omega_c\omega^c+\mathscr{G}_{ab}u^au^b\nonumber\\
&&-\frac{1}{2}\Big{(}R-
K^{(u)}_{ab}K^{(u)ab}-K^{(v)}_{ab}K^{(v)ab}+\theta^{(u)}\theta^{(u)}-\theta^{(v)}\theta^{(v)}\Big{)}\Big{]}\,
.
\end{eqnarray}
Let $X_a=Au_a+Bv_a$ and $Y_a=Bu_a+Av_a$, then, $Y$ is the dual
vector of $X$, and they automatically satisfy $X^eY_e=0$. It's easy
to find
\begin{eqnarray}
\mathcal{L}_{X}\theta^{(Y)}&=&\kappa_X\theta^{(X)}-D_c\left[AD^cB-BD^cA+\omega^c\left(A^2-B^2\right)\right]\nonumber\\
&&
-\mathscr{G}_{ab}X^aY^b-\sigma^{(X)}_{ab}\sigma^{(Y)ab}-\frac{1}{n-2}\theta^{(X)}\theta^{(Y)}\nonumber\\
&&+\theta^{(u)}\mathcal{L}_{X}B+\theta^{(v)}\mathcal{L}_{X}A\, .
\end{eqnarray}
Considering the relation (\ref{linearforY}), we have
$$\mathcal{L}_X\theta^{(Y)}=A\mathcal{L}_X\theta^{(v)}+B\mathcal{L}_X\theta^{(u)}+\theta^{(v)}\mathcal{L}_XA+\theta^{(u)}\mathcal{L}_XB\, ,$$
then, when the codimension-2 surface is compact without boundary, we
get
\begin{eqnarray}
\label{XYdualDeformationuv}
\int\kappa_X\mathcal{L}_{X}\epsilon_q&=&\int\epsilon_q\Big{[}\mathscr{G}_{ab}X^aY^b+\sigma^{(X)}_{ab}\sigma^{(Y)ab}
+\frac{1}{n-2}\theta^{(X)}\theta^{(Y)}\nonumber\\
&&+A\mathcal{L}_X\theta^{(v)}+B\mathcal{L}_X\theta^{(u)}\Big{]}\, .
\end{eqnarray}
We can also get the deformation equation of $\omega_a$, which is
given by
\begin{equation}
\label{deformationomegaxuv}
\mathcal{L}_X\omega_a=D_a\kappa_X+\left(\frac{n-3}{n-2}\right)D_a\theta^{(Y)}-D_{c}\sigma^{(Y)c}_{~a}
+K_{c}\tilde{D}_aY^c+q_a^{~b}Y^c\mathscr{G}_{bc}\, .
\end{equation}
However, it should be noted here: now, $\omega_a$ and $\kappa_X$ are
given in eqs.(\ref{omegaaorthogonal}) and
(\ref{surfacegravityXorth}) respectively. Although the definitions
of $\omega_a$ and $\kappa_X$ are dependent of the selection of the
local frames, the deformation equation of $\omega_a$ has similar
form in different frames if we give an appropriate definition of the
``surface gravity" $\kappa_X$. Actually, from above equation or
eq.(\ref{deformationomegaln}), the combination
$\left(\mathcal{L}_X\omega_a-D_a\kappa_X\right)$ is frame
independent.

From the assumption $X_a=Au_a+Bv_a$, $Y_a=Bu_a+Av_a$, and
eq.(\ref{DtildeXuv}), we can easily get the explicit expression of
the term $K_c\tilde{D}_aY^c$:
\begin{equation}
K_c\tilde{D}_aY^c=-D_aB\theta^{(u)}-D_aA\theta^{(v)}-\theta^{(X)}\omega_a\,
.
\end{equation}
By this, the deformation equation (\ref{deformationomegaxuv})
becomes
\begin{eqnarray}
&&\mathcal{L}_X\omega_a+\theta^{(X)}\omega_a=D_a\kappa_X+\left(\frac{n-3}{n-2}\right)D_a\theta^{(Y)}-D_{c}\sigma^{(Y)c}_{~a}\nonumber\\
&&~+q_a^{~b}Y^c\mathscr{G}_{bc}-\theta^{(u)}D_aB-\theta^{(v)}D_aA\,
.
\end{eqnarray}
Similar to get eq.(\ref{deformangularmomentum}), it's not hard to
find: for $D_a\phi^a=0$ and $\mathcal{L}_X\phi^a=0$, we have
\begin{eqnarray}
\label{deformangularmomentum1}
\mathcal{L}_X\int\epsilon_q\left(\phi^a\omega_a\right)&=&\int\epsilon_q\Bigg{\{}
\frac{1}{2}\left(D^a\phi^b+D^{b}\phi^a\right)\sigma^{(Y)}_{ab}+\phi^aY^b\mathscr{G}_{ab}\nonumber\\
&&+A{\phi}^aD_a\theta^{(v)}+B{\phi}^aD_a\theta^{(u)}\Bigg{\}}\, .
\end{eqnarray}
Certainly, to get above equation, compactness of the codimension-2
surface is also required, and this equation has closed relation to
the balance equation of the angular momentum of the spacetime.

\vspace{.3cm}

 At the end of this section, we give some discussions:

(i). The so called (cross) focusing  equations are just the special
examples of eq.(\ref{FocusXY1}). Principally, for a pair of
arbitrary normal vectors $X$ and $Y$, we can get the value of
$\mathcal{L}_{X}\theta^{(Y)}$ on some given codimension-2 surface
according to eq.(\ref{FocusXY1}). Of course, in this procedure, we
have to consider a foliation of some neighborhood of the
codimension-2 surface such that the projection operator is Lie
dragging along $X$.

(ii). By selecting different frames, the deformation equations of
$\omega_a$'s have similar forms as given in
eqs.(\ref{deformationomegaln}) and (\ref{deformationomegaxuv}).
Further, the combination $\mathcal{L}_X\omega_a-D_a\kappa_X$ is
frame independent. To get this conclusion, we have to define an
appropriate ``surface gravity" $\kappa_X$ to match the definition of
the $SO(1,1)$ connection $\omega_a$. When $X$ is self-dual or
anti-self-dual, i.e., $Y=\pm X$, we get the Damour-Navier-Stokes
equation.

(iii). The situation where $Y$ is the dual of $X$ is very simple and
special.  The compactness of the codimension-2 surface is important.
Without this assumption, we can not get the simple expression for
the integrals of the deformation equations, i.e.
eqs.(\ref{XYdualDeformation}) and (\ref{deformangularmomentum}) (or
eqs.(\ref{XYdualDeformationuv}) and (\ref{deformangularmomentum1})
in the orthogonal frame).

(iv). Until now, we haven't discussed some special hypersurface of
the spacetime. In fact, we have focused on the codimension-2
surfaces. Of course the horizon of the spacetime is kind of
hypersurface. In next section, we will find how to study this kind
of hypersurface by our knowledge on the geometry of the
codimension-2 surface.

\section{Trapping Horizon}

In black hole theory, one of the most important objects is the so
called event horizon which is the boundary of the causal past of the
future null infinity~\cite{HawkingEllis,Wald:1984}. So, to describe
the event horizon, one has to know some global information of the
spacetime, for example, the future null infinity. This kind of
horizon is a null hypersurface of the spacetime. However, the so
called trapping horizon is very different from the event
horizon~\cite{Hayward,Hayward1,Hayward2}. The definition of the
trapping horizon is quasilocal, and it does not depend on the
asymptotical behavior of the spacetime. This kind of horizon can be
null, spacelike or timelike according to different spacetime
structures which are involved.

The codimension-2 spacelike surface with
$\theta^{(\ell)}\theta^{(n)}=0$ is called {\it marginal surface}.
The surface with $\theta^{(\ell)}\theta^{(n)}>0$ is called {\it
trapped}, and $\theta^{(\ell)}\theta^{(n)}<0$ is called {\it
untrapped}.

By the definitions of the trapped surface and the untrapped surface,
we can define two regions in the spacetime: A {\it trapped} ({\it
untrapped}) {\it region} is the union of all trapped (untrapped)
surfaces.

We can give similar definitions by using the extrinsic curvature
vector $K^a$ from the relation
$K^cK_c=-2\theta^{(\ell)}\theta^{(n)}$. Sometime, the formulism
without the null frame is enough. However, to give a detailed study
of the marginal surfaces, for example, to give a classification of
the marginal surfaces, it's inevitable to introduce the null frame
or some similar structure.

A marginal surface is called {\it future} if $\theta^{(\ell)}=0$,
$\theta^{(n)}<0$. In this case, if
$\mathcal{L}_{n}\theta^{(\ell)}<0$, we call the future marginal
surface is {\it outer}. The future marginal surface with
$\mathcal{L}_{n}\theta^{(\ell)}>0$ is called {\it inner}.

The {\it past} marginal surface is defined by $\theta^{(n)}=0$,
$\theta^{(\ell)}>0$. Similarly, the past marginal surface with
$\mathcal{L}_{\ell}\theta^{(n)}>0$ is called outer, and the case
with $\mathcal{L}_{\ell}\theta^{(n)}<0$ is called inner.

In some neighborhood of a given codimension-2 surface, we can
imaging this codimension-2 surface belongs to a foliation of this
neighborhood such that the projection operators are Lie dragging
along the null vector $\ell$. This way, we can calculate
$\mathcal{L}_{\ell}\theta^{(Y)}$ for an arbitrary normal vector $Y$.
By the similar assumption and calculation, we get
$\mathcal{L}_{n}\theta^{(Y)}$. By this consideration, we get the
values for $\mathcal{L}_{\ell}\theta^{(\ell)}$,
$\mathcal{L}_{\ell}\theta^{(n)}$, $\mathcal{L}_{n}\theta^{(\ell)}$
and $\mathcal{L}_{n}\theta^{(n)}$ on the given codimension-2 surface
(Of course, this surface belongs to the different foliations of some
neighborhoods according to the different deformation vector $X$'s).
Actually, similar to get eqs.(\ref{deltaXthetaL}) and
(\ref{deltaXthetaN}), it's easy to find the (cross) focusing
equations:
\begin{eqnarray}
\label{Focusln}
\mathcal{L}_{\ell}\theta^{(\ell)}&=&\kappa_{\ell}\theta^{(\ell)} -\mathscr{G}_{ab}\ell^a\ell^b-\sigma_{ab}^{(\ell)}\sigma^{(\ell)ab}-\frac{1}{n-2}\theta^{(\ell)}\theta^{(\ell)}\, ,\nonumber\\
\mathcal{L}_{n}\theta^{(n)}&=&-\kappa_{n}\theta^{(n)} -\mathscr{G}_{ab}n^an^b-\sigma_{ab}^{(n)}\sigma^{(n)ab}-\frac{1}{n-2}\theta^{(n)}\theta^{(n)}\, ,\nonumber\\
\mathcal{L}_{n}\theta^{(\ell)}&=&\kappa_{n}\theta^{(\ell)}
+\omega_c\omega^c-D_c\omega^c+\mathscr{G}_{ab}\ell^an^b-\frac{1}{2}R-\theta^{(\ell)}\theta^{(n)}\,
,
\nonumber\\
\mathcal{L}_{\ell}\theta^{(n)}&=&-\kappa_{\ell}\theta^{(n)}
+\omega_c\omega^c+D_c\omega^c+\mathscr{G}_{ab}n^a\ell^b-\frac{1}{2}R-\theta^{(\ell)}\theta^{(n)}\,
.
\end{eqnarray}
So, for the future marginal surface, the classification of the outer
and inner is determined by
\begin{equation}
\label{deltaNthetaL}
\mathcal{L}_n\theta^{(\ell)}=\omega_c\omega^c-D_c\omega^c+\mathscr{G}_{ab}\ell^an^b-\frac{1}{2}R
\end{equation}
on the future marginal surface, while for the past marginal surface,
this kind of classification is determined by the value
\begin{equation}
\label{deltaNthetaN}
\mathcal{L}_{\ell}\theta^{(n)}=\omega_c\omega^c+D_c\omega^c+\mathscr{G}_{ab}\ell^an^b-\frac{1}{2}R
\end{equation}
on the past marginal surface. Assuming the marginal surface is
closed, then, for the future outer marginal surface and the past
inner marginal surface, above relations give strong constraints on
the scalar curvature $R$ of the marginal surface. For example,
$\int\epsilon_q R$ should be positive if some energy condition is
imposed.

The so called {\it trapping horizon }is the closure of a
hypersurface foliated by the marginal surfaces. The classification
of the trapping horizon inherits from the classification of the
marginal surfaces~\cite{Hayward,Hayward1,Hayward2}.

So we can imagine that the trapping horizon $\mathcal{H}$ is  a
hypersurface which is foliated by a family of $(n-2)$-dimensional
marginal surfaces $S_{\tau}$, where the $\tau\in R$ is called
foliation parameter of the trapping horizon. Assume $X$ is the so
called ``evolution" vector, i.e., the vector which is tangent to
$\mathcal{H}$ and normal to $S_{\tau}$ and satisfies
$\mathcal{L}_{X}\tau=1$.  The details can be found in, for example,
reference~\cite{Gourgoulhon:2005ch}. Since $X$ is normal to the
codimension-2 surface $S_{\tau}$, it can be expressed as $
X_a=A\ell_a- B n_a $ as before. Of course, if we require that the
projection operator is Lie dragging along $X$, then $X$ is also a
deformation vector. In this case, all the formula in previous
sections can be used for this $X$. In following discussion, we
always assume that the evolution vector $X$ is also a deformation
vector.

Here, it should be noted that the foliation near some marginal
surface $S_{\tau}$ has been given in advance. For instance, the
neighborhood of $S_{\tau}$ characterized by the foliation parameter
from $\tau-\Delta \tau$ to $\tau+\Delta \tau $ is given a priori
(and the foliation structure of this neighborhood is also provided
simultaneously). By requiring that $q_{a}^{~b}$ is Lie dragging
along $X$, we get the value of $\mathcal{L}_X \theta^{(Y)}$ on
$S_{\tau}$. For another vector, for example, $\ell$, we have no a
priori neighborhood of $S_{\tau}$ and the associated foliation
structure. However, we can get the value of $\mathcal{L}_{\ell}
\theta^{(Y)}$ on $S_{\tau}$ by requiring that $q_{a}^{~b}$ is Lie
dragging along $\ell$ although the corresponding neighborhood and
associated foliation structure is not provided explicitly. This
point has been discussed at the end of Sec.3.

For the evolution vector $X$, some discussions are listed in order:

(i). For the reparameter or the relabeling of the foliation
$\tau\rightarrow \tau'(\tau)$, the evolution vector changes as
$X\rightarrow X'=(d\tau'/d\tau)^{-1}X$. As we have pointed out
before: According to eq.(\ref{FocusXY1}), the relation between
$\mathcal{L}_X\theta^{(Y)}$ and $\mathcal{L}_{X'}\theta^{(Y)}$ is
complicated due to the existence of the term
$Y^c\tilde{D}_a\tilde{D}^aX_c$. So generally we have no
$\mathcal{L}_{X'}\theta^{(Y)}=f\mathcal{L}_{X}\theta^{(Y)}$ if
$X'=fX$ (i.e., eq.(\ref{nonlinearbehavior})). However, since that
$\tau$ and $\tau'=\tau'(\tau)$ are both constants on the marginal
surfaces, we really have
$\mathcal{L}_{X'}\theta^{(Y)}=(d\tau'/d\tau)^{-1}\mathcal{L}_{X}\theta^{(Y)}$.
Thus, this kind of reparameter does not effect our discussion (the
classification of the trapping horizons).

(ii). It's easy to find that the position of the marginal surface is
independent of the rescaling of the null frame $\{\ell,
n\}\rightarrow \{\lambda\ell, n/\lambda\}$ with some positive
regular function $\lambda$. However, the value of
$\mathcal{L}_{n}\theta^{(\ell)}$ ($\mathcal{L}_{\ell}\theta^{(n)}$)
on the future (past) trapping horizon really depends on the
rescaling of the null frame. So the classification of the outer and
inner of the trapping horizons is complicated. However, in the case
with enough symmetries, the classification of outer of inner of the
trapping horizon is independent of the rescaling of the null frame.

(iii). One can not naively get $\mathcal{L}_{X}\theta^{(\ell)}$ just
from the linear combination of $\mathcal{L}_{\ell}\theta^{(\ell)}$
and $\mathcal{L}_{n}\theta^{(\ell)}$ listed in eq.(\ref{Focusln}).
The reason is very clear according to eq.(\ref{nonlinearbehavior}).
The values of $\mathcal{L}_{X}\theta^{(\ell)}$ and
$\mathcal{L}_{X}\theta^{(n)}$ are given in eqs.(\ref{deltaXthetaL})
and (\ref{deltaXthetaN}). Certainly, for some special case (for
example, the case of spherically symmetric spacetime),
$\mathcal{L}_{X}\theta^{(\ell)}$ is really a linear combination of
$\mathcal{L}_{\ell}\theta^{(\ell)}$ and
$\mathcal{L}_{n}\theta^{(\ell)}$.

(iv). Since the trapping horizon is a hypersurface foliated by the
marginal surfaces, generally we have $\theta^{(\ell)}=0$ on this
hypersurface. Further, since that the ``evolution" vector $X$ is
tangent to this hypersurface, then we expect
$\mathcal{L}_{X}\theta^{(\ell)}=0$ on this hypersurface. From
eq.(\ref{deltaXthetaL}), we get a complicated differential equation
of the function $B$. So it's difficult to get the explicit
expression of $X$. However, in some cases with enough symmetries
(for example, spherically symmetric spacetime), the problem becomes
very simple. We can find that $X$ can be fixed up to a function
which does not depend on the points of the codimension-2 surface.
This will become clear in the FLRW universe.

\section{Horizon Dynamics with Quasilocal Energy}
To study the dynamics of the horizon,  we can firstly assume some
quasilocal energy, i.e., the energy inside the closed codimension-2
surfaces, and then study the deformation of this energy. Generally
speaking, the deformation of the energy can be reduced into a form
like the first law of thermodynamics if this quasilocal energy is
appropriately selected. Need not to say, this discussion heavily
depends on the definition of the quasilocal energy.

\subsection{General Spherically Symmetric Cases}
The results in previous sections are valid for any spacetime. In
this subsection, as an example, we will discuss the spherically
symmetric spacetime. Here, in this simple case, we will not
introduce the quasilocal energy at first. Instead, we show how the
deformation equation of the expansion can be transformed into an
interest form where the quasilocal energy will appear naturally.
Generally, the metric of an $n$-dimensional spherically symmetric
spacetime can be written as
\begin{equation}
g=\beta_{\mu\nu}(y)dy^{\mu}dy^{\nu}+r(y)^2\gamma_{ij}(z)dz^idz^j\, ,
\end{equation}
where $\gamma_{ij}dz^idz^j$ is the standard metric of an
$(n-2)$-dimension sphere of  radius one. Assuming the codimension-2
surface is just the $(n-2)$-sphere, then we have
\begin{equation}
h_{ab}dx^adx^b=\beta_{\mu\nu}dy^{\mu}dy^{\nu}\, ,\qquad
q_{ab}dx^adx^b=r^2\gamma_{ij}dz^idz^j\, .
\end{equation}
With these identifications, it's easy to find
\begin{equation}
\label{extrinsicspherical}
K_{abc}=-\frac{1}{r}q_{ab}\nabla_cr=\frac{1}{n-2}q_{ab}K_c\, ,\qquad
K_c=-\frac{(n-2)}{r}\nabla_cr\, .
\end{equation}
Assuming $X$ and $Y$ are two normal vectors, so, from
$q_{ab}X^b=q_{ab}Y^b=0$, we get that $X_adx^a=X_{\mu}dy^\mu$ and
$Y_adx^a=Y_{\mu}dy^\mu$, where $X_{\mu}$ and $Y_{\mu}$ are
components of $X$ and $Y$ in coordinate bases. Further, we require
that $X$ satisfies $\mathcal{L}_{X}q_a^{~b}=0$. A short calculation
shows that $X_{\mu}$ only depends on the coordinates $y^{\mu}$.
Thus, according to the definition of $\tilde{D}_a$, it's easy to
find
\begin{equation}
\tilde{D}_aX_b=q_a^{~c}h_b^{~d}\nabla_cX_d=0\, .
\end{equation}
So the two order derivative term of $X$  in eq.(\ref{FocusXY1}) is
vanishing, and the behavior of the usual Lie derivative of $X$ in
eq.(\ref{nonlinearbehavior}) is restored. This means that
eq.(\ref{FocusXY1}) only depends on the direction of $X$ (but not
the norm of $X$). Now, we have
$\mathcal{L}_X\theta^{(Y)}=X^a\nabla_a\theta^{(Y)}$, and considering
$\theta^{(Y)}=-K_cY^c$, we get
\begin{equation}
\mathcal{L}_X\theta^{(Y)}=X^e\nabla_eY^c\left(\frac{n-2}{r}\nabla_cr\right)
+\frac{n-2}{r}Y^cX^e\nabla_e\nabla_cr-\frac{n-2}{r^2}X^eY^c\nabla_er\nabla_cr\,
.
\end{equation}
Substituting above result and eq.(\ref{extrinsicspherical}) into
eq.(\ref{FocusXY1}), we have
\begin{eqnarray}
\label{firstlaw1}
&&\left(\frac{n-2}{r}\right)X^aY^b\nabla_a\nabla_br-\frac{(n-2)(n-3)}{2r^2}\left[1-\nabla_ar\nabla^ar\right]\left(X^eY_e\right)\nonumber\\
&&=-\mathscr{G}_{ab}\left[X^aY^b-h^{ab}\left(X^eY_e\right)\right]\,
.
\end{eqnarray}
Here, we have substituted the value of the scalar curvature of the
sphere: $R=(n-2)(n-3)/r^2$. Now, let's choose $X$ just be the
extrinsic curvature vector (or mean curvature vector) of the sphere,
i.e., $X_a=K_a=-(n-2)\nabla_ar/r$. Obviously, this selection
satisfies the condition that the components of the deformation
vector $X$ are only functions of the coordinates $y$. Considering
\begin{equation}
\nabla_b(\nabla^ar\nabla_ar)=\nabla^ar\nabla_b\nabla_ar
+(\nabla_b\nabla_ar)\nabla^ar=2\nabla^ar\nabla_b\nabla_ar\, ,
\end{equation}
and rearranging the terms in eq.(\ref{firstlaw1}), we get
\begin{equation}
\frac{(n-2)}{2r^{n-2}}Y^b\nabla_b\left[r^{n-3}\left(1-\nabla_ar\nabla^ar\right)\right]=\mathscr{G}_{ab}\left[\nabla^arY^b-h^{ab}\left(\nabla^erY_e\right)\right].
\end{equation}
After integrating this equation on the codimension-2 surface, we get
\begin{equation}
\label{firstlaw2} Y^b\nabla_b\left[\frac{(n-2)\Omega_{n-2}}{16\pi
G}r^{n-3}\left(1-\nabla_ar\nabla^ar\right)\right]=\mathscr{A}\mathscr{T}_{ab}\left[\nabla^arY^b-h^{ab}\left(\nabla^erY_e\right)\right]\,
,
\end{equation}
where $\mathscr{A}=\Omega_{n-2}r^{n-2}$ is the area of a sphere of
radius $r$, and $\mathscr{T}_{ab}$ is the energy momentum tensor of
the spacetime. Here, we have used the Einstein equation
$\mathscr{G}_{ab}=8\pi G \mathscr{T}_{ab}$. By defining
\begin{equation}
\label{Misnersharp} \mathscr{E}=\frac{(n-2)\Omega_{n-2}}{16\pi
G}r^{n-3}\left(1-\nabla_ar\nabla^ar\right)\, ,
\end{equation}
and
\begin{equation}
\psi_a=\mathscr{T}_{ab}\nabla^br + w\nabla_ar\, ,\qquad
w=-\frac{1}{2}h^{ab}\mathscr{T}_{ab}\, ,
\end{equation}
we can put eq.(\ref{firstlaw2}) into a very familiar form:
\begin{equation}
\label{unifiedfirstlaw} \mathcal{L}_{Y}\mathscr{E}=
\mathscr{A}\psi_aY^a+w\mathcal{L}_Y\mathscr{V}\, ,
\end{equation}
where the normal vector $Y$ is arbitrary. $\mathscr{V}$ is the
volume inside the $(n-2)$-sphere which is given by
$\mathscr{V}=\Omega_{n-2}r^{n-1}/(n-1)$. It should be noted: we have
selected a special $X$ which is proportional to the mean curvature
vector of the $(n-2)$-sphere. So we are considering a special
foliation of some neighborhood of the codimension-2 surface. By this
selection of the deformation vector, the deformation equation
(\ref{FocusXY1}) can be transformed into above simple form.

The quantity $\mathscr{E}$ actually is just the so called
Misner-Sharp energy insider the $(n-2)$-sphere~\cite{Misner:1964je}.
While $\psi_a$ is a kind of energy flux which is called energy
supply. The scalar $w$ is an energy density, and $w\mathcal{L}_Y
\mathscr{V}$ in eq.(\ref{unifiedfirstlaw}) is a work term.
Eq.(\ref{unifiedfirstlaw}) is called {\it unified first law}, which
is firstly found by Hayward in GR~\cite{Hayward, Hayward1,
Hayward2}. Certainly, one can get eq.(\ref{unifiedfirstlaw})
directly from the Einstein equation. Here, we have deduced it from
the deformation equation of the codimension-2 surface.

Above discussion is independent of the null frame. If we introduce
the null frame $\{\ell, n\}$, then the mean curvature vector can be
expressed as
\begin{equation}
K_a=\theta^{(\ell)}n_a+\theta^{(n)}\ell_a=-\frac{(n-2)}{r}\nabla_ar\
,
\end{equation}
and other quantities can also be expressed by using the null frame,
for example, the Misner-Sharp energy $\mathscr{E}$:
\begin{equation}
\label{MisnersharpNullFrame}
\mathscr{E}=\frac{(n-2)\Omega_{n-2}}{16\pi
G}r^{n-3}\left[1+\frac{2~r^2}{(n-2)^2}\theta^{(\ell)}\theta^{(n)}\right]\,
.
\end{equation}
Certainly, it's  also easy to get the expression for the energy
supply in this null frame.

Everything is simple when the spherical symmetry exists. However,
for the general case, the situation becomes complicated. Maybe, the
most difficult problem is: how to select an appropriate quasilocal
energy. In the four dimension, to get an equation which is similar
to the unified first law (\ref{unifiedfirstlaw}), Hayward {\it
et.al.} have used the so called ``Hawking
mass"~\cite{Hayward3,Hayward:2004dv,Hayward:2004fz,Bray:2006pz}. In
the next subsection, we will give a similar discussion in the higher
dimension. However, firstly, we have to generalize the Hawking mass
(energy) to the higher dimension ($n\geq 4$).

\subsection{More General Cases in Higher Dimensions}
Eq.(\ref{MisnersharpNullFrame}) depends on the null frame. However,
we can put it into another form
\begin{equation}
\mathscr{E}=\frac{\Omega_{n-2}}{16\pi
G(n-3)}r^{n-1}\left[R-\left(\frac{n-3}{n-2}\right)K^cK_c\right]\, ,
\end{equation}
where $R$ is the scalar curvature of the codimension-2 surface. This
equation still depends on the coordinate function $r$. By using the
area, $\int \epsilon_q$, of the closed codimension-2 surface, we can
transform it into a more general form:
\begin{equation}
\label{GenralizedHawkingmass}
\mathscr{E}=\frac{\left(\int\epsilon_q\right)^{\frac{n-3}{n-2}}}{16\pi
G\left(\Omega_{n-2}\right)^{\frac{1}{n-2}}(n-3)}\Bigg{\{}\frac{\int\epsilon_q
R}{\left(\int\epsilon_q\right)^{\frac{n-4}{n-2}}}-\left(\frac{n-3}{n-2}\right)\frac{\int\epsilon_qK_cK^c}{\left(\int\epsilon_q\right)^{\frac{n-4}{n-2}}}\Bigg{\}}\,
.
\end{equation}
Now, $\mathscr{E}$ does not depend on any applied structure of the
codimension-2 surface. In static case, a similar mass function has
been studied in reference~\cite{Mizuno:2009fj} by using the
$(n-1)+1$ decomposition of the Einstein theory in $n$-dimension. The
energy expression (\ref{GenralizedHawkingmass}) is interesting:

(i). In the four dimension, this energy is nothing but the so called
Hawking mass (energy)~\cite{Hawking:1968}.

(ii). In the general spherically symmetric cases, it reduces to the
Misner-Sharp energy in the higher dimension (The generalized
Minsner-Sharp energy in general Lovelock gravity theory can be found
in~\cite{Maeda:2007uu,Nozawa:2007vq}).

(iii). The term $(\int \epsilon_q
R)/(\int\epsilon_q)^{\frac{n-4}{n-2}}$ has close relation to Yamabe
invariant (see the chapter 4 of~\cite{Besse:1987} for the details of
the Yamabe invariant). For $n=4$, this term is just Euler number,
and its variation vanishes. On the other hand, for $n\geq 4$, an
arbitrary variation of this term also vanishes if the codimension-2
surface is a compact Einstein manifold without boundary. This can be
found as follows: For example, we just take the variation to be
$\mathcal{L}_{X}$, i.e., the deformation operator with a deformation
vector $X$, then, it's easy to find
\begin{eqnarray}
&&\mathcal{L}_X\int \epsilon_q R=\int\epsilon_q
\Bigg{\{}\left(\frac{n-4}{n-2}\right)R\theta^{(X)}\nonumber\\
&&+2\left[D_aD_b\sigma^{(X)ab}-\left(\frac{n-3}{n-2}\right)D_aD^a\theta^{(X)}\right]\Bigg{\}}\,
.
\end{eqnarray}
To simplify the expression, let's introduce
$\mathscr{A}=\int\epsilon_q$. Assuming the codimension-2 surface is
closed, we get
\begin{equation}
\mathcal{L}_X\left(\frac{\int \epsilon_q
R}{\mathscr{A}^{\frac{n-4}{n-2}}}\right)=-\left(\frac{n-4}{n-2}\right)\mathscr{A}^{-\frac{n-4}{n-2}}\int\epsilon_q
\left\{R\left(\frac{\mathcal{L}_X\mathscr{A}}{\mathscr{A}}+K^eX_e\right)\right\}\,
,
\end{equation}
where $\mathcal{L}_{X}\mathscr{A}=\mathcal{L}_{X}\int
\epsilon_q=-\int \epsilon_q (K^eX_e)$. Thus, there are three ways to
set above variation to be vanished:
\begin{enumerate}[(a).]
\item Obviously, the right hand of above equation is identically
vanished in the four dimension because that $\int \epsilon_q
R/(\int\epsilon_q)^{\frac{n-4}{n-2}}$ is just the Euler number of
 some two dimensional closed manifold.
\item If the codimension-2 surface is assumed to be a closed Einstein
manifold ($R$ is a constant), then we also get vanished variation.
\item Selecting a special deformation vector $X$ such that $K^aX_a$ is a
constant on the codimension-2 surface, then we have
$\mathcal{L}_X\mathscr{A}/\mathscr{A}+K^eX_e=0$.
\end{enumerate}
To simplify the discussion, here, we introduce a quantity
\begin{equation}
\mathscr{K}=\frac{1}{16\pi
G\left(\Omega_{n-2}\right)^{\frac{1}{n-2}}(n-3)}\left(\frac{\int\epsilon_q
R}{\left(\int\epsilon_q\right)^{\frac{n-4}{n-2}}}\right)\, .
\end{equation}
Once one of above three conditions is satisfied, we have
$\mathcal{L}_X\mathscr{K}=0$. Of course, generally,
$\mathcal{L}_X\mathscr{K}\ne 0$ when $n > 4$.

From the definition (\ref{GenralizedHawkingmass}), the deformation
of this energy along the normal vector $X$ is given by
\begin{equation}
\label{variationgeneralenergy1}
\mathcal{L}_X\mathscr{E}=\left(\frac{n-3}{n-2}\right)\left(\frac{\mathscr{E}}{\mathscr{A}}\right)\mathcal{L}_{X}\mathscr{A}
+\mathscr{A}^{\frac{n-3}{n-2}}\mathcal{L}_{X}\left(\frac{\mathscr{E}}{\mathscr{A}^{\frac{n-3}{n-2}}}\right)
\, ,
\end{equation}
To get this equation, the requirement $\mathcal{L}_X\mathscr{K}=0$
is not necessary. However, when $\mathcal{L}_X\mathscr{K}=0$, we can
transform above equation into another useful form
\begin{equation}
\label{variationgeneralenergy}
\mathcal{L}_X\mathscr{E}=\left(\frac{n-3}{n-2}\right)\left(\frac{\mathscr{E}}{\mathscr{A}}\right)\mathcal{L}_{X}\mathscr{A}
+\mathscr{A}^{\frac{n-3}{n-2}}\mathcal{L}_{X}\left(\frac{\mathscr{E}}{\mathscr{A}^{\frac{n-3}{n-2}}}-\mathscr{K}\right)\,
.
\end{equation}
After inserting $\mathscr{K}$, the term inside the last bracket in
above equation is proportional to $K^cK_c$.

To find $\psi_a$ and $w$ like quantities as in the spherically
symmetric case, we need further calculations of the deformation of
this energy. Firstly, it's easy to find
\begin{eqnarray}
&&-\mathcal{L}_{X}\left(K^cK_c\right)=\mathcal{L}_X\theta^{(K)}\nonumber\\
&&=-\left(\mathscr{G}_{ab}+K_{cda}K^{cd}{}_{b}\right)\left[K^aX^b-h^{ab}\left(K_eX^e\right)\right]\nonumber\\
&&+\frac{1}{2}\left(R-K_{abc}K^{abc}-K_cK^c\right)\cdot\left(K_eX^e\right)
\nonumber\\
&&-K^e\tilde{D}_c\tilde{D}^cX_e-K_{c}\left(X^e\nabla_{e}K^c\right)\,
,
\end{eqnarray}
then, we get
\begin{eqnarray}
&&-\frac{1}{2}\mathcal{L}_{X}\left(K^cK_c\right)=-K^e\tilde{D}_c\tilde{D}^cX_e+\left(\frac{1}{2}\mathscr{G}_{ab}h^{ab}\right)\cdot\left(K_eX^e\right)
\nonumber\\
&&~~~-\left(\mathscr{G}_{ab}+C_{cda}C^{cd}{}_{b}\right)
\left[K^aX^b-\frac{1}{2}h^{ab}\left(K_eX^e\right)\right]\nonumber\\
&&~~~+\frac{1}{2}\left[R-\left(\frac{n-1}{n-2}\right)K_cK^c\right]\cdot\left(K_eX^e\right)\,
,
\end{eqnarray}
where we have used the fact that
$\mathcal{L}_{X}\left(K^cK_c\right)=2K_c\left(X^e\nabla_eK^c\right)$
and the definition of $C_{ab}^{~~c}$ in eq.(\ref{generalshear}).
Substituting this result into eq.(\ref{variationgeneralenergy1}),
after some algebraic calculations, we get an important formula of
this section:
\begin{eqnarray}
\label{deformationofenergy} &&\mathcal{L}_X\mathscr{E}=
\int\epsilon_q\Bigg{\{}
\left(\frac{\mathcal{E}}{n-2}\right)\left(\frac{\mathcal{L}_X\mathscr{A}}{\mathscr{A}}+K_eX^e\right)\Bigg{\}}\nonumber\\
&&+\frac{1}{8\pi G}\left(\frac{L}{n-2}\right)\int
\epsilon_q\Bigg{\{}-K^e\tilde{D}_c\tilde{D}^cX_e
\nonumber\\
&&-\left(\mathscr{G}_{ab}+C_{cda}C^{cd}{}_{b}\right)
\left[K^aX^b-\frac{1}{2}h^{ab}\left(K_eX^e\right)\right]\nonumber\\
&&+\frac{1}{2}\left(\mathscr{G}_{ab}h^{ab}\right)\cdot\left(K_eX^e\right)\Bigg{\}}\,
,
\end{eqnarray}
where
$L=\mathscr{A}^{\frac{1}{n-2}}/\left(\Omega_{n-2}\right)^{\frac{1}{n-2}}$,
and
\begin{equation}
\mathcal{E}=\frac{L}{16\pi
G(n-3)}\left[R-\left(\frac{n-3}{n-2}\right)K_cK^c\right]\, .
\end{equation}
Here, $\mathcal{E}$ is a quantity like an energy density. In fact,
from the definition of the energy $\mathscr{E}$, we have
$\mathscr{E}=\int \epsilon_q \mathcal{E}$.
Eq.(\ref{deformationofenergy}) gives the deformation of the energy
$\mathscr{E}$ inside the closed codimension -2 surface. Some remarks
are listed in order:
\begin{enumerate}[(i).]
\item If $n=4$, the energy $\mathscr{E}$ is the Hawking
mass, and eq.(\ref{deformationofenergy}) reduces to the one given by
Bray et.al~\cite{Bray:2006pz}. Of course, we can also consider the
cases with cosmological constant as in~\cite{Bray:2006pz}. According
to previous discussion, in this case, we have
$\mathcal{L}_X\mathscr{K}=0$ because $\mathscr{K}$ is actually a
topological quantity now. However, we have to point out: The four
dimension is not so special if we put the deformation of the energy
into the form (\ref{deformationofenergy}).

\item In the general spherically symmetric cases, $C_{ab}^{~~c}$
and $K^e\tilde{D}_c\tilde{D}^c X_e$ are vanishing. Further, the
first term in the right hand of eq.(\ref{deformationofenergy}) is
also vanished, then eq.(\ref{deformationofenergy}) reduces to the
one given in eq.(\ref{firstlaw2}). This is an expectable result.

\item It should be noted that the vector $X$ can be changed
arbitrarily in this equation. So, if we select $X$ such that
$K^eX_e=0$, then we get
\begin{eqnarray}
\label{XYdualDFE} \mathcal{L}_X\mathscr{E}&=&-\frac{1}{8\pi
G}\left(\frac{L}{n-2}\right)\int\epsilon_q\Bigg{\{}
\Big{(}K^e\tilde{D}_c\tilde{D}^cX_e\nonumber\\
&&~~~~+\mathscr{G}_{ab}+C_{cda}C^{cd}{}_{b}\Big{)} K^aX^b
\Bigg{\}}\, .
\end{eqnarray}
Further, when $X$ is just the dual vector of $K_c$, then, $K^cX_c$
automatically vanishes and the term $K^e\tilde{D}_c\tilde{D}^c X_e$
is a total derivative on the closed codimension-2 surface (see
eq.(\ref{totalderivative})). So, in this case,  the term
$K^e\tilde{D}_c\tilde{D}^cX_e$ can be omitted.

\item It's easy to find: the first term in the right hand of eq.(\ref{deformationofenergy}) vanishes if $\mathcal{E}$
is a constant on the codimension-2 surface. This means, on the
equi-$\mathcal{E}$ surface, we always have
\begin{eqnarray}
\label{deformationofenergy1}
\mathcal{L}_X\mathscr{E}&=&\frac{1}{8\pi
G}\left(\frac{L}{n-2}\right) \times\int\epsilon_q\Bigg{\{}
-K^e\tilde{D}_c\tilde{D}^cX_e\nonumber\\
&&-\left(\mathscr{G}_{ab}+C_{cda}C^{cd}{}_{b}\right)
\left[K^aX^b-\frac{1}{2}h^{ab}\left(K_eX^e\right)\right]\nonumber\\
&&+\frac{1}{2}\left(\mathscr{G}_{ab}h^{ab}\right)\cdot\left(K_eX^e\right)
\Bigg{\}}\, .
\end{eqnarray}
However, unfortunately, the marginal surface (on which we have
$K_cK^c=0$) is generally not an equi-$\mathcal{E}$ surface unless it
is an Einstein manifold. Of course, we also have
$\mathcal{L}_X\mathscr{K}=0$ if the codimension-2 surface is an
Einstein manifold.

\item We can get the result (\ref{deformationofenergy1}) by another
method: Tuning the vector $X$ such that $K_cX^c$ is always a
constant on the codimension-2 surface. By this selection, we have
$\mathcal{L}_X\mathscr{K}=0$. The requirement in the item (iii)
($K_cX^c=0$) is just a special case of this method.

\item It's interest to study the monotonicity of this energy
$\mathscr{E}$ as in~\cite{Bray:2006pz,Mizuno:2009fj}. Although
$\mathscr{K}$ is not a topological term for $n>4$, the deformation
of the energy $\mathscr{E}$, i.e., eq.(\ref{deformationofenergy}),
really has a very similar structure as in the four dimension. So,
it's possible to get some monotonicity behavior of this energy.
However, this is far beyond the aims of this paper, and we will not
give further discussions here.

\end{enumerate}

Above discussions tell us: The situation becomes a little bit
complicated when the codimension-2 surfaces are not Einstein
manifolds. In the following discussion, to simplify the problem, we
will firstly discuss the case in which the Einstein condition is
imposed. Since the $(n-2)$-dimension Einstein manifolds are always
constant curvature spaces when $n<6$, so, in lower dimensions, the
generalized Hawking mass (\ref{GenralizedHawkingmass}) has no big
difference from the Misner-Sharp Energy in the spherically symmetric
cases. However, the problem becomes interesting when $n\geq6$
because in these cases the Einstein manifolds may be inhomogeneous.
To discuss the situation without the Einstein condition on the
codimension-2 surface, we have to consider some special deformation
vector $X$. We will study this situation at the end of the next
subsection.

\subsection{Dynamics of Trapping Horizon}
The discussion in previous two subsections have given enough
preliminaries to study the dynamics of the trapping horizon.
Firstly, let's consider the general spherically symmetric cases.

\subsubsection{Spacetime with Spherical Symmetry}
In previous sections, with the spherical symmetry, we have shown
that the deformation equation can be transformed into the form of
eq.(\ref{unifiedfirstlaw}). By choosing $X$ (This corresponds to the
selection of  $Y$  in eq.(\ref{unifiedfirstlaw})) to be the tangent
vector of the trapping horizon  on which $\nabla_ar\nabla^ar=0$.
More precisely, we choose $X$ to be the evolution vector of the
trapping horizon. Then, on the trapping horizon, we have
$\mathcal{L}_X(\nabla_ar\nabla^ar)=0$, so eq.(\ref{firstlaw2})
becomes
\begin{equation}
\mathcal{L}_X\mathscr{E}=\left(\frac{\bar{\kappa}}{2\pi}\right)\mathcal{L}_{X}S\,
,
\end{equation}
where
\begin{equation}
\label{kappabarsphere} \frac{\bar{\kappa}}{2\pi}=\frac{(n-3)}{4\pi
r}\, , \qquad S=\frac{\mathscr{A}}{4G}\, .
\end{equation}
This $\bar{\kappa}$ is just the ``effective surface gravity" studied
in~\cite{Ashtekar:2004cn,Cai:2010sz}. Another interesting surface
gravity is defined by
\begin{equation}
\label{energysupply1}
\mathscr{A}\psi_aX^a=\left(\frac{\kappa}{2\pi}\right)\mathcal{L}_{X}S\,
,
\end{equation}
which has a form
\begin{equation}
\label{surfacegravity1}
\frac{\kappa}{2\pi}=\frac{4G}{n-2}\left[\left(\frac{n-3}{\Omega_{n-2}}\right)\frac{\mathscr{E}}{r^{n-2}}-wr\right]\,
.
\end{equation}
In above expression of the surface gravity $\kappa$, the energy
$\mathscr{E}$ takes value on the trapping horizon. In the four
dimension, this surface gravity is just the one defined by
Hayward~\cite{Hayward}. By this $\kappa$, the evolution of
$\mathscr{E}$ on the trapping horizon becomes
\begin{equation}
\label{firstlawsphere}
\mathcal{L}_X\mathscr{E}=\left(\frac{\kappa}{2\pi}\right)\mathcal{L}_{X}S+w\mathcal{L}_X\mathscr{V}\,
.
\end{equation}
This is a first law like equation. Obviously, $\kappa$ is a constant
on the marginal surface ($D_a\kappa=0$). However, generally, it's
not a constant on the full trapping horizon. The reason is:
generally, we have no $\mathcal{L}_X\kappa= 0$. So $\kappa$ may
evolve on the trapping horizon.

\subsubsection{General Cases with Einstein Condition}

In the more general cases where the codimension-2 surface is
Einstein (so $\mathcal{L}_X\mathscr{K}=0$), we have
eqs.(\ref{variationgeneralenergy}), (\ref{deformationofenergy}) and
(\ref{deformationofenergy1}) which have been found in previous
sections. Eq.(\ref{variationgeneralenergy}) is simple. If we set $X$
to be the evolution vector of the trapping horizon, then we get
\begin{equation}
\label{effectivesurfacegravity}
\mathcal{L}_X\mathscr{E}=\frac{\bar{\kappa}}{2\pi}\mathcal{L}_{X}S\,
,
\end{equation}
where the ``effective surface gravity" $\bar{\kappa}$ and the
entropy $S$ are defined as
\begin{equation}
\frac{\bar{\kappa}}{2\pi}=4
G\left(\frac{n-3}{n-2}\right)\left(\frac{\mathscr{E}}{\mathscr{A}}\right)\,
,\qquad S=\frac{\mathscr{A}}{4G}\, .
\end{equation}
Of course, this kind of surface gravity is effective, and usually it
can not reduce to the surface gravity of the corresponding
stationary spacetime. By this definition, it's easy to find
$(n-3)\mathscr{E}=(n-2)\bar{T}S$ with
$\bar{T}=\bar{\kappa}/2\pi$~\cite{Cai:2010sz}.

For the marginal surface, we have $K^cK_c=0$, so
eq.(\ref{deformationofenergy}) reduces to
eq.(\ref{deformationofenergy1}). This is a kind of energy balance
equation. Actually, from the Einstein equation
$\mathscr{G}_{ab}=8\pi G \mathscr{T}_{ab}$,
eq.(\ref{deformationofenergy1}) becomes
\begin{eqnarray}
&&\mathcal{L}_X\mathscr{E}=\left(\frac{L}{n-2}\right)\times
\int\epsilon_q\Bigg{\{}
-\frac{1}{8\pi G}K^e\tilde{D}_c\tilde{D}^cX_e\nonumber\\
&&-\left(\mathscr{T}_{ab}+\frac{1}{8\pi G}C_{cda}C^{cd}{}_{b}\right)
\left[K^aX^b-\frac{1}{2}h^{ab}\left(K_eX^e\right)\right]\nonumber\\
&&+\frac{1}{2}\left(\mathscr{T}_{ab}h^{ab}\right)\cdot\left(K_eX^e\right)
\Bigg{\}}\, .
\end{eqnarray}
In the general spherically symmetric cases, in above equation, the
terms related to the energy-momentum tensor $\mathscr{T}_{ab}$ give
$A\psi_aX^a+w\mathcal{L}_X\mathscr{V}$. However, in the general
cases, there are two additional terms corresponding to
$C_{ab}^{~~c}$ and $K^e\tilde{D}_c\tilde{D}^cX_e$. Of course,
without enough symmetry, it's also not easy to complete the
integral, and we can only write the contribution of the
energymomentum tensor $\mathscr{T}_{ab}$ into an integral form
\begin{equation}
\label{mattercontribution} \int\epsilon_q \left( \psi^{(m)}_a +
w^{(m)}H_a\right)X^a\, ,
\end{equation}
where $w^{(m)}$ and $\psi^{(m)}$ are defined as
\begin{equation}
w^{(m)}=-\frac{1}{2}\mathscr{T}_{ab}h^{ab}\, ,\quad
\psi^{(m)}_a=\mathscr{T}_{ab} H^b+w^{(m)}H_a\, .
\end{equation}
Obviously, $w^{(m)}$ is the same as the one in the spherically
symmetric case. The vector $H^a$ plays the role of $\nabla^ar$ in
the spherically symmetric case, which is defined as\footnote{By
using the normal vector $H^a$, we can also define a generalized
Kodama vector as: $\epsilon_{ab}H^b$~\cite{Kodama:1979vn}.  In the
spherically symmetric case, it's possible to define the surface
gravity from this vector, for example, see
reference~\cite{Nozawa:2007vq}.}
\begin{equation}
H_a=-\frac{L}{n-2}K_a\, ,
\end{equation}
Eq.(\ref{mattercontribution}) comes from the matter fields. While
the terms represented by $C_{ab}^{~~c}$ and
$K^e\tilde{D}_c\tilde{D}^cX_e$ can be understood as the contribution
from the change of a gravitational field. In some sense, we can
understand them to be some gravitational radiation. It's obvious
that shear tensor $C_{ab}^{~~c}$ provides an energy supply
\begin{equation}\label{psig}
\psi^{(g)}_aX^a=\frac{1}{8\pi G}C_{cda}C^{cd}{}_{b}
\left[H^aX^b-\frac{1}{2}h^{ab} \left(H_eX^e\right)\right]
\end{equation}
as the usual energy-momentum tensor $\mathscr{T}_{ab}$. However, it
has no contribution to $w$. Instead, it seems that the term
$K^e\tilde{D}_c\tilde{D}^cX_e$ plays the role of $w$. Actually, if
$K^eX_e$ is nonvanishing, formally we can define
\begin{equation}
\label{wg} w^{(g)}=\frac{1}{8\pi
G}\left(\frac{K^e\tilde{D}_c\tilde{D}^cX_e}{K^bX_b}\right)\, .
\end{equation}
With these identifications, the contribution of the gravitational
radiation is
\begin{equation}
\int\epsilon_q \left( \psi^{(g)}_a + w^{(g)}H_a\right)X^a\, .
\end{equation}
So the deformation of the energy $\mathscr{E}$ becomes
\begin{equation}
\label{denergypsimg}\mathcal{L}_X\mathscr{E}=\int\epsilon_q
\left\{\left(\psi^{(m)}_a+\psi^{(g)}_a\right)+\left(w^{(m)}+w^{(g)}\right)H_a
\right\}X^a\, .
\end{equation}
Similar to eq.(\ref{energysupply1}), we hope the energy supply could
provide a similar definition of some surface gravity as the one in
eq.(\ref{surfacegravity1}). Assuming $X$ to be the evolution vector
of the trapping horizon, and using
eq.(\ref{effectivesurfacegravity}),  we get
\begin{eqnarray}
&&\int\epsilon_q
\left\{\left(\psi^{(m)}_a+\psi^{(g)}_a\right)X^a\right\}
=-\int\epsilon_q\Bigg{\{}\Bigg{[}\left(\frac{n-3}{n-2}\right)\left(\frac{\mathscr{E}}{\mathscr{A}}\right)\nonumber\\
&&-\left(\frac{L}{n-2}\right)\left(w^{(m)}+w^{(g)}\right)\Bigg{]}\cdot(K_a
X^a)\Bigg{\}}\, .
\end{eqnarray}
Unlike the case with the spherical symmetry, usually, we can not get
a surface gravity which is a constant on the marginal surface.
Actually, by defining
\begin{equation}
\label{surfacegravity2} \frac{\kappa}{2\pi}=\frac{4
G}{n-2}\Bigg{[}\left(n-3\right)\left(\frac{\mathscr{E}}{\mathscr{A}}\right)-L\left(w^{(m)}+w^{(g)}\right)\Bigg{]}
\end{equation}
and considering
$\mathcal{L}_X\epsilon_q=\theta^{(X)}\epsilon_{q}=-(K^aX_a)\epsilon_q$,
at most we have
\begin{equation}
\int\epsilon_q
\left\{\left(\psi^{(m)}_a+\psi^{(g)}_a\right)X^a\right\}=\int
\frac{\kappa}{2\pi}\mathcal{L}_Xs\, ,
\end{equation}
where $s=\epsilon_q/4G$ can be simply understood as the entropy
associated with the area element of the marginal surface. Obviously,
the surface gravity (\ref{surfacegravity2}) reduces to the one in
eq.(\ref{surfacegravity1}) if the spherical symmetry is restored. So
eq.(\ref{denergypsimg}) becomes
\begin{equation}
\label{firstlawgeneral}
\mathcal{L}_X\mathscr{E}=\int\left(\frac{\kappa}{2\pi}\right)
\mathcal{L}_Xs+\int w \mathcal{L}_Xv\, .
\end{equation}
where $w=w^{(m)}+w^{(g)}$ and $\mathcal{L}_Xv=\theta^{(X)}\epsilon_q
L/(n-2)$. In the spherically symmetric case, above equation reduces
to eq.(\ref{firstlawsphere}).

Above discussions are independent of any local frame. It's also
clear that above description does not depend on the relabeling of
the foliation of the trapping horizon ($X\rightarrow f X$ with some
relabeling factor $f$). Since only  $K_cK^c=0$ and
$\mathcal{L}_X(K_cK^c)=0$ (on the trapping horizon) have been used,
so all discussions are valid on any kind of trapping horizon (future
or past, outer or inner).

\subsubsection{Expression in Null Frame}

However, to make the problem easy to understand, it's better to
introduce some null frame.  For an arbitrary null frame $\{\ell,
n\}$ with $\ell^an_a=-1$, we can express $X$ and $K$ to be
$X^a=A\ell^a-Bn^a$ and $K^a=\theta^{(\ell)}n^a+\theta^{(n)}\ell^a$
respectively. It's easy to find
$K^aX_a=-A\theta^{(\ell)}+B\theta^{(n)}$. From eq.(\ref{YtDtDX}), we
get
\begin{eqnarray}
\label{KDDY}
&&K^e\tilde{D}_c\tilde{D}^cX_e=-\theta^{(\ell)}\left(D^cD_cA+2\omega^cD_cA+AD_c\omega^c+A\omega_c\omega^c\right)\nonumber\\
&&~~~~~~+\theta^{(n)}\left(D^cD_cB-2\omega^cD_cB-BD_c\omega^c+B\omega_c\omega^c\right)\,
.
\end{eqnarray}
So, on the future trapping horizon ($\theta^{(\ell)}=0$ and
$\theta^{(n)}<0$), when $K^aX_a\ne 0$, we have
\begin{eqnarray}
\frac{K^e\tilde{D}_c\tilde{D}^cX_e}{K^aX_a}=\frac{1}{B}\left(D^cD_cB-2\omega^cD_cB-BD_c\omega^c+B\omega_c\omega^c\right)\,
.
\end{eqnarray}
Then, eq.(\ref{wg}) becomes
\begin{equation} w^{(g)}=\frac{1}{8\pi
G}\left(\frac{1}{B}\right)\left(D^cD_cB-2\omega^cD_cB-BD_c\omega^c+B\omega_c\omega^c\right)\,
.
\end{equation}
In the null frame, the energy supply (\ref{psig}) is also very
simple:
\begin{equation}
\psi^{(g)}_aX^a=-\frac{1}{8\pi G} \left(\frac{
L}{n-2}\right)A\theta^{(n)}\sigma^{(\ell)}_{ab}\sigma^{(\ell)ab}\, .
\end{equation}
Similarly, the energy supply of the matter fields becomes
\begin{equation}
\psi^{(m)}_aX^a=-\left(\frac{
L}{n-2}\right)A\theta^{(n)}\mathscr{T}_{ab}\ell^a\ell^b\, .
\end{equation}
After substituting above expressions into eq.(\ref{denergypsimg}),
we get
\begin{eqnarray}
\label{LYELNNONNULL} &&\mathcal{L}_{X}\mathscr{E}=-\left(\frac{
L}{n-2}\right)\int
\epsilon_q\Bigg{\{}\theta^{(n)}\Bigg{[}A\left(\mathscr{T}_{ab}\ell^a\ell^b+\frac{1}{8\pi
G}\sigma^{(\ell)}_{ab}\sigma^{(\ell)ab}\right)\nonumber\\
&&+ B\mathscr{T}_{ab}\ell^an^b+\frac{1}{8\pi
G}\left(D^cD_cB-2\omega^cD_cB-BD_c\omega^c+B\omega_c\omega^c\right)
\Bigg{]}\Bigg{\}}\, .
\end{eqnarray}
This result is  valid even in the case where $K^aX_a=0$. If
$K^aX_a=0$, we can't define $w^{(g)}$ as eq.(\ref{wg}). Actually,
now, the deformation of the energy reduces to eq.(\ref{XYdualDFE}).
Further, since $K^aX_a=B\theta^{(n)}$ on the future trapping
horizon, so, to require that $K^aX_a=0$, we have to set
$X^a=A\ell^a$ (i.e., the trapping horizon is null). Thus,
eq.(\ref{KDDY}) implies that $K^e\tilde{D}_c\tilde{D}^cX_e=0$.
Therefore, in the null frame which just has been used,
eq.(\ref{XYdualDFE}) becomes
\begin{eqnarray}
\label{LYELNNULL} \mathcal{L}_{X}\mathscr{E}=-\left(\frac{
L}{n-2}\right)\int
\epsilon_q\theta^{(n)}\Bigg{[}A\left(\mathscr{T}_{ab}\ell^a\ell^b+\frac{1}{8\pi
G}\sigma^{(\ell)}_{ab}\sigma^{(\ell)ab}\right)  \Bigg{]}\, .
\end{eqnarray}
Obviously, we can get this result just by setting $B=0$ in
eq.(\ref{LYELNNONNULL}).

Generally, the evolution vector $X$ has to satisfy
$\mathcal{L}_X\theta^{(\ell)}=0$ on the future trapping horizon.
Then, considering eq.(\ref{deltaXthetaL}) and
$K^aX_a=B\theta^{(n)}$, the relation (\ref{LYELNNONNULL}) becomes
$\mathcal{L}_X\mathscr{E}=\left(\frac{n-3}{n-2}\right)\left(\frac{\mathscr{E}}{\mathscr{A}}\right)\mathcal{L}_{X}\mathscr{A}
$. This is nothing but eq.(\ref{variationgeneralenergy}) taking
value on the trapping horizon. This result implies that the energy
$\mathscr{E}$ does not evolve on the null trapping horizon on which
we have $K^aX_a=B\theta^{(n)}=0$. This point can also be directly
found from eqs.(\ref{deltaXthetaL}) and (\ref{LYELNNULL}).

Above discussion is valid for any null frame $\{\ell, n\}$ which
satisfies the relation $\ell^an_a=-1$. Sometime, one can rescale the
null frame $\{\ell, n\}$ such that $\theta^{(n)}$
satisfies~\cite{Booth:2003ji,Booth:2006bn,Kavanagh:2006qe}
\begin{equation}
\theta^{(n)}=-\frac{n-2}{L}<0\, .
\end{equation}
It's easy to find this requirement also means $\mathcal{L}_nL=-1$ on
the marginal surface of the trapping horizon. Now,
eq.(\ref{LYELNNONNULL}) is very simple:
\begin{equation}
\label{LYELNNONULL1} \mathcal{L}_{X}\mathscr{E}=\int
\epsilon_q\Bigg{[}A\left(\mathscr{T}_{ab}\ell^a\ell^b+\frac{1}{8\pi
G}\sigma^{(\ell)}_{ab}\sigma^{(\ell)ab}\right)+
B\left(\mathscr{T}_{ab}\ell^an^b+\frac{\zeta_a\zeta^a}{8\pi
G}\right) \Bigg{]}\, .
\end{equation}
where $\zeta_c=\omega_c-D_c\ln{B}$. Since the codimension-2 surface
is closed, all possible total derivatives can be omitted. So after
setting $\theta^{(n)}=-(n-2)/L$, we can omit the total derivative
terms in eq.(\ref{LYELNNONNULL}). By this consideration,
eq.(\ref{LYELNNULL}) also becomes simple:
\begin{eqnarray}
\label{LYELNNULL1} \mathcal{L}_{X}\mathscr{E}=\int
\epsilon_q\Bigg{[}A\left(\mathscr{T}_{ab}\ell^a\ell^b+\frac{1}{8\pi
G}\sigma^{(\ell)}_{ab}\sigma^{(\ell)ab}\right)  \Bigg{]}\, .
\end{eqnarray}
However, it should be noted here: After selecting this kind of null
frame, we can not get result (\ref{LYELNNULL1}) just by setting
$B=0$ in eq.(\ref{LYELNNONULL1}). This is because there are two
$D_a\ln {B}$ terms in eq.(\ref{LYELNNONULL1}).

In this null frame, eq.(\ref{LYELNNONULL1}) tells us that the
evolution of the energy $\mathscr{E}$ on the future trapping horizon
is decomposed into two pars: \begin{enumerate}[(i).]
\item The contribution of the usual matter fields ---
$\mathscr{T}_{ab}\ell^a\ell^a$ and $\mathscr{T}_{ab}\ell^an^a$ ;
\item The contribution of the gravitational radiation ---
$\sigma^{(\ell)}_{ab}\sigma^{(\ell)ab}$ and $\zeta_a\zeta^a$.
\end{enumerate}
We will give more discussions on the gravitational radiation at the
end of this section. For the past trapping horizon, it's also easy
to get similar results like eqs.(\ref{LYELNNONULL1}) and
(\ref{LYELNNULL1}). Since the procedure is similar, we will not give
further discussions here.

\subsubsection{Codimension-2 Surface without Einstein Condition}
Without the Einstein condition, the problem becomes complicated even
in  the four dimension. For an arbitrary null frame $\{\ell,n\}$
with $\ell_an^a=-1$, the evolution of the energy becomes
\begin{eqnarray}
\label{DENONEINSTEIN}
&&\mathcal{L}_{X}\mathscr{E}=\int\epsilon_q\Bigg{\{}
\left(\frac{\mathcal{E}}{n-2}\right)\left(\frac{\mathcal{L}_X\mathscr{A}}{\mathscr{A}}+B\theta^{(n)}\right)\Bigg{\}} \nonumber\\
&&-\left(\frac{ L}{n-2}\right)\int
\epsilon_q\Bigg{\{}\theta^{(n)}\Bigg{[}A\left(\mathscr{T}_{ab}\ell^a\ell^b+\frac{1}{8\pi
G}\sigma^{(\ell)}_{ab}\sigma^{(\ell)ab}\right)+ B\mathscr{T}_{ab}\ell^an^b\nonumber\\
&&+\frac{1}{8\pi
G}\left(D^cD_cB-2\omega^cD_cB-BD_c\omega^c+B\omega_c\omega^c\right)
\Bigg{]}\Bigg{\}}\, .
\end{eqnarray}
Generally, the evolution vector $X^a=A\ell^a-Bn^a$ does not satisfy
the requirement that $K^cX_c=B\theta^{(n)}$ is a constant on the
marginal surface. However, in the case of $B > 0$, we can always
rescale the null frame such that
\begin{equation}
B\theta^{(n)}=-\frac{n-2}{L}
\end{equation}
is a constant on the marginal surface (Of course, $L$ is a constant
on the marginal surface). It's easy to find this requirement also
means $\mathcal{L}_XL=1$ (This selection is also used in
~\cite{Hayward:2004dv,Hayward:2004fz}). Thus, by using this special
null frame, the evolution of the energy becomes
\begin{eqnarray}
\label{DENoEinsteinNull} \mathcal{L}_{X}\mathscr{E}&=&\int
\epsilon_q\Bigg{\{}\left(\frac{A}{B}\right)\left(\mathscr{T}_{ab}\ell^a\ell^b+\frac{1}{8\pi
G}\sigma^{(\ell)}_{ab}\sigma^{(\ell)ab}\right)\nonumber\\
&&+ \mathscr{T}_{ab}\ell^an^b+\frac{\zeta_c\zeta^c}{8\pi G}
\Bigg{\}}\, ,
\end{eqnarray}
where $\zeta_c$ is the same as the one in eq.(\ref{LYELNNONULL1}).
Considering that $X$ has to satisfy eq.(\ref{deltaXthetaL}) with
$\theta^{(\ell)}=0$ and $\mathcal{L}_X\theta^{(\ell)}=0$, we have
\begin{equation}
\mathcal{L}_{X}\mathscr{E}=\frac{1}{16\pi G}\mathcal{L}_XL\int
\epsilon_q
R=\left(\frac{n-3}{n-2}\right)\left(\frac{\mathscr{E}}{\mathscr{A}}\right)\mathcal{L}_X\mathscr{A}\,
,
\end{equation}
where $\mathscr{E}$ takes value on the marginal surface. To get
above result, we have inserted the relation $\mathcal{L}_XL=1$. It
should be noted here: when $K^aX_a$ is a constant on the
codimension-2 surface, then we have $\mathcal{L}_X\mathscr{K}=0$, so
eq.(\ref{variationgeneralenergy1}) can be transformed into
eq.(\ref{variationgeneralenergy}). Then, on the trapping horizon,
the second term in the right hand of
eq.(\ref{variationgeneralenergy}) vanishes, and then we get above
equation. Eq.(\ref{DENoEinsteinNull}) also shows the evolution of
the energy inside the marginal surface can also be decomposed into
two parts: The contribution of the matter  and the contribution of
the gravitational radiation .

When the tapping horizon is null, we have $K^aX_a=B\theta^{(n)}=0$.
Eq.(\ref{DENONEINSTEIN}) reduces to eq.(\ref{LYELNNULL}). So, on the
null trapping horizon, the energy $\mathscr{E}$ inside the marginal
surface does not evolve regardless the marginal surface is Einstein
or not.

Here, some remarks are listed in order:

(i). The deformation of the generalized Hawking mass
(\ref{GenralizedHawkingmass}), i.e., eq.(\ref{deformationofenergy}),
has a very similar form as the one in the four dimension. When the
marginal surface is an Einstein manifold, the evolution of the
generalized Hawking mass (\ref{GenralizedHawkingmass}) on the
trapping horizon is given by eq.(\ref{effectivesurfacegravity}).
This result can be decomposed into the form (\ref{firstlawgeneral}).
In the null frame, this result can also be expressed in
eqs.(\ref{LYELNNONULL1}) and (\ref{LYELNNULL1}). In the general case
without the Einstein condition of the marginal surface, by choosing
some special null frame, the evolution of the generalized Hawking
mass can also be decomposed into the matter field part and
gravitational radiation part.

(ii). The gravitational radiation is carried by
$\sigma^{(\ell)}_{ab}$ (or $\sigma^{(n)}_{ab}$ in the past case).
The number of the degrees of freedom of this tensor is $n(n-3)/2$.
This is just the number of graviton polarizations in $n$-dimension.
In the previous discussion, we think that the term which corresponds
to $\zeta_a\zeta^a$ is also a kind of gravitational radiation.
However, we have to point out: the detailed physical meaning of the
term $\zeta_a\zeta^a$ is still unclear. Certainly, this term really
comes from the change of the gravitational field, so it's reasonable
to regard it as a term of gravitational radiation.

(iii). The problem of the angular momentum: The deformation of the
Hawking mass or it's generalized version in
eq.(\ref{GenralizedHawkingmass}) in some sense are not sufficient to
describe the full dynamics of the trapping horizon. One has to
consider the angular momentum
separately~\cite{Hayward:2006ss,Hayward:2006hq}. We can define the
angular momentum to be $J_{\phi}=\int \epsilon_q(\phi^a\omega_a)$
for some tangent vector $\phi$ which satisfies
$\mathcal{L}_X\phi^a=0$ and $D_a\phi^a=0$, then, the deformation of
the angular momentum can be studied by
eq.(\ref{deformangularmomentum}) (or
(\ref{deformangularmomentum1})). In the four dimension, there is
only one possible angular momentum. However, in the higher
dimension, the situation is complicated: It's possible that there
are several (not single) angular momentums associated with the
trapping horizon.

(iv). Generally the surface gravity defined in
eq.(\ref{surfacegravity2}) is not a constant on the marginal
surface, i.e., $D_a\kappa\ne 0$. In the spherically symmetric case,
this kind of surface gravity is really a constant on the marginal
surface. However, generally, it still evolves on the trapping
horizon, i.e., $\mathcal{L}_X\kappa\ne 0$. This is very different
from the static case in which the surface gravity is a constant on
the horizon. Although we can get some energy balance like equation
(for example, eq.(\ref{LYELNNONULL1})), however, frankly speaking,
the definition of the surface gravity of general nonnull trapping
horizon is still an open problem~\cite{Nielsen:2007ac}. Of course,
we hope the definition of the surface gravity can give some
physically acceptable result when the system is almost equilibrium.

(v). For the null trapping horizon, the generalized Hawking mass
does not evolve on the trapping horizon. Further, if null energy
condition is imposed, all terms in the right hand of
eq.(\ref{LYELNNULL1}) have to be vanished. So there are no dynamical
version first laws associated with the null trapping horizons. To
study the dynamics of the null trapping horizon, we have to consider
other method, for example, the phase space version first
law~\cite{Ashtekar:1998sp,Ashtekar:1999wa,Ashtekar:1999yj,Ashtekar:2000hw,Ashtekar:2000sz}.

\section{Horizon Dynamics without Quasilocal Energy}

Without selecting some quasilocal energy inside the codimension-2
surface, we can also study the dynamics of the trapping horizon.
This kind of discussion heavily depends on the deformation equations
of the expansions and the $SO(1,1)$ connection we have get in
section 4. Eqs.(\ref{XYdualDeformation}) and (\ref{LXOmegaa}) are
key equations to construct some first law like equation in this
formalism. Certainly, one of the most important problems in this
formalism is also the definition of the surface gravity. As
mentioned at the end of last section, to make the problem easy to
understand, it's better to discuss some near equilibrium state at
first. So, in this section, we discuss the slowly evolving trapping
horizon proposed by Booth {\it
et.al.}~\cite{Booth:2003ji,Kavanagh:2006qe,Booth:2006bn,Booth:2009ct}.
This theory just describes the trapping horizon which corresponds to
the near equilibrium state. We will generalize the slowly evolving
future outer trapping horizon to the past trapping horizons
(sometime there are no future trapping horizons in a given
spacetime). This is important to study the FLRW universe because the
past trapping horizon is important in this case.

To make the discussion clear, here, we put the deformation equations
of the expansions (After selecting $X$ to be the evolution vector,
these are actually evolution equations), i.e.,
eqs.(\ref{deltaXthetaL}) and (\ref{deltaXthetaN}) into simple forms:
\begin{eqnarray}
\label{simplifieddeform}
&&\mathcal{L}_X\theta^{(\ell)}=-D_cD^cB+2\omega^cD_cB-B\mathcal{L}_n\theta^{(\ell)}+A\mathcal{L}_{\ell}\theta^{(\ell)}\,
,\nonumber\\
&&\mathcal{L}_X\theta^{(n)}=D_cD^cA+2\omega^cD_cA+A\mathcal{L}_{\ell}\theta^{(n)}-B\mathcal{L}_{n}\theta^{(n)}\,
.
\end{eqnarray}
For the future trapping horizon, we have $\mathcal{L}_X
\theta^{(\ell)}=0$, so the relation between $A$ and $B$ is given by
a two order differential equation of $B$. Similarly, for the past
trapping horizon, we have $\mathcal{L}_X\theta^{(n)}=0$, and the
relation between $A$ and $B$ is encoded in a two order differential
equation of $A$.

\subsection{Equilibrium State}

The theory of the so called slowly evolving horizon is proposed to
describe the dynamics of some horizon which corresponds to the near
equilibrium state. So, to study the slowly evolving horizon, firstly
we have to study the horizon which corresponds to an equilibrium
state. Actually, in black hole theory, the equilibrium state is
described by the dynamical behavior of some  horizon which is null.
Of course, this means the vector $X$ is null. For the null future
trapping horizon, we can assume $X_a=A\ell_a$, and then the
evolution equations of the expansions become
\begin{equation}
\label{simplifieddeformFT}
\mathcal{L}_X\theta^{(\ell)}=A\mathcal{L}_{\ell}\theta^{(\ell)}\,
,\qquad
\mathcal{L}_X\theta^{(n)}=D_cD^cA+2\omega^cD_cA+A\mathcal{L}_{\ell}\theta^{(n)}\,
.
\end{equation}
Similarly, for the null past trapping horizon, by setting
$X_a=-Bn_a$, the evolution equations  are simplified to be
\begin{equation}
\label{simplifieddeformPT}
\mathcal{L}_X\theta^{(n)}=-B\mathcal{L}_{n}\theta^{(n)}\,
,\qquad\mathcal{L}_X\theta^{(\ell)}=-D_cD^cB+2\omega^cD_cB-B\mathcal{L}_n\theta^{(\ell)}\,
.
\end{equation}

Firstly, assuming the null energy condition is satisfied, then by
using $\mathcal{L}_X\theta^{(\ell)}=0$ and the (cross) focusing
equations (\ref{Focusln}), on the null future trapping horizon, we
have
\begin{equation}
\sigma^{(\ell)}_{ab}=0\, , \qquad \mathscr{G}_{ab}\ell^a\ell^b=0\, ,
\end{equation}
and on the null past trapping horizon, we have
\begin{equation}
\sigma^{(n)}_{ab}=0\, , \qquad \mathscr{G}_{ab}n^an^b=0\, .
\end{equation}
Above equations tell us: $K^{(\ell)}_{ab}=0$ on the null future
trapping horizon, and $K^{(n)}_{ab}=0$ on the null past trapping
horizon. Further, $\mathscr{G}_{ab}\ell^a\ell^b=0$ and
$\mathscr{G}_{ab}n^an^b=0$ just imply that there are no matter flux
across the codimension-2 surface.

Secondly, if we also require that $q_a^{~c}\mathscr{G}_{cb}\ell^b=0$
on the null future trapping horizon and
$q_a^{~c}\mathscr{G}_{cb}n^b=0$ on the null past trapping horizon,
then, from the Codazzi equations (\ref{codazziL}), (\ref{codazziN})
and relation (\ref{RiemanntoWeylEinstein}), we get
$q_a^{~e}q^{bc}\ell^d\mathscr{C}_{ebcd}=0$ on the null future
tapping horizon and $q_a^{~e}q^{bc}n^d\mathscr{C}_{ebcd}=0$ on the
null past trapping horizon ($q_a^{~c}\mathscr{G}_{cb}\ell^b=0$ or
$q_a^{~c}\mathscr{G}_{cb}\ell^b=0$ can be satisfied if dominant
energy condition is assumed.).

Finally, from above requirements and eq.(\ref{LXOmegaa}), it's easy
to find
\begin{equation}
\mathcal{L}_X\omega_a-D_a\kappa_X=0
\end{equation}
on the null trapping horizon. This result does not depend on the
selection of $A$ ($B$) in $X_a=A\ell_a$ ( $X_a=-Bn_a$). According to
eq.(\ref{deformationomegaxuv}), it's also independent of the
rescaling of the null frame $\{\ell, n\}$. Additionally, if one
requires that $\omega_a$ does not evolve, i.e.,
$\mathcal{L}_X\omega_a=0$, then, from above equation or
eq.(\ref{LXOmegaa}), one gets $D_a\kappa_X=0$ on the codimension-2
surface for both cases (future and past). This means $\kappa_X$ is a
constant on the codimension-2 surface. Furthermore, if
$\mathcal{L}_X\kappa_X=0$ is required, then $\kappa_X$ is a constant
on the null trapping horizon (regardless future or past).

In fact, the requirement that $\kappa_X$ is a constant on the null
trapping horizon gives some constraints on the function $A$ ($B$).
In the case of future, this can be found from following equations:
\begin{eqnarray}
\label{foliationframerelation1}
D_a\kappa_X&=&\kappa_{\ell}D_aA+AD_a\kappa_{\ell}=0\, ,\nonumber\\
\mathcal{L}_X\kappa_X&=&A\left(\kappa_{\ell}\mathcal{L}_{\ell}A+A\mathcal{L}_{\ell}\kappa_{\ell}\right)=0\,
.
\end{eqnarray}
Similarly, for the past null trapping horizon, $B$ has to satisfy
\begin{eqnarray}
\label{foliationframerelation2}
D_a\kappa_X&=&-\kappa_{n}D_aB-BD_a\kappa_{n}=0\, ,\nonumber\\
\mathcal{L}_X\kappa_X&=&B\left(\kappa_{n}\mathcal{L}_{n}B+B\mathcal{L}_{n}\kappa_{n}\right)=0\,
.
\end{eqnarray}
Since,  until now, the null frame $\{\ell, n\}$ can be arbitrarily
rescaled: $\{\ell, n\}\rightarrow\{\lambda\ell, n/\lambda\}$,
generally,  $\kappa_{\ell}$ ($\kappa_{n}$) is not a constant on the
future (past) null trapping horizon. Given a null frame $\{\ell,
n\}$, we can always find some $A$ ($B$) to satisfy
eq.(\ref{foliationframerelation1})
((\ref{foliationframerelation2})). Obviously, this $A$ ($B$) is not
unique. Inversely, given an evolution vector $X^a$,  we can always
find some null frame $\{\ell, n\}$ (also not unique) to satisfy
eqs.(\ref{foliationframerelation1}) and
(\ref{foliationframerelation2}). So
eqs.(\ref{foliationframerelation1}) and
(\ref{foliationframerelation2}) give the relations between the
foliation structure ($X$) and the null frame.

For example, if we have select some $\lambda$ such that
$\kappa_{\ell}$ ($\kappa_n$) fulfills the requirement that
$\kappa_{\ell}$ ($\kappa_n$) is a constant on the future (past) null
trapping horizon, then, from eq.(\ref{foliationframerelation1})
((\ref{foliationframerelation2})), we have to set $A$ ($B$) such
that $D_aA=\mathcal{L}_{\ell}A=0$ ($D_aB=\mathcal{L}_{n}B=0$) on the
null future (past) trapping horizon (we have assume that
$\kappa_{\ell}$ and $\kappa_n$ are both nonvanished). Obviously,
these conditions can be satisfied if we choose that $A$ ($B$) is a
constant. In this case, it's easy to find $\kappa_X$ satisfies (From
eqs.(\ref{omegaln}) and (\ref{XnablaY2}), it's easy to find
$q_a^{~c}X^b\nabla_bX_c=0$.)
\begin{equation}
\label{surfacegravity} X^a\nabla_aX^b=\pm\kappa_XX^b\, ,
\end{equation}
where $``+"$ corresponds to the case of future, while $``-"$
corresponds to the case of past. This is just the usual formula to
define a surface gravity. Obviously, this kind of surface gravity is
defined up to a constant coefficient.

Actually, for a given function $A$ ($B$), we can set $\lambda$ to be
proportional to $A$ with a constant coefficient (or $\lambda$ which
is proportional to $1/B$), and rescale the null frame such that
$X^a\nabla_aX^b=\pm\kappa_XX^b$ (with constant $\kappa_X$) is
satisfied under the resulting null frame.

It should be noted here, we have not discuss the relabeling of the
foliation until now. The vector $X$ is rescaled by a factor
$f(\tau)=(d\tau'/d\tau)^{-1}$ if we relabel the foliation:
$\tau\rightarrow \tau'(\tau)$. For the null trapping horizons, this
just means that $A$ or $B$ is rescaled by the factor $f$. So we can
use the same procedure as before to find a special null frame such
that the relation (\ref{surfacegravity}) is always held.

From above discussion, we find: {\it to make that $\kappa_X$ is a
real surface gravity given by eq.(\ref{surfacegravity}), it's
necessary to select a special null frame to match the given
foliation structure of the null trapping horizon (i.e. $X$).}
Certainly, the physics the horizon should not depend on the
relabeling of the foliation and rescaling of the null frame. So
different $X$'s (with the preferred $\{\ell, n\}$ and  the
corresponding $\kappa_X$) in some sense are physically equivalent.
In summary, the null trapping horizon can be characterized by an
equivalent class which can be expressed by a triplet
$$\Big{[}X\, , \{\ell, n\}\, , \kappa_X\Big{]}\, .$$
Of course, to foliate the null trapping horizon, the simplest way is
select $X^a=\ell^a$ (or $X^a=-n^a$). With this selection,
eq.(\ref{surfacegravity}) is automatically satisfied.

Above statement has close relation to the isolated horizon
(especially in the case with constant $A$ ($B$)) defined by Ashtekar
{\it el.al.}~\cite{Ashtekar:1998sp}. In those cases, one mainly
focuses on the future outer trapping horizon. Here, we also study
the past trapping horizon. Our discussion is independent of the
selecting of the metric  of the null hypersuface (degenerate). In
fact, this is a rough way to reconstruct of the isolated
horizon~\cite{Booth:2006bn}.

Conclusively, on these null trapping horizons, there are no
gravitational radiation and matter flux, and $\kappa_X$'s are
constants. These properties correspond to the equilibrium state of
the thermodynamics of the horizon. Further, now
eqs.(\ref{XYdualDeformation}) and (\ref{deformangularmomentum}) just
mean
\begin{equation}
\left(\frac{\kappa_X}{2\pi}\right)\mathcal{L}_XS=0\, ,\qquad
\mathcal{L}_{X}J_{\phi}=0\, ,
\end{equation}
where $S\sim \int\epsilon_q$ and $J_{\phi}\sim
\int\epsilon_q\left(\phi^a\omega_a\right)$ can be explained as the
entropy and the angular momentum associated with the null trapping
horizons. Since there are no matter flux and gravitational radiation
across the null trapping horizon, these physical quantities do not
change along the trapping horizons as expected. So there are no
dynamical version of the first law associated with the null trapping
horizons. Actually, this point have been found in the Sec.6.: the
energy $\mathscr{E}$ does not evolve on the null trapping horizon.
In fact, one can study the first law of the null trapping horizons
by using the phase space
method~\cite{Ashtekar:1998sp,Ashtekar:1999wa,Ashtekar:1999yj,Ashtekar:2000hw,Ashtekar:2000sz}.

\subsection{Near Equilibrium State} The near equilibrium means that
$X$ is almost a null vector. This suggests that $X_a=A\ell_a-Bn_a$
slightly deviates from a null vector. However, to characterize this
small deviation, it's not sufficient to set one of $A$ or $B$ to be
very small. The reason is that there are two ambiguities for the
vector $X$ we have mentioned several times:

(i). There are freedoms to choose the null frame $\{\ell,
n\}\rightarrow\{\lambda\ell, n/\lambda\} $ for some positive
function $\lambda$, and this makes an ambiguity for $B/A$ or $A/B$
by factors $\lambda^2$ or $1/\lambda^2$.

(ii). The relabeling of the foliation of the trapping horizon
$\tau\rightarrow \tau'(\tau)$. This makes an ambiguity that  $X$ can
be rescaled as $X\rightarrow X'=f X$ with $f=(d\tau'/d\tau)^{-1}$.

We can assume that the norm of $X$ (or $X_aX^a=2AB$) approaches zero
such that $X$ is almost a null vector. This will eliminate the
ambiguity of rescaling of the null frame. However, since $X'$ and
$X$ are essentially equivalent to describe the trapping horizon, so
the norm of the evolution vector will get a factor $|f|$ if we use
$X'$. We need some procedure to carefully treat the rescaling of the
foliation.

For the null trapping horizons, we can always find some special null
frame to satisfy the requirement of $X^b\nabla_bX^a=\pm\kappa_XX^a$
(with a constant $\kappa_X$), and above two ambiguities in some
sense are fixed to get the preferred null frame and the
corresponding surface gravity. However, for the general case of
non-null horizon, the situation is very different. From
eq.(\ref{XnablaY2}), it's not hard to find\footnote{This equation
reduces to the result given in~\cite{Gourgoulhon:2005ch} when
$A=1$.}
\begin{equation}
X^b\nabla_bX_a=\kappa_X
\left(\epsilon_{ab}X^b\right)-D_a\left(AB\right)+\left(\mathcal{L}_{X}A\right)\ell_a-\left(\mathcal{L}_XB\right)n_a\,
.
\end{equation}
Obviously, now, it's impossible to get
$X^b\nabla_bX^a=\pm\kappa_XX^a$ by rescaling the null frame. So,
principally, the strategy of the rescaling of the null frame in the
null horizon case is meaningless for the non-null trapping horizon.
Actually, in this case, we do not know how to select a preferred
null frame to define a surface gravity associated with $X$. However,
mimicking the null case, for a given foliation of the trapping
horizon, by rescaling the null frame, we can always set $A$ (or $B$)
to be a constant for the future case (or the past case). Without
losing generality, we can set $A=1$ (or $B=1$). This procedure is
similar to the null cases. Thus, for the future trapping horizon, we
have
\begin{equation}
\label{futureX} X^a=\ell^a-Cn^a\, ,
\end{equation}
while for the past trapping horizon, we have
\begin{equation}
\label{pastX} X^a=C\ell^a-n^a\, .
\end{equation}
It should be noted here, the functions $C$'s depend on the
relabeling of the foliation of the trapping horizons. Actually, the
most general form of the evolution vector can be expressed as
\begin{equation}
\label{Xbar}
X^a=f(\tau)\bar{X}^a=f(\tau)\left(A\bar{\ell}^a-B\bar{n}^a\right)\,
.
\end{equation}
with some null frame $\{\bar{\ell}, \bar{n}\}$. One can regard that
the factor $f$ comes from the relabeling of the
foliation\footnote{Here, to make the discussion clear, we also
require that $A$ (or $B$) can not be further viewed as a relabeling
factor. For example, $A$ (or $B$) is not a constant on the
codimension-2 surface, so it can not be absorbed into $f$.
Certainly, in some special cases, $A$ and $B$ are both constants on
the codimension-2 surface. In this case, we can absorb  $A$ (or $B$)
into the factor $f$, and the evolution vector is simply assumed to
be $X^a=f(\ell^a-C'n^a)$ (or $X^a=f(C'\ell^a-n^a)$) with some
function $C'$.}. For the future case, redefining the null frame
$\bar{\ell}^a\rightarrow \ell^a=(fA)\bar{\ell}^a$ and
$\bar{n}^a\rightarrow n^a=\bar{n}^a/(fA)$, we can put $X$ into the
form (\ref{futureX}) with
\begin{equation}
\label{Cinbarframe} C=f^2AB\, .
\end{equation}
Similar situation happens in the past case. By this, the relabeling
of the foliation is encoded in the function $C$. So it seems that
one can relabel the foliation of the trapping horizons such that
$C$'s to be arbitrary nonvanished values (Obviously, for null cases,
 $C$'s are always zero and independent of any foliation
structure).

Since for any kind of trapping horizon, we can always take $|C|$'s
to be arbitrarily small value by selecting the foliation parameter
$\tau$, so the assumption of small $|C|$'s is still not enough to
discuss the slowly expanding behavior of the trapping horizons. To
describe the almost null property, we need some quantity which is
independent of the relabeling of the foliation and the rescaling of
the null frame. We will discuss the cases of future and past
separately.

\vspace{.3cm}

$\bullet$ For the future trapping horizon, Booth {\it
et.el.}~\cite{Booth:2003ji,Kavanagh:2006qe,Booth:2006bn,Booth:2009ct}
give three {\it slowly expanding conditions} (here we gives a
generalized $n$-dimension version):

\begin{enumerate}[(F-i).]

\item The so called evolving parameter $\epsilon\ll 1$ with
\begin{equation}
\label{epsilondefinition}
\frac{\epsilon^2}{L^2}=\max{\left[|C|\left(\|\sigma^{(n)}\|^2+(8\pi
G)\mathscr{T}_{ab}n^an^b+\frac{1}{n-2}\theta^{(n)}\theta^{(n)}\right)\right]}\,
;
\end{equation}

\item The Ricci scalar, the $SO(1,1)$ normal connection and the
energy-momentum tensor satisfy
$$|R|\, ,\quad \|\omega_a\|^2\quad\mathrm{and} \quad (8\pi G) \mathscr{T}_{ab}\ell^an^b\preceq \frac{1}{L^2}\, ;$$

\item The derivatives of horizon fields are at most the same order in
$\epsilon$ as the (maximum of the) original fields. For example,
$$\|D_aC\|\preceq \frac{C_{m}}{L}\, ,\qquad \|D_aD_bC\|\preceq\frac{C_{m}}{L^2}\, .$$

\end{enumerate}
Here, $\|\cdot\|$ is the norm of (tangent) tensor fields on the
codimension-2 Riemannian manifold, while $|\cdot|$ is the absolute
value of some scalar. The quantity $L$ is some length scale of the
codimension-2 surface. For example, the radius of the closed $(n-2)$
manifold: $L=(\mathscr{A}/\Omega_{n-2})^{\frac{1}{n-2}}$ which has
been defined just bellow eq.(\ref{deformationofenergy}). $C_m$ is
the maximum value of $|C|$ on the codimension-2 surface. The
relation $E \preceq F$ means $E\leq k_0 F$ for some constant $k_0$
of order one.

Before using these conditions, we give some discussions:

Firstly, $\epsilon$ defined in the condition (F-i) is independent of
the relabeling of the foliation and the rescaling of the null frame.
This can be easy found from the expressions of $\epsilon$ in the
null frame $\{\bar{\ell}, \bar{n}\}$ bellow eq.(\ref{Xbar}):
$$
\frac{\epsilon^2}{L^2}=\max{\left[\left|\frac{B}{A}\right|\left(\|\sigma^{(\bar{n})}\|^2+(8\pi
G)\mathscr{T}_{ab}\bar{n}^a\bar{n}^b+\frac{1}{n-2}\theta^{(\bar{n})}\theta^{(\bar{n})}\right)\right]}\,
.
$$
This is just what we hope to find: The parameter should not depend
on the relabeling of the foliation and the local frame. By this
consideration and the definition of  $\epsilon$, the requirement of
$\epsilon\ll 1$ in condition (F-i) essentially gives some constraint
on the dynamical behavior of the codimension-2 surface.

Secondly, the evolution vector $X$ is not arbitrary, and it has to
satisfy $\mathcal{L}_X\theta^{(\ell)}=0$. This gives a differential
equation of $C$:
\begin{equation}
\label{Cequation}
\mathcal{L}_X\theta^{(\ell)}=-D_cD^cC+2\omega^cD_cC-C\mathcal{L}_n\theta^{(\ell)}+\mathcal{L}_{\ell}\theta^{(\ell)}=0\,
.
\end{equation}
This is just a special case of the first equation in
(\ref{simplifieddeform}). Furthermore, considering $D_af=0$ and the
expression of $\mathcal{L}_{\ell}\theta^{(\ell)}$ in
eq.(\ref{Focusln}), the relation $\mathcal{L}_X\theta^{(\ell)}=0$
does not give any constraint on the relabeling factor $f$. This
means eq.(\ref{Cequation}) is actually a equation of $AB$ in
(\ref{Cinbarframe}). Thus, the function $C/f^2$ is determined by the
geometrical behavior (both intrinsic and extrinsic) of the
codimension-2 surface. Remembering that $\epsilon\ll 1$ has some
requirement on the geometry of the codimension-2 surface in the
trapping horizon, so $C/f^2$ (or $AB$) is also required to satisfy
some condition by the behavior of $\epsilon$. Actually, in some
simple case, we can find the explicit relation between  $AB$ and
$\epsilon$. This situation happens in the case of the FLRW universe,
and we will find it in the next section.

So the relabeling of the foliation is arbitrary until now. However,
if we require
\begin{equation}
\label{Ccondition} |C|\preceq \epsilon^2\, ,
\end{equation}
then, the function $f$ has to satisfy some condition. This
corresponds to some special selection of the foliation parameter
$\tau$, and then we can not relabeling the foliation arbitrarily. In
the following discussion, we will always assume this condition on
$|C|$.

Now, let's discuss the implication of the conditions (F-i), (F-ii) ,
(F-iii) and (\ref{Ccondition}). From these slowly expanding
conditions and eq.(\ref{Focusln}), it's not hard to find on the
future trapping horizon, we have
\begin{equation}
\left|\mathcal{L}_{n}\theta^{(\ell)}\right|\preceq \frac{1}{L^2}\, ,
\end{equation}
then, considering eq.(\ref{Cequation}) and $|C|\preceq \epsilon^2$,
one gets that
\begin{equation}
\left|\mathcal{L}_{\ell}\theta^{(\ell)}\right|=\|\sigma^{(\ell)}\|^2+8\pi
G \mathscr{T}_{ab}\ell^a\ell^b\preceq \frac{\epsilon^2}{L^2}
\end{equation}
is satisfied on the future trapping horizon. If the null energy
condition is assumed, the two terms in the middle of above equation
are both nonnegative. It's also easy to find
\begin{equation}
\label{slowlyevlovF1}
K^{(X)}_{ab}=\sigma^{(\ell)}_{ab}-CK^{(n)}_{ab}=\sigma^{(\ell)}_{ab}+\mathscr{O}(\epsilon^2)\,
,
\end{equation}
Obviously, the first term is the order of $\epsilon$, while the
terms proportional to $C$ is the order of $\epsilon^2$. Since
$\epsilon\ll 1$, we have $K^{(X)}_{ab}\approx\sigma^{(\ell)}_{ab}$.
Similarly we have $K^{(Y)}_{ab}\approx\sigma^{(\ell)}_{ab} $. With
the same discussion, we get
\begin{equation}
\label{slowlyevlovF2}
\mathscr{T}_{ab}X^aY^b=\mathscr{T}_{ab}\ell^a\ell^b-C^2\mathscr{T}_{ab}n^an^b=\mathscr{T}_{ab}\ell^a\ell^b+\mathscr{O}(\epsilon^4)\,
.
\end{equation}
Here, the vector $Y$ is the dual of $X$, i.e., $Y^a=\ell^a+Cn^a$.
For other quantities, one can also get reasonable approximations.

In addition to the null energy condition, let's assume that energy
-momentum tensor also satisfies dominant energy condition. This
means, for every future-pointing causal vector field $Z^a$, the
vector field $-\mathscr{T}_{ab}Z^b$ must be a future pointing causal
vector. This assumption of energy-momentum implies
\begin{equation}
g^{ac}\mathscr{T}_{ab}Z^b\mathscr{T}_{cd}Z^d=\|q_a^{~b}\mathscr{T}_{bc}Z^c\|^2-2(\mathscr{T}_{ab}\ell^aZ^b)(\mathscr{T}_{cd}n^cZ^d)\leq
0\, .
\end{equation}
By selecting $Z^a=\ell^a$, we get $$
\|q_a^{~b}\mathscr{T}_{bc}\ell^c\|^2\leq
2(\mathscr{T}_{ab}\ell^a\ell^b)(\mathscr{T}_{cd}n^c\ell^d)\preceq
\epsilon^2/L^4\, , $$ and then
\begin{equation}
\label{dominantF1} \|q_a^{~b}\mathscr{T}_{bc}\ell^c\|\preceq
\frac{\epsilon}{L^2}\, .
\end{equation}
From the Codazzi equation (\ref{codazziL}) and
(\ref{RiemanntoWeylEinstein}) and above result, we get
\begin{equation}
\label{dominantF2} \|
q_a^{~e}q^{bc}\ell^d\mathscr{C}_{ebcd}\|\preceq
\frac{\epsilon}{L^2}\, .
\end{equation}
For the equilibrium state, the function $C$ is identically vanished.
So $\epsilon$ is also vanished. Obviously, above conditions of the
tensors (on the future trapping horizon) give
$\sigma^{(\ell)}_{ab}=0$, $\mathscr{T}_{ab}\ell^a\ell^b=0$,
$q_a^{~b}\mathscr{T}_{bc}\ell^c=0$ and
$q_a^{~e}q^{bc}\ell^d\mathscr{C}_{ebcd}=0$. These are just the
requirements for the case of the null trapping horizon.

During above discussions, all the conditions and results are focused
on some given marginal surface. However, to study the evolution of
this marginal surface, these conditions are still not enough.
Remembering in the case of future null trapping horizon, to ensure
that some physical quantities (the area and the angular momentum
associated with the horizon) do not evolve, we have required the
condition $\mathcal{L}_X\omega_a=0$ and $\mathcal{L}_X\kappa_X=0$.
These just mean that  $\omega_a$ and $\kappa_X$ do not evolve
respect to the evolution vector $X^a$. Similarly, here there are
also {\it slowly evolving conditions }:

\begin{enumerate}[(F-i').]
\item  $\|\mathcal{L}_X\omega_a\|$ and
$|\mathcal{L}_X\kappa_X|\preceq \epsilon/L^2$;
\item $|\mathcal{L}_X\theta^{(n)}|\preceq \epsilon/L^2 $.
\end{enumerate}
From eq.(\ref{LXOmegaa}) and the restriction condition of the fields
(and their derivatives) on codimension-2 surface, it's easy to find
\begin{equation}
\|D_a\kappa_X\|\preceq \frac{\epsilon}{L^2}\, .
\end{equation}
Considering that the absolute value of $\mathcal{L}_X\kappa_X$ also
satisfies this conditions, then $\kappa_X$ is almost a constant on
the trapping horizon (at least for some finite interval of the
foliation parameter $\tau$). Thus, $\kappa_X$ can be expanded as
\begin{equation}
\label{slowlyevlovF3} \kappa_X=\kappa_o+\mathscr{O}(\epsilon)\, ,
\end{equation}
where $\kappa_o$ is the leading term of the expansion. 

The condition (F-ii') just requires that $\theta^{(n)}$ also evolves
slowly. It should be noted here: generally, in the definition of the
future trapping horizon, there are no requirements on
$\mathcal{L}_X\theta^{(n)}$ although that the sign of $\theta^{(n)}$
has close relation to the classification of the future trapping
horizon.

\vspace{.3cm}

$\bullet$ For the past trapping horizon, we can gives similar
conditions to describe the slowly expanding properties:

\begin{enumerate}[(P-i).]
\item The evolving parameter $\epsilon\ll 1$ with
$$
\frac{\epsilon^2}{L^2}=\max{\left[|C|\left(\|\sigma^{(\ell)}\|^2+(8\pi
G)\mathscr{T}_{ab}\ell^a\ell^b+\frac{1}{n-2}\theta^{(\ell)}\theta^{(\ell)}\right)\right]}\,
;
$$

\item The Ricci scalar, the $SO(1,1)$ normal connection and
the energy-momentum tensor satisfy
$$|R|\, ,\quad \|\omega_a\|^2\quad\mathrm{and} \quad (8\pi G) \mathscr{T}_{ab}\ell^an^b\preceq \frac{1}{L^2}\, ;$$

\item The derivatives of horizon fields are at most the same order in
$\epsilon$ as the (maximum of the) original fields. For example,
$$\|D_aC\|\preceq \frac{C_{m}}{L}\, ,\qquad \|D_aD_bC\|\preceq\frac{C_{m}}{L^2}\, .$$
\end{enumerate}
Similar to the future case, we can choose some foliation parameter
such that $|C|$ is small and satisfy the condition $|C|\preceq
\epsilon^2$.

Substituting these conditions into the expression of
$\mathcal{L}_{\ell}\theta^{(n)}$  in eq.(\ref{Focusln}), it's easy
to find on the past trapping horizon, we have
\begin{equation}
\left|\mathcal{L}_{\ell}\theta^{(n)}\right|\preceq \frac{1}{L^2}\, .
\end{equation}
then, by setting $B=1$ and $A=C$ in eqs.(\ref{simplifieddeform}) and
considering $\mathcal{L}_X\theta^{(n)}=0$ on the past trapping
horizon, we get
\begin{equation}
\left|\mathcal{L}_{n}\theta^{(n)}\right|=\|\sigma^{(n)}\|^2+8\pi G
\mathscr{T}_{ab}n^an^b\preceq \frac{\epsilon^2}{L^2}\, .
\end{equation}
Certainly, by assuming the null energy condition, the two terms in
the middle of above equation are nonnegative separately. It's also
easy to find
\begin{equation}
K^{(X)}_{ab}=CK^{(\ell)}_{ab}-\sigma^{(n)}_{ab}=-\sigma^{(n)}_{ab}+\mathscr{O}(\epsilon^2)\,
,
\end{equation}
and
\begin{equation}
K^{(Y)}_{ab}=CK^{(\ell)}_{ab}+\sigma^{(n)}_{ab}=\sigma^{(n)}_{ab}+\mathscr{O}(\epsilon^2)\,
.
\end{equation}
Here, the vector $Y$ is given by $Y^a=C\ell^a+n^a$.  Considering
$\epsilon\ll 1$, we have $K^{(X)}_{ab}\approx -\sigma^{(n)}_{ab}$
and $K^{(Y)}_{ab}\approx\sigma^{(n)}_{ab} $. With the same
discussion, we get
\begin{equation}
\mathscr{T}_{ab}X^aY^b=C^2\mathscr{T}_{ab}\ell^a\ell^b
-\mathscr{T}_{ab}n^an^b=
-\mathscr{T}_{ab}n^an^b+\mathscr{O}(\epsilon^4)\, .
\end{equation}
Since the null vector $n^a$ is antiself dual, i.e.,
$\epsilon_{ab}n^b=-n_a$, there is a sign difference from the case of
the future trapping horizon. Additionally, if we assume the dominant
energy condition is satisfied, then, we have
\begin{equation}
\label{dominantP} \|q_a^{~b}\mathscr{T}_{bc}n^c\|\preceq
\frac{\epsilon}{L^2}\, ,\qquad \|
q_a^{~e}q^{bc}n^d\mathscr{C}_{ebcd}\|\preceq \frac{\epsilon}{L^2}\,
.
\end{equation}
Similar to the future case, the slowly evolving conditions are
given:
\begin{enumerate}[(P-i').]
\item  $\|\mathcal{L}_X\omega_a\|$ and
$|\mathcal{L}_X\kappa_X|\preceq \epsilon/L^2$;
\item $|\mathcal{L}_X\theta^{(\ell)}|\preceq \epsilon/L^2 $.
\end{enumerate}
With these conditions, one can find that $\kappa_X$ is nearly a
constant on the past trapping horizon. So it can also be expanded as
$\kappa_X=\kappa_o+\mathscr{O}(\epsilon)$.

At the end of this subsection, some discussions are given as
follows:

(i). In the discussion of the null trapping horizons, for some given
foliation structure (some given $X$) of the trapping horizon, we can
always get $X^b\nabla_bX^a=\pm \kappa_X X^a$ by selecting a special
null frame. So, in some sense, we can foliate the null trapping
horizon arbitrarily. However, in the non-null cases, the nonvanished
functions $C$'s carry the information of the foliation, and any
restriction on them is actually a kind of restriction on the
foliations. So, to discuss the slowly evolving trapping horizon, we
have to select an appropriate foliation, i.e., the vector $X$.

(ii). For some special cases in which the codimension-2 surface has
enough symmetry, for example, the spherical symmetry, the problem is
greatly simplified~\cite{Booth:2010eu}. In this case, some
quantities such as the shear tensor $C_{ab}^{~~c}$ and the $SO(1,1)$
connection $\omega_a$ are both vanished. Further, most of the
tensors are independent of the points on the codimension-2 surface.

(iii). Clausiu like equations: For the future slowly evolving
trapping horizon, substituting the results (\ref{slowlyevlovF1}),
(\ref{slowlyevlovF2}) and (\ref{slowlyevlovF3}) into
eq.(\ref{XYdualDeformation}), and considering slowly evolving
condition (F-ii'), we get
\begin{equation}
\label{firstlawfuture}
\left(\frac{\kappa_o}{8\pi G}\right) \mathcal{L}_X\mathscr{A}=\int\epsilon_q\left[\mathscr{T}_{ab}\ell^a\ell^b+\sigma^{(\ell)}_{ab}\sigma^{(\ell)ab}\right]\nonumber\\
\, ,
\end{equation}
Similarly, for the past slowly evolving horizon, we have
\begin{equation}
\label{firstlawpast}
-\left(\frac{\kappa_o}{8\pi G}\right)\mathcal{L}_X\mathscr{A}=\int\epsilon_q\left[\mathscr{T}_{ab}n^an^b+\sigma^{(n)}_{ab}\sigma^{(n)ab}\right]\nonumber\\
\, .
\end{equation}
These are Clausius like equations. The integrals of the matter flux
$\mathscr{T}_{ab}\ell^a\ell^b$ or $\mathscr{T}_{ab}n^an^b$ and
gravitational radiation $\|\sigma^{(\ell)}\|^2$ or
$\|\sigma^{(n)}\|^2$ have the form of $\pm T\mathcal{L}_XS$. Here
$S=\mathscr{A}/4G$ and $T=|\kappa_o|/2\pi$. Certainly, these kind of
Clausius like equations hold up to the second order of the
$\epsilon$. It should be noted here: On the future trapping horizon,
we have $\mathcal{L}_X\mathscr{A}=-C\int \epsilon_q\theta^{(n)}$,
while on the past trapping horizon, we have
$\mathcal{L}_X\mathscr{A}=C\int \epsilon_q\theta^{(\ell)}$. So the
sign of  $\mathcal{L}_X\mathscr{A}$ is determined by the function
$C$ and the type of the trapping horizon. Assume the null energy
condition is satisfied, then, the positive temperature requires that
$\delta_XS\geq 0$. Here, we have defined $\delta_X=\pm
\mathrm{sign}(\kappa_o)\mathcal{L}_X$, and $``+"$ and $``-"$
correspond to the future and past respectively.

\section{Trapping Horizon in FLRW Universe}

In this section, as an example, we study the slowly evolving
trapping horizons in the FLRW universe. We also discuss the dynamics
of these kinds of trapping horizons in the formalism with the
quasilocal energy (the Misner-Sharp energy) at the end of this
section.

\subsection{Classification of the Trapping Horizons in FLRW Universe  }

Firstly, we gives the classification of the trapping horizons in the
FLRW universe. The metric of  the FLRW universe $(\mathcal{M},g)$ is
\begin{equation}
g=-dt^2 +\frac{a^2}{1-kr^2}dr^2 + a^2 r^2 d\Omega_{n-2}^2\, ,
\end{equation}
where $a=a(t)$ is scale factor and $k=0,\pm 1$, while
$d\Omega_{n-2}^2$ is the line element of an $(n-2)$-dimensional
sphere. We can decompose the metric into the form as
eq.(\ref{decompositionofmetric}) by introducing  two null vectors
$\ell$ and $n$
\begin{equation}
\label{FRWnullL} \ell_{a}dx^a=\sqrt{\frac{1}{2}}\left(-dt +
\frac{a}{\sqrt{1-kr^2}}dr\right)\, ,
\end{equation}
\begin{equation}
\label{FRWnullN} n_{a}dx^a=\sqrt{\frac{1}{2}}\left(-dt -
\frac{a}{\sqrt{1-kr^2}}dr\right)\, .
\end{equation}
So we have $h_{ab}=-\ell_an_b-n_a\ell_b$, while $q_{ab}$ is just the
metric for the sphere part, i.e., $$
q_{ab}dx^adx^b=a^2r^2d\Omega_{n-2}^2. $$ Obviously, $\ell$ and $n$
are both future directed. It's also easy to find
$\ell_a\ell^a=n_an^a=0$, $\ell_an^a=-1$ and
$q_{ab}\ell^a=q_{ab}n^a=0$. Of course, there are some freedom to
choose these two null vectors, for example, $\ell\rightarrow \lambda
\ell$ and $n\rightarrow \lambda^{-1} n$ for some positive function
$\lambda$.

From now on, we will only consider the more interesting case of the
four dimension. After a simple calculation, the expansions of the
sphere along these two null directions are given by
\begin{equation}
\label{expansionLFRW}
\theta^{(\ell)}=q^{ab}\nabla_{a}\ell_b=\sqrt{2}\left(H +
\sqrt{\frac{1}{\tilde{r}^2}-\frac{k}{a^2}}\right)\, ,
\end{equation}
\begin{equation}
\label{expansionNFRW}
\theta^{(n)}=q^{ab}\nabla_{a}n_b=\sqrt{2}\left(H -
\sqrt{\frac{1}{\tilde{r}^2}-\frac{k}{a^2}}\right)\, .
\end{equation}
Here $\tilde{r}$ is defined as $\tilde{r}=ar$. It's also easy to
find
\begin{eqnarray}
\label{deltatheta4FRW}
&&\mathcal{L}_{\ell}\theta^{(\ell)}=\dot{H}-\frac{1}{\tilde{r}^2}-H\sqrt{\frac{1}{\tilde{r}^2}-\frac{k}{a^2}}\,
, \nonumber\\
&&\mathcal{L}_{n}\theta^{(\ell)}=\dot{H}+\frac{1}{\tilde{r}^2}-H\sqrt{\frac{1}{\tilde{r}^2}-\frac{k}{a^2}}\,
,\nonumber \\
&&\mathcal{L}_{\ell}\theta^{(n)}=\dot{H}+\frac{1}{\tilde{r}^2}+H\sqrt{\frac{1}{\tilde{r}^2}-\frac{k}{a^2}}\,
,\nonumber \\
&&\mathcal{L}_{n}\theta^{(n)}=\dot{H}-\frac{1}{\tilde{r}^2}+H\sqrt{\frac{1}{\tilde{r}^2}-\frac{k}{a^2}}\,
.
\end{eqnarray}
From the expansions in eqs.(\ref{expansionLFRW}) and
(\ref{expansionNFRW}), it's easy to find that: when $H<0$, we always
have $\theta^{(n)}<0$. So the trapping horizon is given by
$\theta^{(\ell)}=0$, and this implies relations
\begin{equation}
\sqrt{\frac{1}{\tilde{r}^2}-\frac{k}{a^2}}=-H\, ,\qquad
\frac{1}{\tilde{r}^2}=H^2+\frac{k}{a^2}\, .
\end{equation}
After substituting above relations, eq.(\ref{deltatheta4FRW})
becomes
\begin{eqnarray}
&&\mathcal{L}_{\ell}\theta^{(\ell)}=\dot{H}-\frac{k}{a^2}\,
,\qquad\mathcal{L}_{n}\theta^{(\ell)}=\dot{H}+2H^2+\frac{k}{a^2}\,
,\nonumber
\\&&\mathcal{L}_{\ell}\theta^{(n)}=\dot{H}+\frac{k}{a^2}\,
,\qquad\mathcal{L}_{n}\theta^{(n)}=\dot{H}-2H^2-\frac{k}{a^2}\, .
\end{eqnarray}
These mean that the marginal surface is always future. Further, the
outer or inner of the marginal surface is determined by the sign of
$\dot{H}+2H^2+k/{a^2}$. So the marginal surface is outer if
$\dot{H}-k/{a^2}<-2(H^2+k/a^2)$ and inner if
$\dot{H}-k/{a^2}>-2(H^2+k/a^2)$. When the null energy condition is
satisfied, we always have $\dot{H}-k/a^2\le 0$. So the marginal
surface may be inner or outer.

For $H>0$, we always have $\theta^{(\ell)}>0$. When
$\theta^{(n)}=0$, we have
\begin{equation}
\sqrt{\frac{1}{\tilde{r}^2}-\frac{k}{a^2}}=H\, ,\qquad
\frac{1}{\tilde{r}^2}=H^2+\frac{k}{a^2}\, ,
\end{equation}
and now eqs.(\ref{deltatheta4FRW}) becomes
\begin{eqnarray}
&&\mathcal{L}_{\ell}\theta^{(\ell)}=\dot{H}-2H^2-\frac{k}{a^2}\,
,\qquad\mathcal{L}_{n}\theta^{(\ell)}=\dot{H}+\frac{k}{a^2}\,
,\nonumber
\\&&\mathcal{L}_{\ell}\theta^{(n)}=\dot{H}+2H^2+\frac{k}{a^2}\,
,\qquad\mathcal{L}_{n}\theta^{(n)}=\dot{H}-\frac{k}{a^2}\, .
\end{eqnarray}
So the marginal surface is always past. The marginal surface is
outer if $\dot{H}-k/{a^2}>-2(H^2+k/{a^2})$ and inner if
$\dot{H}-k/{a^2}<-2(H^2+k/{a^2})$. When the null energy condition is
satisfied, we always have $\dot{H}-k/{a^2}\le 0 $. So there are some
ranges in which the marginal surface is outer or inner if we do not
impose some additional energy conditions.

It's easy to find, in the cases with $H=0$, we always have
$\theta^{(\ell)}>0$ and $\theta^{(n)}<0$, so there are no trapping
horizons in this case. This is expectable because now the spacetime
is actually a flat spacetime.

The FLRW universe is a typical spherically symmetric spacetime, so,
as discussed before, the components of the evolution vector $X$,
i.e., $A$ and $B$, are functions which only depend on the
coordinates $t$ and $r$. So they are constants on the marginal
surface. It's easy to find, for $H<0$, on the future trapping
horizon, the relation $ \mathcal{L}_X\theta^{(\ell)}=0 $ gives
\begin{equation}
\label{Xfuture}
A\mathcal{L}_{\ell}\theta^{(\ell)}=B\mathcal{L}_{n}\theta^{(\ell)}\,
.
\end{equation}
With the null energy condition, from  eq.(\ref{Focusln}), we always
have $\mathcal{L}_{\ell}\theta^{(\ell)}\le 0$. Thus, for the outer
marginal surface, we have $\mathcal{L}_{n}\theta^{(\ell)}<0$, then,
$A$ and $B$ have same signs. While for the inner marginal surface,
we have $\mathcal{L}_{n}\theta^{(\ell)}>0$, $A$ and $B$ have
opposite signs. In both cases, the relation between $A$ and $B$ is
given by
\begin{equation}
\label{BFRW}
B=A\left(\frac{\mathcal{L}_{\ell}\theta^{(\ell)}}{\mathcal{L}_{n}\theta^{(\ell)}}\right)=A\left(\frac{\dot{H}-k/{a^2}}{\dot{H}+2H^2+k/{a^2}}\right)\,
.
\end{equation}

In the cases $H>0$, we have past trapping horizon. It's easy to find
the relation $\mathcal{L}_X\theta_{(n)}=0 $  on the trapping horizon
gives
\begin{equation}
\label{Xpast}
A\mathcal{L}_{\ell}\theta^{(n)}=B\mathcal{L}_{n}\theta^{(n)}\, .
\end{equation}
By using eq.(\ref{Focusln}), with the null energy condition, we
always have $\mathcal{L}_{n}\theta^{(n)}\le 0$. So, for the outer
marginal surface, we have $\mathcal{L}_{\ell}\theta^{(n)}>0$, and
then $A$ and $B$ have opposite signs. While for the inner marginal
surface, we get $\mathcal{L}_{\ell}\theta^{(n)}<0$, so $A$ and $B$
have same signs. In both cases, $A$ is given by
\begin{equation}
\label{AFRW}
A=B\left(\frac{\mathcal{L}_{n}\theta^{(n)}}{\mathcal{L}_{\ell}\theta^{(n)}}\right)=B\left(\frac{\dot{H}-k/{a^2}}{\dot{H}+2H^2+k/{a^2}}\right)\,
.
\end{equation}

Conclusively, in FLRW universe, for the future trapping horizon, we
can set $X_a=A(\ell_a-\alpha n_a)$ with $\alpha$ given by
\begin{equation}
\label{Cfunction}
\alpha=\frac{\dot{H}-k/{a^2}}{\dot{H}+2H^2+k/{a^2}}\, .
\end{equation}
For the past trapping horizon, $X$ has form
$X_a=B(\alpha\ell_a-n_a)$ with the same $\alpha$ given in above
equation. The classification of ``outer" and ``inner" are indicated
by the sign of $\alpha$.

The future trapping horizon is outer when $\alpha >0$, and inner
when $\alpha<0$. Since in the future cases, we have
$X^aX_a=2A^2\alpha$, so the  future outer trapping horizon is a
spacelike hypersurface, while the  future inner trapping horizon is
a timelike hypersurface.

Similarly, the past trapping horizon is outer when $\alpha <0$ and
inner when $\alpha>0$. In this case, we have $X^aX_a=2B^2\alpha$, so
the  past outer trapping horizon is timelike, while the  past inner
trapping horizon is spacelike.

We have to mention some special cases: For the future trapping
horizon, if $\mathcal{L}_{\ell}\theta^{(\ell)}=0$, then from
eq.(\ref{Xfuture}), we have $\alpha=0$. The trapping horizon reduces
to a null hypersurface. Similarly, the past trapping horizon also
reduces to a null hypersurface when $\alpha=0$.

So, for the future and the past trapping horizons, the general form
of the vector $X^a$ is given by $A(\ell^a-\alpha n^a)$ and
$B(\alpha\ell^a-n^a)$ respectively. Certainly, different $A$'s and
$B$'s correspond to the different foliations of the trapping
horizons. However, it should be noted here:  $A$ and $B$ are both
constants on the codimension-2 surface. So, for the future trapping
horizon, we can view the factor $A$ as a factor $f$ provided by some
relabeling of the foliation. Similarly, for the past trapping
horizon, $B$ can be viewed as a relabeling factor by selecting some
foliation parameter $\tau$. By this consideration, the evolution
vector $X^a$ has the form
\begin{equation}
X^a=f(\ell^a-\alpha n^a)
\end{equation}
in the future case, and
\begin{equation}
X^a=f(\alpha\ell^a-n^a)
\end{equation}
in the past case. The null vector still has the form
(\ref{FRWnullL}) and (\ref{FRWnullN}). In the following discussion,
we will fix $f$ to be unit. We will find this selection of  $f$
consists with the requirement $|C|=|f^2\alpha| \preceq \epsilon^2$.
Here, $\alpha$ plays the role of  $AB$ in eq.(\ref{Cinbarframe}).

\subsection{Null Trapping Horizons in FLRW Universe}
For the null trapping horizon, the function $\alpha$ is identically
vanished. From the expression of $\alpha$ in (\ref{Cfunction}), we
have
$$\dot{H}-k/a^2=0$$ on the trapping horizons. The surface gravity
associated with $\ell^a$ for the null future trapping horizon is
given by
\begin{equation}
\label{frwnullkappal} \kappa_{\ell}=\frac{H}{\sqrt{2}}\, .
\end{equation}
The condition that  $\kappa_{X}$ (now $X^a=\ell^a$) is a constant on
the null future trapping horizon gives
$\mathcal{L}_{\ell}\kappa_{\ell}=\dot{H}/2=0$ (So we can only
consider the case with $k=0$). This means $H$ is a negative constant
on the horizon. It's also easy to find
\begin{equation}
\theta^{(n)}=2\sqrt{2}H\, ,\quad
\mathcal{L}_{n}\theta^{(\ell)}=2H^2\, ,\quad
\mathcal{L}_{\ell}\theta^{(n)}=0\, ,\quad
\mathcal{L}_{n}\theta^{(n)}=-2H^2\, .
\end{equation}
So there are only null future inner trapping horizon.

Similarly, for the past case, the surface gravity is given by
\begin{equation}
\label{frwnullkappan} \kappa_{n}=-\frac{H}{\sqrt{2}}\, .
\end{equation}
and  $\mathcal{L}_n\kappa_n=0$ also gives that $\dot{H}=0$. So $H$
is a positive constant on the past null trapping horizon. We also
have
\begin{equation}
\theta^{(\ell)}=2\sqrt{2}H\, ,\quad
\mathcal{L}_{\ell}\theta^{(n)}=2H^2\, ,\quad
\mathcal{L}_{n}\theta^{(\ell)}=0\, ,\quad
\mathcal{L}_{\ell}\theta^{(\ell)}=-2H^2\, .
\end{equation}
This means there are only null past outer trapping horizon.

Conclusively, on the null trapping horizons (future and past), the
Hubble parameter $H$ is always a constant. These kinds of horizons
exist only when $k=0$. Further, only inner horizon exists in the
future case, and only outer horizon exists in the past case. In the
following discussions, we always set $k=0$. It should be noted here:
the surface gravity (\ref{frwnullkappal}) (or (\ref{frwnullkappan}))
is defined up to a positive constant  coefficient.

\subsection{Slowly Evolving Trapping Horizon in FLRW Universe}

Since the spherical symmetry, it's very simple to  study  the slowly
evolving properties of the trapping horizon in the FLRW unverse. In
this case, most of the scalars on the codimension-2 surface are
constants. For example, from the definition, the evolving parameter
$\epsilon$ in the condition (F-i) becomes (we only consider the four
dimension case, and choose  $L$ to be the radius $\tilde{r}=1/|H|$
for $k=0$.)
\begin{equation}
\label{epsilonfrw1}
\frac{\epsilon^2}{\tilde{r}^2}=|\alpha|\left(\mathscr{G}_{ab}n^an^b+
\frac{1}{2}\theta^{(n)}\theta^{(n)}\right)\, .
\end{equation}
Similarly, for the past trapping horizon, the condition (P-i) is
given by
\begin{equation}
\label{epsilonfrw2}
\frac{\epsilon^2}{\tilde{r}^2}=|\alpha|\left(\mathscr{G}_{ab}\ell^a\ell^b+
\frac{1}{2}\theta^{(\ell)}\theta^{(\ell)}\right)\, .
\end{equation}
Since  $f$ is chosen to be unit, so we have $|C|=|\alpha|$.
Straightforward calculation shows: in both cases, on the trapping
horizons,  $\epsilon$'s are given by
\begin{equation}
\label{epsilonandC}
\epsilon^2=|\alpha|\left(4-\frac{\dot{H}}{H^2}\right)\, .
\end{equation}
Remembering that we are considering the trapping horizons which are
near the null future inner trapping horizon or the null past outer
trapping horizon, so the functions $\alpha$'s are assumed to be
negative. By defining
\begin{equation}
\label{paremeters} s = -\frac{\dot{H}}{H^2} > 0\, ,
\end{equation}
then, from the expression of $\alpha$ in eq.(\ref{Cfunction}), we
have
\begin{equation}
\label{Cdelta} \alpha=-\frac{s}{2-s}\, .
\end{equation}
To ensure that $\alpha<0$, we have to require $s<2$. The evolution
parameter $\epsilon$ now has a simple form
\begin{equation}
\label{epsilondelta} \epsilon^2=s \left(\frac{4+s}{2-s}\right)\, .
\end{equation}
It should be emphasized: the result of $\epsilon$ in above equation
is independent of the rescaling of the null frame and the relabeling
of the foliation. Under the relabeling of the foliation, we have
$\alpha\rightarrow f^2 \alpha$, while the term inside the round
brackets in eq.(\ref{epsilonfrw1}) (or (\ref{epsilonfrw2})) will
accept a factor $1/f^2$. So $\epsilon$ is invariant.

By eq.(\ref{epsilondelta}), we can express $s$ as a function of
$\epsilon$. Considering $\epsilon\ll 1$, we have
\begin{equation}
\label{FRWdotH-ka2} s =- \frac{\dot{H}}{H^2} \approx
\frac{\epsilon^2}{2} \ll 1\, .
\end{equation}
So eq.(\ref{Cdelta}) gives $|\alpha|\approx \epsilon^2/4$. Actually,
from eqs.(\ref{epsilondelta}) and (\ref{Cdelta}), we can express
$\alpha$ as a function of $\epsilon$ explicitly. On the other hand,
if we do not choose $f=1$, then we have $|C|=|f^2\alpha|\approx
f^2\epsilon^2/4$. To ensure the relation (\ref{Ccondition}), $f^2$
has to be finite and order one. Certainly, the selection of $f=1$
satisfies this requirement.

It's easy to find that
$$\mathcal{G}_{ab}\ell^an^b=\frac{3}{\tilde{r}^2}+\dot{H}=\frac{3-s}{\tilde{r}^2}\, .$$
So, when $s\ll 1 $, the conditions (F-ii) and (P-ii) are easily
satisfied, i.e., $\mathcal{G}_{ab}\ell^an^b\preceq 1/\tilde{r}^2$.
Obviously, the conditions (F-iii) and (P-iii) are trivially
satisfied. Thanks to the spherical symmetry and the Weyl flat of the
FLRW universe, eqs.(\ref{dominantF1}), (\ref{dominantF2}) and
(\ref{dominantP}) are also trivially satisfied even without any
energy condition (In the general case, the dominant energy condition
is required).

For the future inner trapping horizon ($H<0$), the slowly evolving
conditions (F-i') and (F-ii') reduce to
 $|\mathcal{L}_X\kappa_X|\preceq \epsilon/\tilde{r}^2$ and
 $|\mathcal{L}_X\theta^{(n)}|\preceq \epsilon/\tilde{r}^2 $.
It's easy to find
\begin{equation}
\mathcal{L}_X\theta^{(n)}= -H^2s \left(1+\frac{2+s}{2-s}\right)\, .
\end{equation}
Therefore, from eq.(\ref{FRWdotH-ka2}), we have
$$|\mathcal{L}_X\theta^{(n)}|\preceq
\epsilon^2H^2=\epsilon^2/\tilde{r}^2\, .$$ So, when $\epsilon\ll 1$,
$|\mathcal{L}_X\theta^{(n)}|\preceq \epsilon/\tilde{r}^2 $ is
automatically satisfied. After substituting
$\kappa_{\ell}=H/\sqrt{2}$ and $\kappa_{n}=-H/\sqrt{2}$, we find
\begin{equation}
\label{Kappaxfuture}
\kappa_X=\frac{H}{\sqrt{2}}\left(1-\frac{s}{2-s}\right)<0\, .
\end{equation}
After a simple calculation, we have
\begin{equation}
\mathcal{L}_X\kappa_X=-\frac{2 H^2s}{\left(2-
s\right)^3}\left[2-s+s^2+\left(\frac{\ddot{H}}{\dot{H}H}\right)\right]\,
.
\end{equation}
Thus, if we require
\begin{equation}
\label{slowevolvekappafrw} \left|\frac{\ddot{H}}{H^3}\right|\preceq
\epsilon \, ,
\end{equation}
then we have  $|\mathcal{L}_X\kappa_X|\preceq
\epsilon/\tilde{r}^2=\epsilon H^2$.  The condition
(\ref{slowevolvekappafrw}) means $\dot{H}$ is also required to be
slowly evolving.

In the past outer case ($H>0$), the conditions (P-i') and (P-ii')
reduce to $|\mathcal{L}_X\kappa_X|\preceq \epsilon/\tilde{r}^2$ and
$|\mathcal{L}_X\theta^{(\ell)}|\preceq \epsilon/\tilde{r}^2 $. A
simple calculation shows: on the past outer trapping horizon, we
have
\begin{equation}
\mathcal{L}_X\theta^{(\ell)}= H^2s \left(1+\frac{2+s}{2-s}\right)\,
.
\end{equation}
So the condition $|\mathcal{L}_X\theta^{(\ell)}|\preceq
\epsilon/\tilde{r}^2 $ is also automatically satisfied when
$\epsilon\ll 1$.  Now, it's easy to find the surface gravity
$\kappa_X$ is given by
\begin{equation}
\label{Kappaxpast}
\kappa_X=\frac{H}{\sqrt{2}}\left(1-\frac{s}{2-s}\right)>0\, ,
\end{equation}
and
\begin{equation}
\mathcal{L}_X\kappa_X=\frac{2 H^2s}{\left(2-
s\right)^3}\left[2-s+s^2+\left(\frac{\ddot{H}}{\dot{H}H}\right)\right]\,
.
\end{equation}
So the condition on $|\mathcal{L}_X\kappa_X|$ gives same constraint
as the one of the future inner trapping horizon, i.e.
$|\ddot{H}|\preceq \epsilon H^3$.

Conclusively, the requirement of the evolving parameter $\epsilon\ll
1$ automatically implies that $s=-\dot{H}/H^2$ is very small. While
the slowly evolving condition of $\kappa_X$ requires that
$|\ddot{H}/H^3|$ is also a small quantity.

For the slowly evolving past trapping horizon ($H>0$),
$s=-\dot{H}/H^2$ is very small. This is just one of the condition of
the slow-roll inflation. Additionally, the slowly evolving condition
also requires that $|\ddot{H}/H^3|$ is small to ensure that the
surface gravity $\kappa_X$ changes slowly on the trapping horizon.
This way, the system is near an equilibrium state. This requirement
may has some relation to the second condition of the slow-roll
inflation scenario. Of course, here, we have not introduce any
scalar field and the corresponding potential.

For the past horizon, from eq.(\ref{firstlawpast}), we have
\begin{equation}
\label{firstlawfrw1} -\frac{\kappa_o}{8\pi
G}\mathcal{L}_X\mathscr{A}=\int\epsilon_q \mathscr{T}_{ab}n^an^b\, .
\end{equation}
Up to second order of $\epsilon$ (or the first order of $s$).
Actually, it's easy to find
\begin{equation}
\mathcal{L}_X\mathscr{A}=
-\sqrt{2}\left(\frac{8\pi}{H}\right)\left(\frac{s}{2-s}\right)\approx
-\sqrt{2}\left(\frac{4\pi}{H}\right)s + \mathscr{O}(s^2)
\end{equation}
So we have $\mathcal{L}_X\mathscr{A}<0$. This just means that
$\mathscr{A}$ decreases along $X$ direction. Noted that $X^a=-n^a$
if $s=0$, so $X$ is past pointing. By this consideration, the
negative $\mathcal{L}_X\mathscr{A}$ just means the area of the
marginal sphere of the past trapping horizon increases along the
future direction. The leading order of $\kappa_X$ is $H/\sqrt{2}$,
so we have $ -(\kappa_o/8\pi G)\mathcal{L}_X\mathscr{A}=s/2G
+\mathscr{O}(s^2) $. It's also easy to find
\begin{equation}
\int\epsilon_q \mathscr{T}_{ab}n^an^b=\frac{s}{2G}\, .
\end{equation}
Thus, eq.(\ref{firstlawpast}) holds up to the second order of
$\epsilon$. It should be emphasized here: the surface gravity
$\kappa_X$ is defined up to a constant factor. Actually, the factor
$\sqrt{2}$  in $\kappa_X$ comes from the selection of the null
vectors (\ref{FRWnullL}) and (\ref{FRWnullN}). From this Clausius
relation like equation, we have
$$T=\frac{\kappa_X}{2\pi}\sim \frac{H}{2\pi
}\left(1-\frac{s}{2}\right)+\mathscr{O}(\epsilon^4)\, .
$$

\subsection{Quasilocal Energy and Horizon Dynamics}
To compare with the results for the slowly evolving trapping horizon
in the FLRW universe, in this subsection, we also consider the case
where $k=0$. The quasilocal energy (\ref{Misnersharp}) (or the
energy (\ref{GenralizedHawkingmass})) inside the sphere with the
radius $\tilde{r}$ is
\begin{equation}
\mathscr{E}= \frac{\tilde{r}}{2G}\left(1-\nabla_a\tilde{r}\nabla^a
\tilde{r}\right)\, .
\end{equation}
It's easy to find the effective surface gravity $\bar{\kappa}$ in
eq.(\ref{kappabarsphere}) is given by $\bar{\kappa}=1/2
\tilde{r}=|H|/2$ which has the same form of the surface gravity of
Schwarzschild black hole. While the surface gravity $\kappa$ in
eq.(\ref{surfacegravity1}) becomes
\begin{equation}
\frac{\kappa}{2\pi}=-\frac{|H|}{2\pi}\left(1-\frac{s}{2}\right)\, ,
\end{equation}
where $s$ is defined in eq.(\ref{paremeters}). So the temperature of
the past outer trapping horizon is
$T=|\kappa|/2\pi=H\left(1-s/2\right)/2\pi$.

If we omit the factors $\sqrt{2}$ in eqs.(\ref{Kappaxfuture}) and
(\ref{Kappaxpast}), the temperatures corresponding to $\kappa$ and
$\kappa_X$ are coincided up to the second order of $\epsilon$. In
the description with the quasilocal energy, the Clausius relation is
given by (corresponding to eq.(\ref{firstlawfrw1}))
\begin{equation}
A\psi_aX^a=\frac{\kappa}{2\pi}\mathcal{L}_XS\, ,
\end{equation}
where $S=\mathscr{A}/4G$ and $X$ is the evolution vector of the
trapping horizon. Of course, above equation is exactly held on the
trapping horizon. More detailed discussions of the thermodynamics of
the FLRW universe in various gravity theories can be found
in~\cite{Frolov:2002va,Danielsson:2004xw,Bousso:2004tv,Calcagni:2005vn,CK,CCH08,Cai07,Pad02,Maeda:2009ds}
and related topics can be found in a review paper~\cite{Pad09}.

\section{Conclusion and Discussion}

In this paper, we have studied the deformation of some spacelike
submanifold with an arbitrary codimension. By requiring the
projection operator is Lie dragging along a given normal vector $X$,
we get eq.(\ref{DeltaXKY}). In the case of codimension-1, it reduces
to the evolution equation of the extrinsic curvature of the
spacelike hypersurface (\ref{codimension12}) (and
eq.(\ref{codimension11})). Even in this simple case, there are also
some interest applications on thermodynamics. For example, in static
case, the evolution equation (\ref{codimension11}) becomes
$$
N\mathscr{R}_{ab}u^au^b=D^aD_aN\, .
$$
By defining the so called Tolman-Komar  mass $M_K=\int \epsilon_S
N\mathscr{R}_{ab}u^au^b $ inside some $(n-2)$-closed surface $S$
embedded in the hypersurface, one can give some discussion on the
thermodynamics associated with the surface $S$. More details can be
found in~\cite{Padmanabhan:2010xh,Verl,Abreu:2010sc}. In the more
interesting case of codimension-2, eq.(\ref{DeltaXKY}) reduces to
eq.(\ref{FocusXY1}), i.e.,
\begin{eqnarray*}
\mathcal{L}_X\theta^{(Y)}&=&-\left(\mathscr{G}_{ab}+K_{cda}K^{cd}{}_{b}\right)\left[X^aY^b-h^{ab}\left(X_eY^e\right)\right]\nonumber\\
&&+\frac{1}{2}\left(R-K_{abc}K^{abc}-K_cK^c\right)\cdot\left(X_eY^e\right)
\nonumber\\
&&-Y^e\tilde{D}_c\tilde{D}^cX_e-K_{c}\left(X^e\nabla_{e}Y^c\right)\,
.
\end{eqnarray*}
This result is frame independent, and it reduces to the well known
(cross) focusing  equations after selecting a local null frame. The
deformation of the $SO(1,1)$ connection is given by
$$
\mathcal{L}_X\omega_a-D_a\kappa_X=\left(\frac{n-3}{n-2}\right)D_a\theta^{(Y)}-D_{c}\sigma^{(Y)c}_{~a}
+K_{c}\tilde{D}_aY^c+q_a^{~b}Y^c\mathscr{G}_{bc}
$$
with $Y_a=\epsilon_{ab}X^b$ if we define an appropriate ``surface
gravity" $\kappa_X$. It's just a Damour-Navier-Stokes like equation
if $X$ is self-dual or anti-self-dual. We have investigated the
relation between these deformation equations and the dynamics of the
trapping horizon in two different formalisms: with and without
introducing some quasilocal energy. In the first formalism, we have
proposed a generalized energy (\ref{GenralizedHawkingmass}) in the
higher dimension Einstein gravity theory, i.e.,
$$
\mathscr{E}=\frac{\left(\int\epsilon_q\right)^{\frac{n-3}{n-2}}}{16\pi
G\left(\Omega_{n-2}\right)^{\frac{1}{n-2}}(n-3)}\Bigg{\{}\frac{\int\epsilon_q
R}{\left(\int\epsilon_q\right)^{\frac{n-4}{n-2}}}-\left(\frac{n-3}{n-2}\right)\frac{\int\epsilon_qK_cK^c}{\left(\int\epsilon_q\right)^{\frac{n-4}{n-2}}}\Bigg{\}}\,
.
$$
This energy reduces to the Hawking energy in the four dimension. In
the case of general spherical symmetry, it reduces to the
Misner-Sharp energy in the higher dimension. We have also studied
the deformation of this quasilocal energy, and the deformation
equation has been given in eq.(\ref{deformationofenergy}). Once the
requirement $\mathcal{L}_X\mathscr{K}=0$ is fulfilled, on the
trapping horizon, we always have
$$\mathcal{L}_X\mathscr{E}=\left(\frac{n-3}{n-2}\right)\left(\frac{\mathscr{E}}{\mathscr{A}}\right)\mathcal{L}_{X}\mathscr{A}\, .$$
This result is the total variation (evolution) of the energy on the
trapping horizon, and it can be decomposed into two parts as in the
four dimension: contributions from the matter fields
($\mathscr{T}_{ab}$) and the contribution from gravitational
radiation ($\|\sigma\|^2$ and $\|\zeta\|^2$). When the marginal
surface is Einstein, we also study the first law like equation of
the trapping horizon. It has a similar form as the one with the
spherical symmetry. However, generally, it's impossible to define a
surface gravity which is a constant on the marginal surface.
Further, it also evolves on the trapping horizon even in the
spherically symmetric case. This means the system is generally
nonequilibrium (even far from equilibrium point) if we regard the
temperature is proportional to the surface gravity. To make the
problem clear, we have studied some near equilibrium state by
considering the slowly evolving trapping horizon proposed by Booth
in the formalism without the quasilocal energy.

To study the past trapping horizon in the FLRW universe, we
generalize the definition of the slowly evolving future outer
trapping horizon to the past trapping horizon. We find, for the
slowly evolving past trapping horizon, the Clausius like equation is
modified to be
$$ -\left(\frac{\kappa_o}{8\pi G}\right)\mathcal{L}_X\mathscr{A}
=\int\epsilon_q\left[\mathscr{T}_{ab}n^an^b+\|\sigma^{(n)}\|^2\right]\,
.
$$
After classifying the trapping horizon in the FLRW universe, as an
example, we study the slowly evolving trapping horizon in this
spacetime. We find, to require  the past trapping horizon in the
FLRW universe to be slowly evolving, the Hubble parameter has to
satisfy $-\dot{H}/H^2\approx \epsilon^2/2\ll 1$. Further,
$|\ddot{H}|/H^3$ is required to be (at most) the order of
$\epsilon$. These conditions have close relation to the scenario of
the slow-roll inflation. We also compare the temperature defined in
the formalism with the quasilocal energy ($T=|\kappa/2\pi|$) and the
temperature defined in the formalism without the quasilocal energy (
$T=|\kappa_X/2\pi|$). We find these two temperatures are essentially
the same up to the second order of the slowly evolution parameter
$\epsilon$.

\section{Acknowledgement}

The author would like to thank Prof. Rong-Gen Cai and Prof.
Nobuyoshi Ohta for their long term encouragement and kind help. The
author would like to thank Prof. K.i.Maeda, Prof. S.Mukohyama and
Prof. T.Shiromizu for their useful discussions on the surface
gravity and the generalized energy form. The author also thanks
Dr.Masato Minamitsuji, Dr.Masato Nozawa and Dr.Chul-Moon Yoo for
their useful discussions and kind help. The author would like to
thank Dr.Booth for his valuable suggestions and comments.  This work
is supported by JSPS fellowship No.P09225.

\vspace{.3 cm}

\begin{center}
{\bf Appendix A}
\end{center}
In this appendix, we give a detailed derivation of
eq.(\ref{DeltaXKY}). It's easy to find the first term in
eq.(\ref{LXKYAB}) is given by
\begin{eqnarray}
\label{first1}
&&q_{a}^{~c}q_{b}^{~d}X^e\nabla_eK^{(Y)}_{cd}=q_{a}^{~c}q_{b}^{~d}X^e\nabla_e\left(q_{c}^{~f}q_{d}^{~g}\nabla_{f}Y_g\right)\nonumber\\
&&=q_{a}^{~c}q_{b}^{~g}X^e\nabla_eq_{c}^{~f}\nabla_{f}Y_g
+q_{a}^{~f}q_{b}^{~d}X^e\nabla_eq_{d}^{~g}\nabla_{f}Y_g
+q_{a}^{~f}q_{b}^{~g}X^e\nabla_e\nabla_{f}Y_g\nonumber\\
&&=\left(q_{b}^{~g}\nabla_{f}Y_g \right)\tilde{D}_aX^f
+\left(q_{a}^{~f}\nabla_{f}Y_g\right)\tilde{D}_{b}X^g
+q_{a}^{~f}q_{b}^{~g}X^e\nabla_e\nabla_{f}Y_g \nonumber\\
&&=\left(q_{b}^{~g}\nabla_{f}Y_g \right)\left(\tilde{D}_aX^f\right)
+\left(\tilde{D}_aY_g\right)\left(\tilde{D}_{b}X^g\right)\nonumber\\
&&
~~+q_{a}^{~f}q_{b}^{~g}X^eY^h\mathscr{R}_{efgh}+\underline{q_{a}^{~f}q_{b}^{~g}X^e\nabla_f\nabla_{e}Y_g}\,
.
\end{eqnarray}
Here, from second line to third line, we have used
eq.(\ref{XnablaY1}). Again, by using eq.(\ref{XnablaY1}) or
eq.(\ref{XnablaY2}), we have
\begin{eqnarray}
&&q_{a}^{~f}q_{b}^{~g}X^e\nabla_f\nabla_{e}Y_g=-q_{a}^{~f}\nabla_f(Y_c\tilde{D}_bX^c)\nonumber\\
&&-q_{a}^{~f}q_{b}^{~g}\nabla_fX^e\nabla_{e}Y_g-q_{a}^{~f}\nabla_fq_{b}^{~g}X^e\nabla_{e}Y_g\nonumber\\
&&=-\left(\tilde{D}_aY_c\right)\left(\tilde{D}_{b}X^c\right)-\underline{q_{a}^{~f}Y_c\nabla_f(\tilde{D}_bX^c)}\nonumber\\
&&-K^{(X)}_{~~a}{}^{c}K^{(Y)}_{cb}-\left(q_{b}^{~g}\nabla_{f}Y_g
\right)\left(\tilde{D}_aX^f\right)-q_{a}^{~f}\nabla_fq_{b}^{~g}X^e\nabla_{e}Y_g\,
.
\end{eqnarray}
It's also easy to find
\begin{eqnarray}
&&q_{a}^{~f}Y_c\nabla_f(\tilde{D}_bX^c)=K^{(Y)}_{~~a}{}^{c}K^{(X)}_{cb}
+q_{a}^{~f}q_{b}^{~e}Y^g\nabla_f\nabla_{e}X_g\nonumber\\
&&~~+q_{a}^{~f}\nabla_fq_{b}^{~g}Y^e\nabla_{g}X_e\, .
\end{eqnarray}
So we get
\begin{eqnarray}
&&q_{a}^{~f}q_{b}^{~g}X^e\nabla_f\nabla_{e}Y_g=-q_{a}^{~f}\nabla_f(Y_c\tilde{D}_bX^c)\nonumber\\
&&~~~-q_{a}^{~f}q_{b}^{~g}\nabla_fX^e\nabla_{e}Y_g-q_{a}^{~f}\nabla_fq_{b}^{~g}X^e\nabla_{e}Y_g\nonumber\\
&&=-\left(\tilde{D}_aY_c\right)\left(\tilde{D}_{b}X^c\right)-K^{(Y)}_{~~a}{}^{c}K^{(X)}_{cb}-K^{(X)}_{~~a}{}^{c}K^{(Y)}_{cb}\nonumber\\
&&~~~-\left(q_{b}^{~g}\nabla_{f}Y_g
\right)\left(\tilde{D}_aX^f\right)~-q_{a}^{~f}\nabla_fq_{b}^{~g}Y^e\nabla_{g}X_e\nonumber\\
&&~~~-q_{a}^{~f}\nabla_fq_{b}^{~g}X^e\nabla_{e}Y_g
-q_{a}^{~f}q_{b}^{~e}Y^g\nabla_f\nabla_{e}X_g\, .
\end{eqnarray}
Then, after substituting above equation into eq.(\ref{first1}), we
find
\begin{eqnarray}
&&q_{a}^{~c}q_{b}^{~d}X^e\nabla_eK^{(Y)}_{cd}=q_{a}^{~f}q_{b}^{~g}X^eY^h\mathscr{R}_{efgh}-q_{a}^{~f}q_{b}^{~e}Y^g\nabla_f\nabla_{e}X_g\,
,\nonumber\\
&&~~~-K^{(X)}_{~~a}{}^{c}K^{(Y)}_{cb}-K^{(X)}_{~~b}{}^{c}K^{(Y)}_{ca}+
K^{(Z)}_{ab}\, .
\end{eqnarray}
where $Z_a=\mathcal{L}_XY_a$. Substituting above equations into
eq.(\ref{LXKYAB}), we arrive at
\begin{equation}
\label{LXKYAB1} \mathcal{L}_X K^{(Y)}_{ab}=
K^{(Z)}_{ab}+q_{a}^{~f}q_{b}^{~g}X^eY^h\mathscr{R}_{efgh}-\underline{q_{a}^{~f}q_{b}^{~e}Y^g\nabla_f\nabla_{e}X_g}\,
.
\end{equation}
However, since we have
\begin{eqnarray}
&&q_{a}^{~f}q_{b}^{~e}Y^g\nabla_f\nabla_{e}X_g=q_{a}^{~f}Y^g\nabla_f\left(q_{b}^{~e}\nabla_{e}X_g\right)-q_{a}^{~f}\nabla_fq_{b}^{~e}Y^g\nabla_{e}X_g\nonumber\\
&&=q_{a}^{~f}Y^g\nabla_f\left(q_{g}^{~h}q_{b}^{~e}\nabla_{e}X_h+h_{g}^{~h}q_{b}^{~e}\nabla_{e}X_h\right)\nonumber\\
&&~~~-q_{a}^{~f}\nabla_fq_{b}^{~e}Z_e
+q_{a}^{~f}\nabla_fq_{b}^{~e}X^g\nabla_{g}Y_e\nonumber\\
&&=q_{a}^{~f}Y^g\nabla_f\left(K^{(X)}_{bg}+\tilde{D}_bX_g\right)+q_{a}^{~f}q_{b}^{~e}\nabla_fZ_e
\nonumber\\
&&~~~+q_{a}^{~f}\nabla_fq_{b}^{~e}(q_e^{~h}+h_{e}^{~h})X^g\nabla_{g}Y_h\nonumber\\
&&=-q_{a}^{~f}\nabla_fY^gK^{(X)}_{bg}+Y^g\tilde{D}_a\tilde{D}_bX_g+K^{(Z)}_{~ab}
\nonumber\\
&&~~~+K_{acb}\left(X^g\nabla_{g}Y^c\right)+K_{abc}\left(X^g\nabla_{g}Y^c\right)\nonumber\\
&&=-K^{(Y)}_{a}{}^gK^{(X)}_{bg}+Y^g\tilde{D}_a\tilde{D}_bX_g+K^{(Z)}_{~ab}
\nonumber\\
&&~~~-K_{acb}\left(Y_e\tilde{D}^cX^e\right)+K_{abc}\left(X^g\nabla_{g}Y^c\right)\,
,
\end{eqnarray}
and after substituting this result into eq.(\ref{LXKYAB1}), we
finally get the deformation equation (\ref{DeltaXKY}).

\begin{center}
{\bf Appendix B}
\end{center}
Remembering that $\epsilon_{ab}$ is defined as
\begin{equation}
\epsilon_{ab}=n_a\ell_b-\ell_an_b\, \qquad \mathrm{or} \qquad
\epsilon_{ab}=u_av_b-v_au_b\, ,
\end{equation}
it's easy to find
\begin{equation}
\epsilon_{a}{}^{c}\epsilon_{cb}=h_{ab}\, ,\qquad
\epsilon_{ab}\epsilon_{cd}=h_{ad}h_{bc}-h_{ac}h_{bd}\, .
\end{equation}
This also suggests that we can introduce the tensor $\epsilon_{ab}$
without using the local frames. By defining
\begin{equation}
\bar{X}^a=\epsilon^{ab}X_b\, ,\qquad \bar{Y}^a=\epsilon^{ab}Y_b\, ,
\end{equation}
then, we have
\begin{equation}
X^a=\epsilon^{ab}\bar{X}_b\, ,\qquad Y^a=\epsilon^{ab}\bar{Y}_b\,
,\qquad \bar{X}_a\bar{Y}^a=-X_aY^a\, .
\end{equation}
It's also easy to find $\ell_a=\epsilon_{ab}\ell^b$ and
$n_a=-\epsilon_{ab}n^b$. This means:  $\ell_a$ is self-dual, but
$n_a$ is anti-self-dual.
%

\begin{center}
{\bf Appendix C}
\end{center}
From the definition of the covariant derivative $\tilde{D}$, we have
\begin{eqnarray}
&&\tilde{D}_aX_b=h_{b}^{~d}q_a^{~c}\nabla_cX_d=\varepsilon_{IJ}e^I_be^{Jd}q_a^{~c}\nabla_cX_d\nonumber\\
&&=\varepsilon_{IJ}e^I_bq_a^{~c}\nabla_c\left(e^{Jd}X_d\right)-\left(\varepsilon_{IJ}e^I_bq_a^{~c}\nabla_ce^{Jd}\right)X_d\nonumber\\
&&=D_aX^J\varepsilon_{IJ}e^I_b+\omega_{abd}X^d\nonumber\\
&&=D_aX^J\varepsilon_{IJ}e^I_b+\omega_{a}\epsilon_{bd}X^d\, ,
\end{eqnarray}
and then
\begin{eqnarray}
&&\tilde{D}_c\tilde{D}_aX_b=q_c^{~e}q_a^{~f}h_{b}^{~g}
\nabla_e\left(D_fX^J\varepsilon_{IJ}e^I_g+\omega_{f}\epsilon_{gd}X^d\right)\,
,\nonumber\\
&&=\left(D_cD_aX^J\right)\varepsilon_{IJ}e^I_b+
D_aX^J\omega_{c}\epsilon_{bd}\varepsilon_{IJ}e^{Id}\nonumber\\
&&+D_c\omega_{a}\epsilon_{bd}X^d+D_cX^J\omega_{a}\epsilon_{bd}\varepsilon_{IJ}e^{Id}+\omega_a\omega_c\epsilon_{bd}\epsilon^d_{~e}X^e\,
.
\end{eqnarray}
So we get eq.(\ref{YtDtDX}). Here, we have used the fact that
$\tilde{D}_{a}\epsilon_{bc}=0$.

\end{document}